\documentclass[11pt]{report}

\newcommand{\beq}{\begin{equation}}
\newcommand{\eeq}{\end{equation}}
\newcommand{\beqs}{\begin{eqnarray}}
\newcommand{\eeqs}{\end{eqnarray}}

\def\tr{\;{\rm tr}\;}
\def\sgn{\;{\rm sgn}\;}
\newcommand{\dd}[2]{\frac {\partial #1}{\partial #2}}
\newcommand{\pdr}{\partial}

\newcommand{\half}{\frac{1}{2}}
\newcommand{\eps}{\epsilon}
\newcommand{\ov}[1]{\frac{1}{#1}}

\newcommand{\La}{\Lambda}

\newcommand{\masssq}{~(mass)^2~}
\newcommand{\msqovgsq}{{~m^2 \over \tilde g^2~}}

\newcommand{\gr}{grassmannian~}
\newcommand{\z}{\zeta_-}

\newcommand{\W}{\tilde W}

\newcommand{\fl}{\noindent}

\newcommand{\N}{{1 \over N}}\newcommand{\Nsq}{{1 \over N^2}}
\newcommand{\Ntr}{{{\rm tr} \over N}}

\begin{document}
\input{epsf}


\pagenumbering{roman}

\thispagestyle{empty}

\begin{Large}

\begin{center}

{\bf Large$-N$ Limit as a Classical Limit: Baryon in
Two-Dimensional QCD and Multi-Matrix Models}

\vskip 0.3in

by

\vskip 0.3in

{\bf Govind Sudarshan Krishnaswami}

\vskip .8in

Submitted in Partial Fulfillment\\
of the\\
Requirements for the Degree\\
Doctor of Philosophy \vskip 0.2in

Supervised by\\
{\bf Professor Sarada G. Rajeev} \vskip 0.7in

Department of Physics and Astronomy\\
The College\\
Arts \& Sciences \vskip 0.2in

University of Rochester\\
Rochester, New York\\
\vskip 0.1in

2004

\end{center}

\end{Large}

\pagebreak

\pagestyle{myheadings}


\centerline{\Large\bf Curriculum Vitae}
\bigskip

The author was born in Bangalore, India on June 28, 1977. He
attended the University of Rochester from 1995 to 1999 and
graduated with a Bachelor of Arts in mathematics and a Bachelor of
Science in physics. He joined the graduate program at the
University of Rochester as a Sproull Fellow in Fall 1999. He
received the Master of Arts degree in Physics along with the
Susumu Okubo prize in 2001. He pursued his research in Theoretical
High Energy Physics under the direction of Professor Sarada G.
Rajeev.

\pagebreak



\centerline {\Large\bf Acknowledgments} \smallskip

It is a pleasure to thank my advisor Professor S. G. Rajeev for
the opportunity to be his student and learn how he thinks. I have
enjoyed learning from him and working with him. I also thank him
for all his support, patience and willingness to correct me when I
was wrong.

I am grateful to Abhishek Agarwal, Levent Akant, Leslie Baksmaty and
John Varghese for their friendship, our discussions and
collaboration. Leslie's comments on the introductory sections as I
wrote this thesis, were very helpful. I thank Mishkat Bhattacharya,
Alex Constandache, Subhranil De, Herbert Lee, Cosmin Macesanu, Arsen
Melikyan, Alex Mitov, Jose Perillan, Subhadip Raychaudhuri and many
other students for making my time in the department interesting, and
for their help and friendship.

Thanks are also due to the Physics and Astronomy department for all
the support I have received as a student at Rochester. Prof. Arie
Bodek helped with his advice and discussions on comparison of our
results with experimental data. Prof. Nick Bigelow, Ashok Das, Joe
Eberly, Yongli Gao, Richard Hagen, Bob Knox, Daniel Koltun, Susumu
Okubo, Lynne Orr, Judy Pipher, Yonathan Shapir, Ed Thorndike, Steve
Teitel, Emil Wolf, Frank Wolfs and Wenhao Wu have taught me courses
and been generous with their advice and support as have all the
other faculty in our department. I would also like to acknowledge
Prof. Eric Blackman, Tom Ferbel, Steve Manly and Jonathan
Pakianathan, who have been on my exam committees. I have also learnt
from and had useful discussions with many of the faculty in the
mathematics department, including Fred Cohen, Michael Gage, Steve
Gonek, Allan Greenleaf, Naomi Jochnowitz and Douglas Ravenel.

The staff at the Physics and Astronomy department has been a great
help. I would like to thank Connie Jones, Barbara Warren, Judy Mack,
Diane Pickersgill, Shirley Brignall, Sue Brightman, Pat Sulouff,
Marjorie Chapin, Dave Munson, Mary Beth Vogel and all the others for
going out of their way for me.

I would also like to thank my parents Leela and Krishnaswami, my
sister Maithreyi, and my wife Deepa for all their love and support.
The Bartholomeusz', Bhojs, Govenders, Rajamanis, Rajeevs, Savi and
Madu have been family away from home while I was a student at
Rochester.

\pagebreak


\centerline {\Large\bf Abstract}
\smallskip

In this thesis, I study the limit of a large number of colors ($N$)
in a non-abelian gauge theory. It corresponds to a classical limit
where fluctuations in gauge-invariant observables vanish. The
large-dimension limit for rotation-invariant variables in atomic
physics is given as an example of a classical limit for vector
models.

The baryon is studied in Rajeev's reformulation of two-dimensional
QCD in the large-$N$ limit: a bilocal classical field theory for
color-singlet quark bilinears, whose phase space is an infinite
grassmannian. In this approach, 't Hooft's integral equation for
mesons describes small oscillations around the vacuum. Baryons are
topological solitons on a disconnected phase space, labelled by
baryon number. The form factor of the ground-state baryon is
determined variationally on a succession of increasing-rank
submanifolds of the phase space. These reduced dynamical systems are
rewritten as interacting parton models, allowing us to reconcile the
soliton and parton pictures. The rank-one ansatz leads to a
Hartree-type approximation for colorless valence quasi-particles,
which provides a relativistic two-dimensional realization of
Witten's ideas on baryon structure in the $1/N$ expansion. The
antiquark content of the baryon is small and vanishes in the chiral
limit. The valence-quark distribution is used to model parton
distribution functions measured in deep inelastic scattering. A
geometric adaptation of steepest descent to the grassmannian phase
space is also given.

Euclidean large-$N$ multi-matrix models are reformulated as
classical systems for $U(N)$ invariants. The configuration space of
gluon correlations is a space of non-commutative probability
distributions. Classical equations of motion (factorized loop
equations) contain an anomaly that leads to a cohomological
obstruction to finding an action principle. This is circumvented by
expressing the configuration space as a coset space of the
automorphism group of the tensor algebra. The action principle is
interpreted as the partial Legendre transform of the entropy of
operator-valued random variables. The free energy and correlations
in the $N \to \infty$ limit are determined variationally. The
simplest variational ansatz is an analogue of mean-field theory. The
latter compares well with exact solutions and Monte-Carlo
simulations of one and two-matrix models away from phase
transitions.


%
%








\pagebreak


\thispagestyle{myheadings}

\tableofcontents

\thispagestyle{myheadings}

\pagebreak


\thispagestyle{headings}

\listoffigures


\pagebreak


\pagenumbering{arabic}

\pagestyle{myheadings}

\chapter{Introduction}
\label{ch-intro}

\thispagestyle{myheadings}

\markboth{}{Chapter \ref{ch-intro}. Introduction}

\section{QCD and the Large-$N$ Classical Limit}
\label{s-intro-qcd-large-N-classical}


In this section, we motivate and provide the historical context for
the issues studied in this thesis. We use analogies with celestial
mechanics and atomic physics, since they are familiar and more
mature physical theories. However, the analogies are not to be taken
too literally. A summary of the thesis and a short survey of the
literature is also included in this introduction.

\smallskip

\fl {\bf \large The Standard Model: Strong and Electroweak
Interactions}

\smallskip

At the turn of the twenty-first century, the standard model of
particle physics for electroweak and strong interactions is at an
intermediate stage of development, one that is similar to a stage in
mechanics after Newton's force laws had been discovered, but
hamiltonian and lagrangian mechanics were still being developed. The
latter are not merely elegant repackagings of the force laws in
terms of single functions, but provide a way to go beyond the two
body problem in a systematic manner, and passage to continuum
mechanics. Moreover, it is difficult to imagine doing statistical
mechanics without the idea of a hamiltonian. What is more, these
developments played a key role in the discovery of the ``next
theory'', quantum mechanics. Drawing an analogy with atomic physics,
we are at a stage after the discovery of quantum mechanics, but
before its reformulation in terms of path integrals. Path integrals
are indispensable to the theoretical development that followed
quantum mechanics, quantum field theory.

Spectacular experimental discoveries, such as patterns found in the
hadronic spectrum \cite{gell-mann,pdg}, electroweak gauge bosons
\cite{rubbia}, and scaling in deep inelastic scattering
\cite{friedman-kendall}, parallel the discovery of the detailed
orbits of planets and moons, or of discrete atomic spectra. On the
other hand, equally deep theoretical discoveries such as the gauge
principle \cite{yang-mills}, electroweak unification
\cite{weinberg-salam}, asymptotic freedom in non-abelian gauge
theories \cite{asym-freedom}, and perturbative renormalizability of
unbroken and spontaneously-broken gauge theories
\cite{thooft-veltman-renor-ym} remind us of the discovery of the
inverse-square force law between the sun and planets, or of the
uncertainty principle and the Schrodinger equation for electrons in
atoms.

Based on experiment and guiding principles such as gauge invariance
and renormalizability, the lagrangian of the standard model has now
been almost completely defined. The standard model is based on
non-abelian gauge theories (Yang-Mills theories) for the force
carriers: gluons for the strong force, and the $W$ and $Z$
electroweak gauge bosons, which are generalizations of the photon of
electromagnetism. These bosons couple to fermionic matter particles,
the quarks and leptons (which include the electron). Finally, an
untested part of the standard model consists of scalar fields that
are expected to provide spontaneous breaking of the electroweak
gauge symmetry.

Quantum Chromodynamics (QCD) is the unbroken non-abelian gauge
theory describing the interactions of quarks and gluons. In a gauge
theory, the basic gluon and quark fields have some redundant degrees
of freedom that are not observable. The true dynamical degrees of
freedom (the ``hadrons'') are determined by the principle of gauge
invariance. In QCD, electric charge is replaced by color charge of
quarks and gluons. Unlike photons, which are electrically neutral,
gluons carry color charge. QCD also has the property of asymptotic
freedom which, at high momentum transfers, suggests that quarks and
gluons behave as free particles (up to logarithmic corrections).
This is in contrast to the electrostatic force between charges or
the gravitational force between planets and the sun, which grow
stronger at short distances. On the other hand, the strong force
does not fall off at large distances, but leads instead to the
phenomenon of confinement. In fact, the ``fundamental particles'',
quarks and gluons, have never been isolated. Unlike the electron,
which can be removed from an atom by ionization, quarks and gluons
seem to be confined inside (colorless, gauge-invariant) bound states
of the strong interactions. These are called mesons (such as the
pion) and baryons (such as the proton and neutron). In addition,
there are glueballs, which are gauge-invariant bound-states of
gluons alone. However, there is no quantum number that distinguishes
glueballs from mesons. These are collectively referred to as
hadrons.

\smallskip

\fl {\bf \large The success of perturbation theory around the
$\hbar \to 0$ classical limit}

\smallskip

Non-abelian gauge theories, being quantum field theories, can be
illuminated through perturbation around a classical limit. A
classical limit is one in which the fluctuations in {\it some}
observables vanish. In the limit of vanishing Planck's constant, the
fluctuations in {\em all} observables of a gauge theory, not just
those that are gauge-invariant, vanish. So it is not surprising that
the loop expansion around the free-field classical limit of $\hbar
\to 0$ (``perturbation theory'') has been tremendously successful in
making quantitative predictions about electroweak interactions. The
latter are not just weak at currently accessible energies, but also
involve a spontaneously-broken gauge symmetry. Thus, {\em all}
states, not just the gauge-invariant ones, and in particular, the
gauge bosons $W^\pm, Z^0$, are observable.

As for the strong interactions, because of asymptotic freedom,
perturbation theory makes accurate predictions about some aspects of
their high-energy behavior. In particular, logarithmic scaling
violations, and other phenomena where quarks and gluons can be
treated as almost free, are well described by perturbation theory
around the $\hbar \to 0$ limit. Indeed, QCD is the only known
renormalizable four-dimensional theory with the correct high-energy
behavior. This is primarily why it is believed to be the correct
theoretical model for strong interactions. (The quantum theory can
also be studied around non-trivial classical solutions in the $\hbar
\to 0$ limit, such as solitons and instantons \cite{rajaraman}).

\smallskip

\fl {\bf \large Why study QCD? What is left to do?}

\smallskip

The discovery of the basic lagrangian, and the verification of its
predictions for short distances, is only the first step in
understanding dynamics. The finite-time and asymptotic behavior of
solutions of the differential equations of classical mechanics hold
a lot of information and surprises not evident over short times. In
celestial mechanics, elliptical orbits of planets could be derived
easily from Newton's equations of motion. But the effects of Jupiter
on the motion of other planets, or the estimation of the perihelion
shift of Mercury, required reformulations of the theory and the
development of new approximation methods, such as classical
perturbation theory. These eventually led to the discovery of
general relativistic corrections to newtonian gravity. Another
example is the study of the stability of the solar system, which in
turn led to the development of chaotic dynamics and
Kolmogorov-Arnold-Moser theory \cite{arnold}.

We have a similar situation in QCD. Perturbation theory around the
trivial vacuum in the $\hbar \to 0$ limit allows us to examine the
strong interaction at only short distances, where quarks and gluons
are almost free. This tells us very little about the long-distance
behavior or about bound states of quarks and gluons. In fact, the
problem is quite severe, since {\em all} observed particles are
hadronic bound states. A mechanism that explains both qualitatively
and quantitatively the confinement of quarks and gluons within
hadrons has not been found. The empirically deduced long-range
linear potential between quarks has also not been derived from QCD.
How quarks and gluons bind to form hadrons, in other words the
structure functions of these bound states (such as the proton),
cannot be understood by merely perturbing around the vacuum. Though
gluons are massless, the lightest observed particle of the theory is
massive. Understanding this mass gap in the hadronic spectrum is an
outstanding challenge. There are other peculiarities in the mass
spectrum of hadrons. For example, the linearly rising Regge
trajectories, where the angular momenta of hadronic resonances
(short-lived particles) are linearly related to the squares of their
masses, suggests that quarks are held together by a string with
constant tension \cite{greenschwarzwitten}. Chiral symmetry-breaking
is yet another phenomenon one would like to establish in QCD.
Perhaps its most dramatic manifestation is that practically massless
quarks (up and down, $5-7$ MeV) bind to form the proton, which is
over a hundred times as heavy (938 MeV). The situation is such that,
even though QCD is believed to be the correct theory, it has not
been possible, aside from large-scale numerical simulations and
certain non-relativistic situations involving mesons with two heavy
quarks, to calculate the mass or wave function (structure function)
of even a single observed particle in the theory! However, there are
many simplified models such as dimensional reductions, strong
coupling expansions, low energy effective theories or supersymmetric
generalizations that exhibit some of these features. One hopes that
further progress can be made by combining new physical and
mathematical ideas. Of course, there is a wealth of experimental
data \cite{pdg} and increasingly accurate numerical simulations (see
for example \cite{creutz,teper}) with which to compare. Finally, we
can hope that a deeper understanding of QCD will lead us to physical
principles and mathematical structures, that are needed for any
future physical theory that observation may require us to invent.
For example, a deep appreciation for the hamiltonian and Poisson
bracket formulation of classical mechanics led Dirac to invent the
canonical formalism, which is more convenient for quantum field
theory than the original Schrodinger equation of single particle
quantum mechanics.

\smallskip

\fl {\bf \large Alternative Classical Limits}

\smallskip

Quantum theories often have more than one classical limit in which
some observables do not fluctuate. Usually, each is characterized by
assuming a limiting value for some parameter. Different classical
limits are useful for formulating different phenomena characterizing
the same underlying quantum theory. For instance, in atomic physics,
the $\hbar \to 0$ classical limit gives no indication of why the
atom is stable. However, there are other classical limits in atomic
physics. The limit of a large number of dimensions is a classical
limit where rotationally invariant variables do not fluctuate. This
limit provides a ``classical'' explanation for why the electron does
not fall to the minimum of the Coulomb potential (see \S
\ref{s-d-to-infty-atom}). Atomic Hartree-Fock theory is yet another
classical limit that provides a way of studying many-electron atoms
\cite{rajeev-2dqhd}.

QCD has a different classical limit from $\hbar \to 0$. This is the
limit of a large number of colors $N$, where the structure group of
the gauge theory is $SU(N)$. There appear to be three colors in
nature. Fluctuations in gauge-invariant observables {\it alone}
vanish in the large-$N$ limit. Thus, we should expect the large-$N$
classical limit to be a better approximation than the $\hbar \to 0$
limit for unbroken gauge theories with strongly-coupled gauge
fields, where it is only the gauge invariant states that are
observed. This is the regime of interest for the strong-interaction
phenomena mentioned previously. The large-$N$ limit, as an
approximation for non-abelian gauge theories, was originally
proposed by 't Hooft in a perturbative context \cite{thooft-planar}.
In this limit, planar Feynman diagrams were found to dominate. 't
Hooft calculated the spectrum of mesons in the large-$N$ limit of
two-dimensional QCD by summing such planar diagrams
\cite{thooft-2d-meson}, and found an infinite tower of bound states,
and the analogs of linearly-rising Regge trajectories. The planar
diagrams also suggested a connection to a string model of hadrons
which was expected from the linearly rising potential and Regge
trajectories. We will see later that summing such planar diagrams
only gives the linear approximation around the vacuum of the
large-$N$ limit. Nevertheless, several other empirical facts about
the strong interaction also seemed likely to be accommodated in the
large-$N$ limit (see Witten \cite{wittenN}). For example, Zweig's
rule, an empirical deduction concerning, for instance, the mixing of
mesons with glue or exotic states, is exact in the linear
approximation to the large-$N$ limit where mesons are pure $q \bar
q$ states \cite{wittenN}. Thus, there are theoretical and
phenomenological reasons to expect that $N = \infty$ is a good
starting point for an approximation to QCD, though only $N = 3$
corresponds to nature.

Despite the success of the diagrammatic point of view, it obscures
the classical nature of the large-$N$ limit. Though it is believed
that the large-$N$ limit of QCD is a classical limit, its
formulation as a classical dynamical system is not well understood.
In particular, we would like to identify the classical configuration
space, the equations of motion and action or hamiltonian, and also
develop the approximation methods required to understand the theory.
In this thesis, we present some investigations of the large-$N$
limit as a classical limit for two simplified models of QCD: the
quark structure of baryons in two-dimensional QCD, and matrix models
as a simple model for gluon structure.

\smallskip

\fl {\bf \large More on why QCD is difficult: Divergences and
Gauge Invariance}

\smallskip

Three of the many aspects that make QCD hard to understand are the
large number of degrees of freedom, gauge invariance, and
ultraviolet divergences. As currently formulated, QCD is a divergent
quantum field theory. The only known way to make finite predictions
in four space-time dimensions, aside from the numerical lattice-QCD
approach, is by the procedure of perturbative renormalization
\cite{thooft-veltman-renor-ym}. While very successful, this
procedure is inherently tied to the perturbative solution of the
theory. Rather than address the important question of finding an
alternative finite formulation of the theory, we will work with
regularized versions of the theory. One such regularization is a
matrix model, which reduces the number of degrees of freedom. We
will also work with a reduction to two dimensions, where the theory
is ultraviolet finite without any need for regularization. This will
allow us to focus attention on the difficulties arising from gauge
invariance.

In a gauge theory, we have the unusual situation where a physical
theory is formulated in terms of unobservable particles, the quarks
and gluons. To accommodate the properties of the observed hadrons,
it is then necessary to rewrite the theory in terms of
gauge-invariant observables. The principle of gauge invariance is no
longer relevant when we work with a gauge-invariant formulation of
the theory. We must look for other geometric, probabilistic, or
algebraic principles that play as important a role. This is still a
formidable task in the full $3+1$ dimensional theory. So we address
this question in the simpler contexts mentioned above. Quarks
transform as $N$-component vectors under color, the $SU(N)$
structure group, while gluons transform in the adjoint
representation as $N \times N$ hermitian matrices. The components of
these vectors and the matrix elements are the so-called ``gauge
degrees of freedom'' (quarks and gluons) that carry the color
quantum number, and are not directly observable. Only
color-invariant combinations are observable. Furthermore,
gauge-invariant observables (the Wilson loop and meson observables)
are non-local, which means that we need to make the passage from
local gauge fields to non-local loop or string-like variables.

\smallskip

\fl {\bf \large The Baryon in the Large-$N$ Limit of 2d QCD}

\smallskip

Vector models have fewer degrees of freedom, making them easier to
deal with than matrix models. We therefore devote the first half of
this thesis to the vector model of quarks in two-dimensional QCD,
interacting via a linear potential due to longitudinal gluons. 't
Hooft's work on two-dimensional QCD left a puzzle as to how to
handle baryons. An early proposal of Skyrme was that baryons must
arise as solitons in a theory of mesons \cite{Skyrme3}. However, 't
Hooft's equation for mesons was linear, and did not support soliton
solutions. Rajeev discovered a bilocal (along a null line)
reformulation of two-dimensional QCD in terms of color-invariant
quark bilinears. In the large-$N$ classical limit, it is a
hamiltonian dynamical system whose phase space is an infinite
dimensional grassmannian manifold. The phase space is a curved
manifold because the quark density matrix is a projection operator,
making this theory strongly interacting even in the large-$N$ limit.
This is an indication of the type of geometric ideas that may play a
role in a gauge-invariant reformulation of QCD. 't Hooft's meson
spectrum is recovered as a linear approximation to the equations of
motion on the grassmannian around the naive vacuum. Moreover, the
phase space is disconnected, with components labelled by an
integer-valued baryon number. The baryon arises as a topological
soliton, and we study this in this thesis. In particular, we
determine the ``shape'', or more precisely, the form factor, which
tells us, roughly, how quarks are distributed inside the proton,
though in a color invariant manner, without explicit reference to
color carrying quarks.

We also examine another puzzle regarding the structure of the
proton: the relation between the soliton and parton pictures. The
parton model is a complementary view to the soliton picture of the
baryon \cite{feynman-photon-hadron,bjorken-paschos}. Deep inelastic
scattering experiments \cite{friedman-kendall} indicate that the
constituents of the proton, called partons, are point-like.  This is
because the proton structure function is roughly scale invariant,
i.e., independent of the length scale at which it is observed. This
was accommodated in the parton model by postulating that the proton
is made of point-like partons. The latter had to be non-interacting
in order to match the observed high-energy behavior. Partons were
identified with quarks and gluons after the discovery of QCD. Their
non-interacting behavior at high energies was regarded as a
consequence of their asymptotic freedom. The QCD-improved parton
model (perturbative QCD) made accurate predictions for how scale
invariance is broken, i.e., the logarithmic scale dependence of
structure functions. However, the dependence of structure functions
on the momentum fraction carried by a parton, which is akin to an
atomic wave function, is non-perturbative and remained inaccessible.
In Part \ref{part-baryon-large-N} of this thesis, we derive an
interacting parton model from the solitonic point of view, thereby
reconciling the two disparate points of view in two space-time
dimensions. We also determine the non-perturbative momentum-fraction
dependence of the quark structure function within two-dimensional
QCD. This two-dimensional theory provides a good approximation to
Deep Inelastic Scattering, and we use it to model the quark
distributions measured in experiment (\S \ref{s-appl-to-DIS}).

\smallskip

\fl {\bf \large Large-$N$ Matrix Models}

\smallskip

Our analysis of quark structure still leaves the gluon distribution
undetermined. Gluons are the force carriers between quarks, just as
photons transmit the force between electric charges in QED. However,
QED is an abelian gauge theory while QCD is a non-abelian gauge
theory, leading to strong self interactions between gluons. One
manifestation of this is in the ability of gluons (unlike photons)
to form hadronic bound states, referred to as glueballs. The
non-abelian nature of gluons also leads to another phenomenon that
makes the strong interactions very different from the quantum
mechanics of an atom. The electron wave function determines the
shape of an atom, and, within the non-relativistic theory, the
photon does not carry any momentum of an atom. But for a proton,
when the momentum transferred by the probe is about $1$ GeV, it is
an experimental fact that quarks and gluons carry roughly equal
portions of its total momentum \cite{cteq}. The contribution from
gluons only grows with the probe's momentum transfer \cite{cteq}.

Of the four possible states of polarization of the gluon in four
dimensions, only two are dynamical i.e. its transverse states. The
longitudinal and time components can be eliminated by gauge fixing
and solving the resulting constraint equation. Within
two-dimensional QCD, there are no transverse gluons, and they were
ignored previously. Now we turn to dynamical gluons, namely, the
propagating gluon degrees of freedom which cannot be eliminated by
gauge fixing. In Part \ref{part-large-N-mat-mod-class-systems} of
this thesis, we study euclidean multi-matrix models in the large-$N$
classical limit. They are a regularized version of gluon dynamics.
Understanding gluon dynamics is much harder than quark dynamics. The
reason is that there are many more gauge-invariant gluon observables
than pure quark observables. The dot products of pairs are the only
invariants of a set of vectors, while the traces of arbitrary
products of a collection of matrices (the gluon correlation tensors)
are all unitary invariants.

It is believed that the large-$N$ limit of matrix models comprises a
classical theory. In this thesis, we identify this classical theory
in a manifestly unitary invariant manner. We show that the
configuration space is a space of non-commutative probability
distributions. The gluon field is an $N\times N$ matrix-valued
random variable. These matrices at different points of space-time do
not commute, and as a consequence we are dealing with a
non-commutative version of probability theory (Chapter
\ref{ch-free-alg-non-comm-prob}). The coordinates on the
configuration space are gluon correlations. Their fluctuations
vanish in the large-$N$ limit, and satisfy the factorized
Schwinger-Dyson (or loop) equations. We identify these as the
classical equations of motion (\S \ref{s-loop-eq-class-eq-mot}).
However, a classical action is very elusive. There is an anomaly in
the equations arising from the transformation from matrix elements
to invariants, manifested as a cohomological obstruction to finding
an action on the configuration space (\S
\ref{s-anomaly-as-cohomology}). We circumvent this obstruction by
expressing the configuration space as a coset space of a
non-commutative analogue of the diffeomorphism group by an isotropy
subgroup. We then find a classical action on the group that is
invariant under the action of the isotropy subgroup. So, the action
really lives on the quotient, i.e., the configuration space (\S
\ref{s-config-space-coset-space}). Our search for principles that
determine a gauge-invariant formulation of large-$N$ matrix models
leads us to the automorphism group of the free algebra and its first
cohomology!

What is more, we find that the entropy of non-commutative
probability theory, developed as a branch of operator algebra by
Voiculescu and collaborators \cite{Voiculescu}, plays a central
role. Whenever we restrict the allowed observables of a physical
system, we should expect an entropy due to our lack of knowledge of
those observables. For example, entropy in statistical mechanics
arises because we do not measure the velocities of individual gas
molecules but only macroscopic variables such as pressure and
density. Similarly, confinement of the color degrees of freedom
should lead to an entropy in the strong interactions. In a matrix
model, expressing the action in terms of unitary invariant gluon
correlations rather than unobservable matrix elements, leads to an
entropy. We find (\S \ref{s-chi-as-entropy}) a variational principle
for the solution of the classical equations of motion for gluon
correlations by maximizing entropy subject to certain constraints.

\smallskip

\fl {\bf \large Approximation Methods}

\smallskip

Progress in newtonian mechanics came about through at least two
ways: (1) general approximation methods, or alternative formulations
of the theory, and (2) exact solutions of some special systems.
Examples of (1) are hamiltonian and lagrangian mechanics,
Hamilton-Jacobi theory, perturbation theory and variational
principles. Examples of (2) are Jacobi's solution of the rigid body,
and the theory of integrable systems.

A common theme throughout this thesis is the use of approximation
methods to solve the classical theories we get in the large-$N$
limit. A classical limit is itself an approximation. However, both
the classical theories we get, dynamics on an infinite grassmannian
for quarks in two-dimensional QCD, and the maximization of entropy
in multi-matrix models, are highly non-linear and non-local
classical theories. They require the development of new
non-perturbative approximation methods. Here, we take inspiration
from the approximation methods of classical mechanics, atomic
physics, and many-body theory. We look for analogs of variational
principles, mean-field theory, Hartree-Fock theory, steepest
descent, and the loop expansion.

Finding the ground-state of the baryon involves minimizing its
energy on an infinite dimensional curved phase space. The first
method we develop is a geometric adaptation of steepest descent for
a curved phase space (\S \ref{s-steepest-descent}). Then we find a
variational approximation method that replaces the full phase space
with finite dimensional submanifolds, and we study the dynamical
system on reduced phase spaces (\S \ref{s-sep-ansatz-formulation},
\S \ref{s-beyond-sep-ansatz}). In the simplest case, we get an
analogue of mean-field theory (Hartree-Fock theory) for a system of
interacting colorless quasi-particles. This is how we are able to
derive the interacting valence parton picture from the exact soliton
description of the baryon (\S \ref{s-quantize-rank-1}, \S
\ref{s-hartree-approx}). We also show how to go beyond this, and
include anti-quarks (\S \ref{s-rank-three-ansatz}). Though the
emphasis is on approximate solutions, along the way we also find the
exact form factor of the baryon in the large-$N$ limit of
two-dimensional QCD for massless current quarks (see
(\ref{e-exact-exp-sol-for-M})). We compare our approximate solution
for the quark distribution function with numerical calculations
\cite{hornbostel} and also measurements from Deep Inelastic
Scattering, and find good agreement (see \S \ref{s-appl-to-DIS}).

Since the 1980s significant progress was made in finding exact
solutions for partition functions and special classes of
correlations of carefully chosen matrix models, such as two matrix
models with specific interactions, matrix chains associated with
Dynkin diagrams of simply-laced Lie algebras etc. The methods often
originated in the theory of integrable systems or conformal field
theory (see for example
\cite{mehtaAB,mehta-book,Staudacher-two-mat,kazakovABAB,eynard-two-mat}).
However, there is a lack of approximation methods to handle generic
matrix models in the large-$N$ limit. Even the analogs of simple
methods such as mean-field theory and variational principles were
previously not known. We find a variational principle that allows us
to determine the gluon correlations from a finite-parameter family
that best approximate the correlations of a given matrix model (\S
\ref{s-classical-action-ppl}). We use this variational principle to
find an analog of mean-field theory for large-$N$ multi-matrix
models (\S \ref{s-var-approx-mat-mod}), and also indicate how one
can go beyond mean-field theory (\S
\ref{a-non-lin-change-of-var-1-mat}). These approximation methods
compare favorably with exact solutions, and with a Monte-Carlo
simulation away from divergences in the free energy (\S
\ref{s-mehta-model}, \S \ref{s-ym-2-mat}). For other approaches to
solving matrix models, see for instance the work of J. Alfaro et.
al. \cite{alfaro}.

\smallskip

\fl {\bf \large Literature on Matrix Models in High Energy Physics}

\smallskip

Since the 1970s, there has been a great deal of work done on
large-$N$ matrix models. Some of the earlier papers are reproduced
in the collection of Ref. \cite{brezin-wadia}. We mention a few of
the many developments. Brezin, Itzykson, Parisi and Zuber studied
the euclidean one-matrix model using the saddle point method in the
large-$N$ limit. They found an important relation between the
quantum mechanics of a single matrix in the large-$N$ limit and a
system of free fermions \cite{brezin-et-al}. Migdal and Makeenko
found that the Wilson loops of a large-$N$ gauge theory satisfy a
closed set of ``factorized loop'' equations \cite{migdal-makeenko}.
Yaffe's coherent states approach \cite{Yaffe}, and Sakita's and
Jevicki's \cite{sakita-book,jevicki-sakita} work on the collective
field formalism of large-$N$ field theories was an important step in
understanding an anomaly in the hamiltonian of the large-$N$ limit.
The anomaly is one of the main differences relative to the $\hbar
\to 0$ classical hamiltonian. Our recent work shows that this
anomaly is in fact the non-commutative analogue of Fisher
information of probability theory \cite{information}. The papers of
Cvitanovic and collaborators \cite{Cvitanovic,Cvitanovic-et-al} on
planar analogues of some of the familiar methods of field theory,
but with non-commutative sources was helpful in our algebraic
formulation of the problem. Eguchi's and Kawai's \cite{eguchi-kawai}
proposal on reducing a matrix field theory to a matrix model with a
finite number of degrees of freedom, but in the large-$N$ limit, has
been a recurring theme ever since.

Important breakthroughs in the study of random surfaces,
two-dimensional string theory, and two-dimensional quantum gravity
coupled to matter, were made after the mid 1980s (see Ref.
\cite{ginsparg-moore} for a review). The planar Feynman-graph
expansion of large-$N$ matrix models was used as a way of
discretizing a two-dimensional surface. Models with one or a finite
number of matrices, and the $c=1$ quantum mechanics of a single
matrix, were of importance in these developments. The double scaling
limit was developed to study surfaces obtained in the continuum
limit. In the double scaling limit, the coupling constants are tuned
to critical values as $N \to \infty$. This limit is not a classical
limit, unlike the 't Hooft large-$N$ limit. Fluctuations in
observables remain large in the double scaling limit. However, the
double scaling limit allows one to include contributions from all
genera in the topological expansion of Feynman diagrams.

The work of Seiberg and Witten in the early 1990s on
electric-magnetic duality allowed the elucidation of vacuum
structure of a large class of supersymmetric gauge theories along
with the mechanisms of chiral symmetry-breaking and confinement in
these cases \cite{seiberg-witten}.

In the mid 1990s, supersymmetric matrix models were proposed as
non-perturbative definitions of M-theory and superstring theory
\cite{BFSS,IKKT}. From the late 1990s onwards, there has been a
great deal of work on the large-$N$ limit of supersymmetric gauge
theories, catalyzed by the AdS/CFT correspondence of Maldacena
\cite{maldacena}. Large-$N$ matrix models are also used to study the
effective super potentials for ${\cal N} = 1$ supersymmetric gauge
theories with adjoint chiral super fields, following the work of
Dijkgraaf and Vafa \cite{dijkgraaf-vafa,chiral-rings}. Matrix models
also find applications to the problem of determining the anomalous
dimensions of operators in ${\cal N} = 4$ supersymmetric Yang-Mills
theory \cite{minahan-zarembo,abhi-raj-sym-mat-mod}.

Random-matrix theory has also been applied to the spectrum of the
QCD Dirac operator (see especially the work of Verbaarshot et. al.
\cite{verbaarschot}).

\smallskip

\fl {\bf \large Random matrices in other areas of Physics and
Mathematics}

\smallskip

Remarkably, random matrices have found applications in many areas of
physics and mathematics outside of particle and high-energy physics.
We list a few of them.

Random matrix theory originally arose from the suggestions of
Wigner and Dyson in the 1950s, that the statistical properties of
the spectra of complicated nuclei could be modelled by a random
hamiltonian \cite{mehta-book}.

Spin systems on random two-dimensional lattices have been studied
using the large-$N$ limit of matrix models. For example, Kazakov
studied the Ising model on a random two-dimensional lattice with
fixed coordination number \cite{Kazakov-rand-ising}.

Random matrices also have deep connections to statistical properties
of zeros of the Riemann zeta function \cite{mehta-book}. Montgomery
and Dyson discovered that the pair correlation of scaled zeros of
the Riemann zeta function is asymptotic to that of eigenvalues of a
large random unitary matrix \cite{montgomery-dyson}. More recently,
the universal part of the moments of the zeta function on the half
line have also been found to be related to those of the
characteristic polynomial of a large random unitary matrix
\cite{keating}.

Some chaotic quantum systems have been modelled by universal
properties of large-$N$ matrix models, as have the universal
correlations in some mesoscopic and disordered systems
\cite{quant-chaos,brezinzee-univ}.

The work of mathematicians, including Voiculescu and collaborators,
on von Neumann algebras led to the development of the field of
non-commutative probability theory \cite{Voiculescu}. The
correlations of large-$N$ matrix models give natural examples of
non-commutative probability distributions.

\section{QCD, Wilson Loop, Gluon Correlations}
\label{s-qcd-wilson-loop-gluon-correlations}


\fl {\bf \large Classical Chromodynamics}

\smallskip

Classical Chromodynamics (the $\hbar \to 0$ limit of QCD) in $3+1$
space-time dimensions is  a non-abelian gauge theory with structure
group $SU(N)$, where the number of colors is $N = 3$ in nature. The
matter fields $q_\alpha^a(x)$ are quarks, spin $\half$ fermions
transforming under $N_f$ copies of the fundamental representation of
$SU(N)$. `$\alpha$' is a flavor index and `$a, b$' are color
indices. Quarks come in $N_f$ flavors, where $N_f = 6$ (up, down,
strange, charm, bottom and top in order of increasing ``current''
quark mass; though quarks have not been isolated and ``weighed'',
one can define their mass using their interactions with electroweak
currents). At low energies, only up and down are important for the
proton and neutron. For the most part, we will ignore flavor
dependence in this thesis. The bosonic gauge (gluon) fields
$[A_\mu(x)]^a_b,~ \mu = 0,1,2,3$ are four $N \times N$ hermitian
matrix-valued fields. They are the components of a connection
one-form on Minkowski space-time ${\mathrm R}^{3,1}$, valued in the
Lie algebra of $SU(N)$. Let ${\cal A} = \{A_\mu(x)\}$ denote the
space of connections. The theory is defined by the action
    \beq
        S_0 = -{N \over 4 g^2} \int d^4x \tr F^{\mu \nu} F_{\mu \nu}
        + \sum_{\alpha=1}^{N_f} \int \bar q^a_\alpha [-i \gamma^\mu [D_\mu]^b_a
        - m_\alpha \delta^b_a ]q_{\alpha b} d^4x
    \eeq
$\gamma^\mu$ are the Dirac gamma matrices. Here the Yang-Mills
field strength is the anti-hermitian matrix field $F_{\mu \nu}(x)
= \pdr_\mu A_\nu - \pdr_\nu A_{\mu} + [A_\mu,A_\nu]$. The
covariant derivative is $ D_\mu = ({\bf 1} \pdr_\mu - i A_\mu)$
where ${\bf 1}$ is the $N \times N$ unit matrix. The coupling
constant $g$ is dimensionless in $3+1$ dimensions. The classical
solutions are field configurations $(A_\mu(x),~ q(x), \bar q(x))$
that extremize the action, the solutions of the partial
differential equations ${N \over g^2} D_\mu F^{\mu \nu} = {N \over
g^2} (\pdr_\mu F^{\mu \nu} - i [A_\mu,F^{\mu \nu}]) = j^\nu.$ Here
$[j^\nu]^a_b = \bar q^a \gamma^\nu q_b$ is the quark current.

Notice that the action $S_0$ is Lorentz invariant. However, not
all observables need to be Lorentz invariant. Lorentz
transformations (which are isometries of Minkowski space
preserving the metric $ds^2 = (dx^0)^2 - (dx^1)^2 + (dx^2)^2 +
(dx^3)^2$) merely relate observables in different reference
frames.

Now, the action $S_0$ is also invariant under the group of local
gauge transformations ${\cal G} = \{U(x)\}$
    \beq
        A_\mu(x) \mapsto U A_\mu U^{-1} + U \pdr_\mu
        U^{-1}; ~~~~
        q(x) \mapsto U(x) q(x)
    \eeq
where $U(x)$ is a map from space-time to the structure group $SU(N)$
that tends to the identity at infinity. The principle of gauge
invariance states that not just the action, but every observable of
the theory must be invariant under gauge transformations! The most
famous gauge-invariant observable is the parallel transport along a
closed curve $\gamma:[0,1] \to {\mathrm R}^{3,1}; ~ \gamma(0) =
\gamma(1)$, the so-called Wilson loop observable
    \beq
        W(\gamma) = \Ntr {\cal P} \exp\bigg\{-i \oint_0^1 A_\mu(\gamma(s))
        ~\dot \gamma^\mu(s)~ ds \bigg\}
    \eeq
where ${\cal P}$exp stands for the path ordered exponential. One
can also consider an open string or meson observable with quarks
at the end points (here $\gamma$ is not a closed curve)
    \beq
        M(\gamma) = \ov{N} \bar q(\gamma(0))^a \bigg[{\cal P}
            \exp\bigg\{-i \int_0^1 A_\mu(\gamma(s))
        ~\dot \gamma^\mu(s)~ ds \bigg\}\bigg]^b_a q(\gamma(1))_b.
        \label{e-meson-observables}
    \eeq
We shall give the physical interpretation of the Wilson loop when
we discuss its expectation value in the quantum theory.

\smallskip

\fl {\bf \large Quantum Chromodynamics}

\smallskip

So far, we have been discussing Classical Chromodynamics. In the
path integral approach to quantization, $(A_\mu(x),~ q(x), ~ \bar
q(x))$ become random variables. Quantum Chromodynamics (QCD) is the
assignment of expectation values to gauge-invariant functions of
these random variables. Naively, they are obtained by averaging over
the quark and gluon fields with a weight given by $e^{iS_0/\hbar }$.
However, on account of the gauge invariance of the action, this
functional integral is ill-defined. Rather than average over the
entire space of connections ${\cal A}$, we should be averaging over
the space of connections modulo gauge transformations ${\cal A / G}$
with the measure induced by the Lebesgue measure on ${\cal A}$.
${\cal A / G}$ parameterizes the true degrees of freedom according
to the gauge principle. The idea is to choose a representative for
each orbit of ${\cal G}$ in ${\cal A}$ (gauge fixing) and integrate
over the coset representatives. In effect, the gauge field has only
two independent components (for example, the transverse polarization
states) after taking into account the relations imposed by gauge
transformations. However, in order to maintain manifest Lorentz
covariance, it is sometimes more convenient to retain all four
components of the gauge field, and introduce Fadeev-Popov ghost
fields (a pair of grassmann-valued hermitian matrix fields,
$c^a_b(x), \bar c^a_b(x)$) that act as negative degrees of freedom.
The standard implementation of this idea \cite{itzykson-zuber} in
the so-called covariant gauges leads to the gauge fixed action
    \beq
        S(A,q,\bar q,c,\bar c) = S_0 + \int d^4x \bigg[
        - \ov{2\xi} \tr (\pdr^\mu A_\mu(x))^2 + \tr \pdr_\mu \bar c(x) D^\mu c(x)
        \bigg]
        \label{e-gauge-fixed-action}
    \eeq
The two additional terms in the action come respectively from the
gauge fixing and the Fadeev-Popov determinant, which is the jacobian
determinant for the induced measure. The expectation values of
gauge-invariant observables ${\cal O}(A,q,\bar q)$ are independent
of the gauge parameter $\xi$ and are given by
    \beq
        \langle {\cal O}(A,q,\bar q) \rangle = {\int d(A,q,\bar q,c,\bar
        c) e^{i S/ \hbar}~ {\cal O}(A,q,\bar q) \over \int d(A,q,\bar q,c,\bar
        c) e^{i S/ \hbar}}
        \label{e-gauge-fixed-expectations}
    \eeq
For example, if ${\cal O}$ is the Wilson loop observable, then
    \beq
        \langle W(\gamma) \rangle = \sum_{n=0}^\infty {(-i)^n \over n!}
        G_{\mu_1 \cdots \mu_n} (\gamma(s_1), \cdots, \gamma(s_n))
        ~
        \theta(0 \leq s_1 \leq s_2 \leq \cdots \leq s_n \leq 1)
    \eeq
where the gluon correlation tensors are
    \beq
        G_{\mu_1 \cdots \mu_n}(x_1, \cdots, x_n) = \langle\Ntr
        A_{\mu_1}(x_1) \cdots A_{\mu_n}(x_n) \rangle
    \eeq
Though the gluon correlation tensors are not gauge-invariant, once
they have been determined in any specific gauge, they may be used to
compute the expectation values of other gauge-invariant quantities
such as the Wilson loop. Since the Wilson loop $W(\gamma)$ and meson
observables $M(\gamma)$ are gauge invariant, their expectation
values can be calculated in any convenient gauge.

The gauge fixed action (\ref{e-gauge-fixed-action}), is no longer
invariant under local gauge transformations, but rather under
global unitary transformations $q(x) \mapsto U q(x),~ \bar q(x)
\mapsto \bar q(x)U^\dag$, $A_\mu(x) \mapsto U A_\mu(x) U^\dag$, $~
c(x) \mapsto U c(x) U^\dag,~ \bar c(x) \mapsto U \bar c(x)
U^\dag$. In other words, we have a matrix field theory (of both
bosonic ($A_\mu$) and fermionic ($c,\bar c$) adjoint fields)
coupled to a vector model. As it stands, this definition leads to
ultra-violet divergent expectation values in $3+1$ dimensions, and
we have to supplement this with rules for regularization and
renormalization. In particular, the coupling ``constant'' is
replaced by a ``running coupling constant'' $g^2(Q^2)$, which
depends on the momentum scale $Q$. For large $Q^2$, we have the
perturbative result that the coupling vanishes logarithmically:
$g^2 \sim 1/{\log({Q^2 / \Lambda_{QCD}^2})}$. Renormalization
introduces the dimensional parameter $\Lambda_{QCD}$ which sets
the scale for the strong interactions and is to be determined
experimentally ($\Lambda_{QCD} \sim 200$ MeV).

In this thesis, we study a pair of finite truncations of QCD. In
Part I of this thesis we will study the vector model alone, though
in two dimensions and in the null gauge where there are no ghosts,
gluons can be eliminated and there are no ultra-violet
divergences. In Part II we will study hermitian bosonic matrix
models, which are matrix field theories where space-time has been
regularized to have only a finite number of points.

\smallskip
\fl {\bf \large Wilson's Area Law Conjecture:} We conclude this
section with Wilson's area law conjecture on the asymptotic
expectation value of Wilson loop observables:
    \beq
    \langle W(\gamma) \rangle \sim e^{-\alpha'
    {\rm Ar}(\gamma)} {\rm ~as~} {\rm Ar}(\gamma) \to \infty
    \eeq
where ${\rm Ar}(\gamma)$ is the minimal area of a surface whose
boundary is $\gamma$. This is a statement about the vacuum of the
pure gauge theory with no dynamical quarks. However, we can
heuristically interpret the conjecture as an asymptotically linear
potential between quark sources. To see this, consider a current
$j^\mu(x)$ which is concentrated on the curve $\gamma^\mu(s)$. It
is regarded as the current density of a quark-antiquark pair that
are produced and annihilated and whose combined trajectory is the
closed curve $\gamma$. Then
    \beq
        \oint A_\mu(\gamma)~ \dot \gamma^\mu(s)~ ds =
            \int A_\mu(x) j^\mu(x) d^4 x
    \eeq
and $\langle W(\gamma) \rangle$ is the probability amplitude for
this process. In particular, consider a rectangular loop in the
$x^1-x^0$ plane with length $L$ and time $T$. Suppose further, that
the potential energy between quarks is asymptotically linear, $E
\sim \alpha' L$. $\alpha'$ is called the string tension since the
linear potential corresponds to quarks being held together by a
string with constant tension. Then the amplitude for this process is
$e^{-ET} \sim e^{-\alpha' LT} = e^{-\alpha' Ar(\gamma)}$. Thus, the
Wilson area law conjecture is a criterion for confinement of
non-dynamical quarks by an asymptotically linear potential.

\section{$d \to \infty$ as a Classical Limit in Atomic Physics}
\label{s-d-to-infty-atom}


To motivate the idea of the large-$N$ limit as a classical limit of
matrix models and gauge theories, let us first describe a much
simpler but analogous idea in the more familiar area of atomic
physics: the problem of determining the ground-state energy and wave
function of electrons in an atom.

The $\hbar \to 0$ classical limit is {\em not} a good approximation,
since the atom is not stable in this limit. The Coulomb potential is
not bounded below and the electrons would fall into the nucleus. Of
course, in the quantum theory, this is prevented by the uncertainty
principle: momentum would grow without bound if we tried to
concentrate the electron wave function at the minimum of the Coulomb
potential. However, there is another classical limit, where the
number of spatial dimensions $d \to \infty$, in which the atom has a
stable ground-state! This classical limit can even be used as the
starting point for an approximation method for $d = 3$.

We illustrate this idea for a hydrogenic atom of atomic number
$Z$. The analogue of unitary invariance in matrix models is
rotational invariance for the atom. So we will consider the
problem of minimizing the energy
    \beq
        E = \int \bigg[{\hbar^2 \over 2m} \dd{\psi^*}{x_i} \dd{\psi}{x_i} - {Ze^2 \over
        r}|\psi({\bf x})|^2        \bigg] d^d x
    \eeq
subject to the unit norm constraint $\int |\psi(${\bf x}$)|^2 d^d x
= 1$, in the zero angular momentum sector $\psi({\bf x}) = \psi(r)$
where $r = (\Sigma_{i=1}^d x_i x_i)^{\half}$. Note that we work in
$d$ spatial dimensions but retain the three-dimensional Coulomb
potential. While fluctuations in all observables vanish in the
$\hbar \to 0$ classical limit, only the fluctuations in rotationally
invariant observables vanish in the $d \to \infty$ classical limit
that we study below.

We first transform from  $x_i \mapsto r$, which introduces a
jacobian $d^d x = \Omega_d r^{d-1} dr$. $\Omega_d$ is the surface
area of the unit sphere $S^{d-1}$. Thus
    \beq
        E = \Omega_d \int \bigg[ {\hbar^2 \over 2 m} {\dd{r}{x_i}}{\dd{r}{x_i}}|\psi'(r)|^2 -
        {Z e^2 \over r} |\psi(r)|^2 \bigg] r^{d-1} dr
    \eeq
and $<\psi|\psi> = \int \psi^*(r) \psi(r) \Omega_d r^{d-1} dr$. We
absorb the jacobian by defining a {\em radial} wave function
$\Psi(r) = \sqrt{\Omega_d} r^{(d-1)/2} \psi(r)$ so that it is
normalized in a simple way $<\Psi|\Psi> = \int \Psi^*(r) \Psi(r)
dr$. In terms of the radial wave function:
    \beqs
        E &=& \int \bigg[{\hbar^2 \over 2m} \bigg\{ |\Psi'(r)|^2 +
            {(1-d)^2 \over 4 r^2}|\Psi(r)|^2 + {(1-d) \over 2}
            \ov{r} {d \over dr} |\Psi(r)|^2 \bigg\}
            - {Ze^2 \over r} |\Psi(r)|^2 \bigg] dr \cr
        &=& \int \bigg[{\hbar^2 \over 2m} |\Psi'(r)|^2 + \bigg\{
        {\hbar^2 (d-1)(d-3)\over 8mr^2} - {Ze^2 \over r} \bigg\} |\Psi(r)|^2 \bigg] dr
    \eeqs
where we have integrated by parts and ignored the surface term which
vanishes for square integrable wave functions. We see that a portion
of the kinetic energy coming from derivatives of the jacobian and
integration by parts now manifests itself as a correction to the
three-dimensional Coulomb potential. The hamiltonian is
    \beq
        H = -{\hbar^2 \over 2 m} {d^2 \over dr^2} + {\hbar^2 \over 8 m} {(d-1)(d-3) \over
        r^2} - {Z e^2 \over r}
    \eeq
To take the $d \to \infty$ limit, we must re-scale to variables
that have a finite limit: $\rho = {r \over \sqrt{d}};~ \pi = -{i
\over d} {d \over d \rho};~ \tilde H = {H \over d};~ \alpha = {e^2
\over d^{3/2}}$. The hamiltonian and commutation relation are
    \beq
        \tilde H = {\hbar^2 \over 2m} \pi^2 + {\hbar^2 \over 8 m
        \rho^2} {(d-1)(d-3) \over d^2} - {Z\alpha \over \rho} ; ~~~
        [\pi,\rho]= -{i \over d}
    \eeq
We see that the large $d$ limit holding $\hbar$ finite is a
classical limit, $\pi$ and $\rho$ commute and fluctuations in the
rotationally invariant variable $\rho$ are small. In this limit,
we get a classical mechanical system whose phase space is the half
plane $\{ \rho > 0, \pi \}$ with hamiltonian and Poisson bracket
    \beq
        \tilde H = {\hbar^2 \over 2m} \pi^2 + {\hbar^2 \over 8 m \rho^2} - {Z \alpha \over
        \rho}; ~~ \{ \rho,\pi \}_{P.B.} = 1.
    \eeq
The main difference between this classical limit and the usual
$\hbar \to 0$ limit is the appearance of a centrifugal barrier to
the Coulomb potential, even for zero angular momentum! Thus we
have a ``classical'' explanation for the stability of the atom in
the $d \to \infty$ limit.

The ground-state is given by the static solution $\pi = 0$, $\rho =
\rho_0 = {\hbar^2 \over 4m Z \alpha}$, which is the minimum of
$V_{eff}(\rho) = {\hbar^2 \over 8 m \rho^2} - {Z \alpha \over
\rho}$. In other words, in the $d \to \infty$ classical limit, the
electron wave function is concentrated at a radial distance of
$\rho_0$ from the nucleus. To recover fluctuations in the electron
position we must quantize this classical theory. The ground-state
energy is $\tilde E = V_{eff}(\rho_0) = -{2 m Z^2 \alpha^2 \over
\hbar^2}$. To quantitatively compare with the known answer in $3$
dimensions we revert to the old variables $E = d \tilde E$ and $e^2
= \alpha~ d^{3 \over 2}$ and set $d = 3$
    \beqs
        E_{exact} &=& -\half {m Z^2 e^4 \over \hbar^2} \cr
        E_{approx} &=& - {2 \over 9} {m Z^2 e^4 \over \hbar^2}
    \eeqs
Alternatively, we could have compared the energy per dimension. Thus
we have a crude first approximation to the ground-state energy of
hydrogenic atoms. Similarly, $\rho_0$ provides a first approximation
to the mean distance of the electron from the nucleus.

This idea has been extended to many electron atoms as well. One way
of implementing the Pauli exclusion principle for many electron
atoms is to let the number of spin states of the electron tend to
infinity along with the number of dimensions. Then the wave function
is taken to be totally anti-symmetric in spin quantum numbers. For
example, D. Herschbach and collaborators \cite{herschbach} have
calculated atomic energy levels in a $\ov{d}$ expansion. The leading
term is less accurate than other methods such as Hartree-Fock
theory. However, its advantage is that the complexity does not grow
as fast with the number of electrons, since the problem can be
reduced to the algebraic minimization of an effective potential.
Even more impressive is the spectacular accuracy they achieved by
including the corrections in an asymptotic series in inverse powers
of $d$. By resumming this series, they obtained an accuracy of more
than 9 significant figures for the ground-state of the Helium atom
\cite{herschbach-2-elec}.

Thus we see that the $d \to \infty$ classical limit allows us to
understand certain features of the theory that the $\hbar \to 0$
classical limit misses. It also serves as a starting point for an
approximation method. We expect the $N \to \infty$ classical limit
to play a similar role in matrix models and gauge theories.

The system we have just studied can also be considered as a
non-relativistic $O(d)$-vector model in one dimension (time), where
the position $x_i(t), i=1,\cdots,d$ is a $d$-component vector. The
restriction to zero angular momentum corresponds to requiring all
observables to be $O(d)$ invariant. More generally, the large-$N$
limit of $O(N)$-vector models are of much interest especially in
statistical physics. $N$ denotes the number of spin projections for
example in a Heisenberg ferromagnet. The large-$N$ limit was first
studied in the context of the spherical model by Stanley
\cite{stanley}. The $O(N)$ non-linear sigma model in three spatial
dimensions describes phase transitions in three dimensions
\cite{itzykson-drouffe}.

\section{The Large-$N$ limit: Planar Diagrams and Factorization}
\label{s-planar-diag-large-N}


The large-$N$ limit was introduced into the study of gauge theories
and matrix models by 't Hooft, who showed that in the large-$N$
limit, with appropriately scaled coupling constants, the planar
Feynman diagrams (or those that can be drawn on a sphere) dominate
\cite{thooft-planar}. Indeed, the usual perturbative sum over
Feynman diagrams can be reorganized according to the genus of the
Riemann surface on which the diagram can be drawn.

Let us indicate how this works for a matrix field theory. The
dynamical variable is an $N \times N$ hermitian matrix-valued
scalar field $A^a_b(x)$, where $a,b$ are ``color'' indices. The
partition function is
    \beq
        Z = \int dA exp\bigg[- {N\over \hbar} \int d^dx \bigg\{
        \half \tr \pdr_\mu A(x) \pdr^\mu A(x) +
        \half \tr A^2(x) + \sum_{k \geq 3} g_k \tr A^k(x) \bigg\}
        \bigg]
    \eeq
We have kept an over all factor of $N$ multiplying the action. The
limit $\hbar \to 0$ holding $N$ fixed is the usual classical
limit. Letting $N \to \infty$ holding $\hbar$ fixed is a different
classical limit. They are not the same because the matrix $A(x)$
depends on $N$, though not on $\hbar$. Let us concentrate on the
factors of $N$ that appear in a Feynman diagram due to the
matrix-valued nature of the field. We will suppress factors of
$\hbar$ and all space-time dependence, symmetry factors etc. It is
convenient to use a double line notation, where propagators of
$A^a_b$ are denoted by oppositely directed double lines, each
carrying one of the two color indices. If we also had vectors
(like quarks $q^a$), we would denote them by single lines carrying
the single color index. Consider a connected Feynman graph in the
perturbative expansion of the logarithm of the partition function
(connected vacuum diagrams, no external legs). Suppose it has $L$
loops and $E$ propagators (edges). Vertices are due to the cubic
and higher interactions. Suppose there are $V_k$ vertices of
coordination number $k \geq 3$, then the total number of vertices
is $V = \sum_{k \geq 3} V_k$.

Now let us think of each loop as the boundary of a face, its
orientation is given by the direction in which the color index
flows on the loop which is its boundary. The propagators that make
up the boundaries are edges. Since we are considering connected
diagrams with no external legs, we get an oriented polyhedral
surface. In other words, a regularization of a Riemann surface.
The number of edges, faces and loops are related to the number of
handles ($=$ genus $G$) by the formula for the Euler
characteristic
    \beq
        V - E + L = \chi = 2 - 2G
    \eeq
Each loop involves a sum over colors and contributes a factor of
$N$. (In the double line notation, a loop involves an inner line
which forms a closed curve with colors summed and an outer line
which is not a closed curve) Each propagator (inverse of the
quadratic term in the action) gives a factor of $1/N$. The $V_k$
$k$-valent vertices contribute a factor of $g_k N$ each. Thus the
factors of $g_k$ and $N$ associated with any Feynman diagram is
    \beq
        N^L (1/N)^E \prod_{k \geq 3} ( g_k N )^{V_k}
        = N^{L-E} N^{\sum_{k \geq 3} V_k} \prod_{k \geq 3}
        g_k^{V_k} = N^{2-2G} \prod_{k \geq 3}
        g_k^{V_k}
    \eeq
Thus diagrams with a common power of $N$ have a fixed genus.
Moreover, in the limit $N \to \infty$ holding $g_k$ fixed, the
leading diagrams are those that can be drawn on a sphere (genus $G =
0$, known as planar diagrams). Suppressed by $1/N^2$, are diagrams
that can be drawn on a torus ($G = 1$, sphere with one handle) and
so forth. Moreover, we see that the logarithm of the partition
function is proportional to $N^2$, so that we should define the
large-$N$ limit of the free energy as $F = - \lim_{N \to \infty}
\ov{N^2} \log Z$.

Suppose we had vectors (quarks) in addition to the matrices. It
turns out that diagrams with $P$ internal quark loops are
suppressed by $(1/N)^P$ compared to the planar diagrams involving
the matrix fields (gluons) alone. This is because a quark loop,
being a single line loop, does not involve a sum over colors.
Diagrams with $P$ internal quark loops correspond to Riemann
surfaces with $P$ punctures.

\smallskip

\fl {\bf \large Large-$N$ Factorization:} These selection rules
regarding the dominant diagrams as $N \to \infty$ continue to hold
even when we consider diagrams with external legs, (i.e. expectation
values of $U(N)$ invariants). However, there is a further
simplification beyond planarity, when we consider expectation values
of products of $U(N)$ invariants: they factorize. Consider a
hermitian multi-matrix model. The matrices are $A_i, ~i=1,\cdots M$
where the $i$'s can be thought of as labelling points in space time.
The action is a polynomial $S(A) = S^{i_1 \cdots i_n} A_{i_1} \cdots
A_{i_n}$ and correlations are given by
    \beq
        Z = \int dA e^{-N \tr S(A)}; ~~~~
        \langle f(A) \rangle = \ov{Z} \int dA e^{-N \tr S(A)} ~
        f(A)
    \eeq
Then the expectation values of $U(N)$ invariants factorize. For
example, let $\Phi_{i_1 \cdots i_n} = \Ntr A_{i_1} \cdots
A_{i_n}$. Then
    \beq
        \langle \Phi_{i_1 \cdots i_n}  \Phi_{j_1 \cdots j_m}
        \rangle = \langle\Phi_{i_1 \cdots i_n}\rangle
        \langle\Phi_{j_1 \cdots j_m} \rangle + {\cal O} ({1 \over N^2})
    \eeq
Factorization can be proven perturbatively, (i.e. in powers of the
cubic and higher order coupling constants $S^{i_1 i_2 \cdots i_n}$
for $n \geq 3$). Let us give an example of factorization in the
very simplest case of the gaussian $S(A) = \half \sum_i A_i^2$.
Then the basic two-point correlation is
    \beq
    \langle [A_i]^a_b ~ [A_j]^c_d \rangle = \ov{N} \delta_{i j} ~
    \delta^a_d ~\delta^c_b~
    \eeq
Wick's theorem says that any correlation is a sum over all
pairwise contractions. For example ($\delta^a_a = N$),
    \beqs
        \langle \Phi_{ij} ~\Phi_{kl} \rangle &=& \langle\N [A_i]^a_b
        [A_j]^b_a~ \N [A_k]^c_d [A_l]^d_c \rangle \cr
         &=& \N \N \delta_{ij} \delta^a_a \delta^b_b ~
         \N \N \delta_{kl} \delta^c_c \delta^d_d
          + \N \N \delta_{ik} \delta^a_d \delta^c_b ~
         \N \N \delta_{jl} \delta^d_a \delta^b_c \cr && +
         \N \N \delta_{il} \delta^a_c \delta^d_b ~
         \N \N \delta_{jk} \delta^c_a \delta^b_d \cr
        &=& \delta_{ij} \delta_{kl} + ~\ov{N^2} \delta_{ik}
            \delta_{jl} + ~\ov{N^2} \delta_{il} \delta_{jk}
        = \langle \Phi_{ij} \rangle \langle \Phi_{kl} \rangle
            + {\cal O}(\ov{N^2})
    \eeqs
Of the three terms on the right, the first corresponds to a planar
diagram, the second is non-planar (and so suppressed by
$\ov{N^2}$) and the third is planar, but suppressed by $\ov{N^2}$
because it involves a contraction between matrices in two
different traces.

The factorization of $U(N)$ invariant observables in the large-$N$
limit implies that they do not have any fluctuations. Thus the
large-$N$ limit is a classical limit for these variables.
Factorization also holds for the invariant observables of a vector
model and also for the meson observables (\ref{e-meson-observables})
of a model with both vector and matrix-valued fields.

\pagebreak

\part{Baryon in the Large-$N$ Limit of 2d QCD}
\label{part-baryon-large-N}

\thispagestyle{myheadings}

\markboth{}{Part \ref{part-baryon-large-N}. Baryon in Large-$N$ 2d
QCD}


\fl {\bf \Large Baryon in the Large-$N$ Limit of 2d QCD}

\medskip

In two dimensions the gluon field has two polarization states. They
can be taken as the null and time components. The null component can
be set equal to zero by a choice of gauge. The time component is not
a propagating degree of freedom. It is eliminated by solving its
equation of motion. This leads to a linear potential between quarks.
Thus, 2d QCD with $N$ colors is a relativistic vector model of
interacting fermions. By summing the planar diagrams of the
large-$N$ limit, 't Hooft obtained a linear integral equation for
meson masses and wave functions \cite{thooft-2d-meson}. However, it
was not clear how baryons arose. Witten proposed that baryons be
described by a Hartree-type of approximation in the large-$N$ limit,
though in a non-relativistic context \cite{wittenN}.

In the null gauge, when quarks are null separated, the parallel
transport operator in the meson variable (\ref{e-meson-observables})
is the identity. Thus, two-dimensional QCD can be formulated as a
bilocal theory of quark bilinears. Rajeev \cite{rajeev-2dqhd} found
a bosonization in terms of bilocal meson variables that satisfy a
quadratic constraint. The latter is the projection operator
constraint on the quark density matrix. We review the derivation of
this theory from 2d QCD and its classical large-$N$ limit in Chapter
\ref{ch-QCD-to-QHD}. The phase space of this theory is an infinite
grassmannian, a disconnected manifold with connected components
labelled by an integer, the baryon number. In Chapter
\ref{ch-grnd-state-baryon}, we study the baryon, which is a
topological soliton, the minimum of energy on the component of the
phase space with baryon number equal to one. The ground-state form
factor of the baryon is determined variationally by restricting the
dynamics to finite-rank submanifolds of the phase space. This leads
to a derivation of an interacting parton model. In the simplest
case, this interacting parton model corresponds to a Hartree-type
approximation to an $N$-boson system in the large-$N$ limit. The $N$
bosons are the quasi-particles corresponding to valence-quarks whose
wave function is already totally antisymmetric in color. In the
large-$N$ and chiral limits, the exact ground-state of the baryon
occurs on a rank-one submanifold of the phase space, corresponding
to a configuration containing only valence-quarks. We use the
valence-quark distribution to model the proton structure function
measured in Deep Inelastic Scattering at low momentum transfers,
where transverse momenta can be ignored as a first approximation.

\chapter{From QCD to Hadrondynamics in Two Dimensions}
\label{ch-QCD-to-QHD}

\thispagestyle{myheadings}

\markboth{}{Chapter \ref{ch-QCD-to-QHD}. From 2d QCD to
Hadrondynamics}


In this chapter we review Rajeev's reformulation
\cite{rajeev-2dqhd,rajeev-derive-str-fn} of $1+1$ dimensional QCD in
the null gauge in terms of color-singlet quark bilinears:
two-dimensional Quantum Hadrondynamics (QHD). The large-$N$ limit
(where $N$ is the number of colors) of the color-singlet sector of
two-dimensional QCD is a classical limit of QHD and is a nonlinear
field theory of a constrained bilocal meson variable valid at all
energies. The classical phase space is a curved manifold, the
infinite dimensional grassmannian whose connected components are
labelled by baryon number. The Poisson algebra of observables is a
central extension of the infinite dimensional unitary Lie algebra.
Hamilton's equations of motion are nonlinear integral equations for
the meson field. Free mesons are small fluctuations around the
vacuum. 't Hooft's linear integral equation for meson masses arises
as the linear approximation. Even in the classical large-$N$ limit,
Hadrondynamics contains the interactions of mesons. Moreover,
baryons arise as topological solitons.

\section{Large-$N$ Limit of 2d QCD}
\label{s-large-N-2d-QCD}


Two-dimensional QCD is a non-abelian gauge theory of quarks coupled
to gluons. The quarks are two-dimensional Dirac spinors transforming
as vectors in the fundamental representation of the structure group
$SU(N)$. The gluons are vector bosons, one-forms valued in the Lie
algebra of $SU(N)$: traceless $N \times N$ hermitian matrices. The
number of colors $N$ is $3$ in nature. The action for a single
flavor of quarks is
    \beq
        S = -{N \over 4 g^2} \int \tr F^{\mu \nu} F_{\mu \nu} d^2x
        + \int \bigg[ \bar q^a \gamma^\mu (-i \delta^b_a \partial_\mu
        - A_{\mu a}^b)q_b - m \bar q^a q_a \bigg] d^2x
    \eeq
where $F_{\mu \nu} = \partial_\mu A_\nu - \partial_\nu A_{\mu} +
[A_\mu, A_\nu]$. Here $a,b$ are color indices, $g$ is a coupling
constant with the dimensions of mass and $m$ is the {\it current}
quark mass. 2d QCD is a finite quantum field theory, there are no
ultra-violet divergences. Apparent infra-red divergences occur
when $m$ is set to zero, but these can be avoided by considering
the $m=0$ case as a limiting case of massive current quarks.

It is convenient to work in null coordinates $t = x^0 - x^1, ~ x =
x^1$ in terms of which the metric is $ds^2 = dt(dt + 2dx)$.
$\partial_t$ is time-like and $\partial_x$ is a null vector and the
initial values of fields are given on the null line $t = 0$. $x$ is
regarded as space and $t$ is regarded as time. The components of
momentum $E dt + p dx$ are energy $E = p_0 $ and null momentum $p =
p_0 + p_1$. The mass shell condition $m^2 = p_0^2 - p_1^2$ becomes
$m^2 = p(2E-p)$. So the energy of a free particle is $E = \half(p +
{m^2 \over p})$. We see an important advantage of null coordinates,
energy and null momentum have the same sign. So quarks have positive
null momentum while anti-quarks have negative momentum. Under a
Lorentz transformation of rapidity $\theta$ ( $\tanh{\theta} = v/c$
where $v$ is the boost velocity), $t \to e^\theta t, ~~ x \to
-(\sinh{\theta}) t + e^{-\theta} x$ and $p \to e^\theta p, ~~ E \to
e^{-\theta} E + (\sinh{\theta})p$. Thus, a scaling of null momentum
is just a Lorentz boost.

The time and null components of the gauge field $A = A_t dt + A_x
dx$ are $A_t = A_0$ and $A_x = A_0+A_1$. We work in the null gauge
$A_x = 0$. The reason to use the null gauge is that the parallel
transport along a null line is the identity. So meson observables
(\ref{e-meson-observables}) simplify. The two-dimensional quark
spinor is $q
= \ov{\sqrt{2}} \left(\begin{array}{c}  \eta \\  \chi \\
\end{array} \right)$. The Dirac matrices in null coordinates
satisfy $(\gamma^t)^2 = 0$, $(\gamma^x)^2 = -1$,
$\{\gamma^t,\gamma^x\}_+ = 2$ and a convenient representation is
    \beq
    \gamma^t = \left( \begin{array}{cc}  0 & 2 \\  0 & 0 \\
    \end{array} \right) , ~~ \gamma^x = \left( \begin{array}{cc}
    0 & -1 \\  1 & 0 \\ \end{array} \right), ~~ \bar q = q^\dag \left( \begin{array}{cc}
    0 & 1 \\  1 & 0 \\ \end{array} \right)
    \eeq
where $\bar q = q^\dag C$ where the charge conjugation matrix
satisfies $C \gamma^\mu C^{-1} = (\gamma^\mu)^T$. The action
becomes
    \beq
    S = \int dt dx \bigg[ {N \over 2g^2} \tr (\pdr_x A_t)^2 + \chi^\dag (\hat E -
    A_t) \chi + \half (\eta^\dag \hat p \eta - \chi^\dag
    \hat p \chi) - {m \over 2} (\eta^\dag \chi + \chi^\dag \eta)
    \bigg]
    \eeq
where $\hat E = -i \pdr_t,~ \hat p = -i \pdr_x$. $\chi$ and
$\chi^\dag$ are therefore canonically conjugate and satisfy
canonical anticommutation relations
    \beq
        [\chi^{\dag a}(x), \chi_b(y)]_+ = \delta(x-y) \delta^a_b~ ,
        ~~ [\chi_a(x), \chi_b(y)]_+ = 0 = [\chi^{\dag a}(x), \chi^{\dag b}(y)]_+
    \eeq
Neither $\eta$ nor $A_t$ has time derivatives, so they are not
dynamical and may be eliminated by solving their equations of
motion,
    \beqs
        \hat p ~ \eta =  m \chi, && - \partial_x^2 A_{tb}^a(x) =
        {g^2 \over N} :\chi^{a\dag}(x) \chi_b(x): \cr
        \eta(x) = {m \over \hat p} \chi(x), && A_{tb}^a(x) =
        -{g^2 \over N} \int dy :\chi^{\dag a}(y) \chi_b(y): \half
        |x-y|
    \eeqs
since $\pdr_x^2 \half |x-y| = \delta(x-y)$. So eliminating the
longitudinal gluons leads to a linear potential between quarks.
Normal ordering is with respect to the Dirac vacuum
    \beq
        \tilde \chi^{\dag a}(p) |0> = 0 {\rm ~for~} p < 0, ~~
        \tilde \chi_b(p) |0> = 0 {\rm ~for~} p > 0
    \eeq
where $\tilde \chi(p) = \int \chi(x) e^{-ipx} dx$. The resulting
hamiltonian is
    \beq
        H = \int dx \chi^{\dag a} \half (\hat p + {m^2 \over \hat p})
        \chi_a
        - {g^2 \over 2N} \int :\chi^{\dag a}(x) \chi_b(x): :\chi^{\dag b}(y)
        \chi_a(y): \half |x-y| dx dy
    \eeq
Introduce the color-singlet meson operator (In
(\ref{e-meson-observables}) the parallel transport between $x$ and
$y$ is the identity in the null gauge)
    \beq
        \hat M(x,y) = - {2 \over N} :\chi^{\dag a}(x) \chi_a(y):
    \eeq
Since the meson observable is gauge-invariant, we may calculate it
in the null-gauge which is convenient for us.

The hamiltonian and momentum can be written in terms of $\hat
M(x,y)$ after rearranging the operators in the normal ordered
products (Note: $[dp] = {dp \over 2\pi}$)
    \beqs
    {H \over N} = -\half \int \half(p + {\mu^2 \over p})
    \tilde{\hat M}(p,p) [dp] &+& {g^2 N \over 8}
    \int \hat M(x,y) \hat M(y,x) {|x-y| \over 2} dx dy \cr
    {P \over N} &=& -\half \int p \tilde {\hat M}(p,p) [dp]
    \eeqs
where $\tilde {\hat M}(p,q) = \int dx dy e^{-i(px-qy)} \hat
M(x,y)$. $\mu^2 = m^2 - {g^2 N\over \pi}$ is a finite
renormalization of current quark mass, a kind of self energy
coming from the rearrangement of currents in the potential energy
term. The Lorentz invariant squared mass is $P(2H-P)$. The
commutation relations can be expressed as
    \beq
    [\hat M(x,y), \hat M(z,u)] = {1 \over N} \bigg[\delta(y-z)
    (\eps(x,u)+\hat M(x,u)) - \delta(x-u)(\eps(z,y) + \hat M(z,y)) \bigg]
    \eeq
where $\eps(x,y)$ is the kernel of the Hilbert transform
    \beq
        \eps(x,y) \equiv \eps(x-y) = {\cal P} \int \sgn(p) e^{ip(x-y)} [dp].
    \eeq
$\ov{N}$ plays the role of $\hbar$ in a conventional quantum system
and the large-$N$ limit holding $\tilde g^2 = g^2 N$ fixed is a
classical theory of the color-invariant dynamical variable $M(x,y)$
with Poisson brackets $-iN[~,~] \mapsto \{~,~\}$
    \beq
    \{M(x,y),M(z,u)\} = -i\bigg[\delta(y-z)
    (\eps(x,u)+ M(x,y)) - \delta(x-u)(\eps(z,y) + M(z,y)) \bigg]
    \label{e-pb-of-M}
    \eeq
From these, one can derive Hamilton's classical equations of motion.
What are the allowed values of $M(x,y)$, i.e. the classical phase
space? From the definition in terms of quark fields, the classical
variable $M$ is a hermitian operator with integral kernel $M(x,y)^*
= M(y,x)$. The fact that quarks are fermions survives the
bosonization and is encoded in a quadratic constraint on $M(x,y)$.
The large-$N$ limit of the quark density matrix $\rho(x,y) = \ov{N}
\chi^{\dag i}(x) \chi_i(y)$ must be a projection when restricted to
color-singlet states: $\rho^2 = \rho$. After normal ordering,
$M(x,y) = \delta(x-y) - \eps(x,y) - 2 \rho(x,y)$ satisfies the
quadratic constraint $(\eps + M)^2 = 1$. Moreover, baryon number $B
= -\half \tr M$ which must be an integer is a topologically
conserved quantum number. Thus the phase space is a disconnected
manifold with connected components labelled by baryon number. It is
the infinite dimensional (restricted) grassmannian \cite{segal}. In
the next section we will quickly review this classical theory on the
grassmannian, 2d Classical Hadrondynamics, which is equivalent to
the color-singlet sector of the large-$N$ limit of two-dimensional
QCD. The quantization of this classical theory, 2d Quantum
Hadrondynamics has been shown by Rajeev \cite{rajeev-2dqhd} to be
equivalent to the color-singlet sector of 2d QCD for all energies
and values of $N$ and $\hbar$. We will see that this large-$N$
classical theory is very different from the $\hbar \to 0$ limit,
Classical Chromodynamics.

\section{Classical Hadron Theory on the Grassmannian}
\label{s-CHD-grassmannian}


The color-singlet sector of two-dimensional QCD is an interacting
theory of Dirac fermions (quarks) which come in $N$ colors, with the
constraint that only states that are $SU(N)$ invariant are
permitted. In the null coordinates we are using, the sign of kinetic
energy and null momentum are the same. So consider the direct sum
decomposition of the one particle complex Hilbert space ${\cal H} =
L^2(R) = {\cal H}_- \oplus {\cal H}_+$ into positive and negative
momentum functions ${\cal H}_+ = \{\psi | \tilde \psi(p) = 0 $~for~$
p < 0 \}$ and ${\cal H}_- = \{\psi | \tilde \psi(p) = 0 {\rm ~for~}
p > 0 \}$. Here $\psi(x) = \int_{-\infty}^\infty \tilde \psi(p)
e^{ipx} [dp]$. The operator $\eps$ introduced earlier is the sign of
momentum, $\eps \tilde \psi(p) = \pm \tilde \psi(p) {\rm ~for~} \psi
\in {\cal H}_{\pm}$. Thus $\eps^\dag = \eps$ and $\eps^2 = 1$. In
momentum space we can regard $\eps$ as a large diagonal matrix with
eigenvalues $-1$ for the negative momentum states and eigenvalues
$+1$ for positive momentum states. Thus, $\eps$ represents the Dirac
vacuum where negative momentum states are completely occupied and
positive ones empty. The novelty is that, since quarks come in $N$
colors, each completely occupied state is occupied by a
color-invariant combination of $N \to \infty$ quarks.

More generally, for any orthogonal decomposition ${\cal H} = L_-
\oplus L_+$, there is a hermitian operator $\Phi$ which is $-1$ on
elements of $L_-$ and $+1$ on those of $L_+$. As before, $\Phi^2 =
1$. $\Phi$ represents a more general state than the Dirac vacuum,
it has eigenvalue $\pm 1$ on states that are completely empty or
filled.

Loosely, the phase space of large-$N$ QCD, the restricted
grassmannian of ${\cal H}$, is the set of all its subspaces that do
not differ too much from ${\cal H}_-$. By the above correspondence,
it is also the set of hermitian idempotent operators $\Phi$ which do
not deviate too much from the vacuum $\eps$.

More precisely, since observables are measured by their deviation
from their values in the Dirac vacuum, we work with the hermitian
operator $M = \Phi - \eps$. Then the restricted grassmannian is
the infinite dimensional manifold
    \beq
    Gr({\cal H},\eps) = \{M^\dag = M, (\eps+M)^2 = 1, \tr M^\dag M < \infty\}
    \eeq
This is the phase space of Classical Hadrondynamics. We recognize
that $M$ is just the large-$N$ limit of the meson operator $\hat
M(x,y) = -{2 \over N} :\chi^{\dag a}(x) \chi_a(y):$ of the previous
section. The Grassmannian describes only those states of the
multi-particle Fock space of the quantum theory
$\Sigma_{r,s=0}^\infty \La^s {\cal H}_-' \otimes \La^r {\cal H}_+$
that are wedge products of single particle states (up to a scalar
multiple). These are the coherent states, ones with a good classical
limit. The more general states of the Fock space which are linear
combinations of such wedge products are not included. Thus the
grassmannian is embedded in the projective space of the fermionic
Fock space with a large co-kernel: the Plucker embedding
\cite{chern,mickelsson}. We will return to this when deriving the
interacting quark model from a variational approximation to the
classical theory on the grassmannian (see \S
\ref{s-fock-space-den-matrix-rank-1},~ \S
\ref{s-fock-space-rank-3}).

As we saw earlier, the grassmannian is a disconnected manifold,
each connected component being labelled by the topologically
invariant integer $B = - \half \tr M$ which is baryon number.
Though neither $\Phi$ nor $\eps$ has a finite trace, the trace of
the difference is well defined.

The restricted grassmannian carries an action of the restricted
unitary group $U({\cal H},\eps)$. $\Phi \mapsto U \Phi U^\dag$
preserves the conditions $\Phi^2 =1, \Phi^\dag = \Phi, \tr(\Phi -
\eps)^2 < \infty$ if
    \beq
        g ~\in~ U({\cal H},\eps) = \{U | U^\dag U = U U^\dag = 1, \tr
        [\eps,U]^\dag [\eps,U] < \infty\}
    \label{e-restricted-unitary-group}
    \eeq
The action on $M$ is
    \beq
    M \mapsto U M U^\dag + U\eps U^\dag - \eps.
    \label{e-action-of-restr-unitary-on-M}
    \eeq
Moreover, any $\Phi$ may be diagonalized by a unitary $U$ and
brought to a standard form with eigenvalues $\pm 1$. However, points
on disconnected components of the phase space cannot be connected by
a unitary transformation since the latter preserve $B$, which is
integer-valued. Thus, the action is transitive on each connected
component. Indeed, the grassmannian is the homogeneous space
${U({\cal H},\eps) / U(H_+) \times U(H_-)}$. We will use this
transitive action of the unitary group to show how to include
anti-quarks and `sea' quarks in a simple way, starting from
valence-quarks alone (see \S \ref{s-bogoliubov}).

The Lie algebra of the restricted Unitary group is easily obtained
by taking $U = e^u \simeq 1 + u$, we get
    \beq
        \underline U({\cal H},\eps) = \{u| u = -u^\dag, \tr [\eps,u]^\dag
        [\eps,u] < \infty \}
    \label{e-restricted-unitary-lie-alg}
    \eeq
The infinitesimal action on $M$ is $M \mapsto [u,\eps + M]$. This
brings us to the Poisson structure on the phase space. The fermion
bilinears $M$ form a representation of a central extension of this
restricted unitary Lie algebra, as we see from the Poisson
brackets (\ref{e-pb-of-M}). This is, in fact, the natural Poisson
bracket on the grassmannian: it is invariant under the action of
the unitary group. This requirement can be used to obtain the
Poisson brackets of non-linear functions of $M$ as well:
    \beq
     \{ f(M),M \} = -i [f'(M), \eps + M]
    \eeq
The dynamics is specified by the hamiltonian we obtained from the
large-$N$ limit of 2d QCD, which may be regarded as a real-valued
quadratic function on the classical phase space
    \beqs
    E(M) = \lim_{N \to \infty} {H \over N} &=& -\half \int \half(p + {\mu^2 \over p})
    \tilde M(p,p) [dp] \cr  && + {\tilde g^2 \over 8}
    \int M(x,y) M(y,x) {|x-y| \over 2} dx dy
        \label{e-energy-of-CHD}
    \eeqs
Total momentum is
    \beq
        \bar P = P/N = -\half \int p \tilde M(p,p) [dp]
        \label{e-momentum-of-CHD}
    \eeq
Time evolution is given by Hamilton's equations
    \beq
    {d M \over dt} = \{E(M), M\} = -i [E'(M),\eps + M]
    \eeq
where the integral kernel of $E'(M)$ is (Note: ${\cal FP}$ is a
finite part integral, a rule for integrating the $\ov{r^2}$
singularity, see Appendix \ref{a-finite-part})
    \beq
    E'(M)(p,q) = -\ov{4} 2\pi \delta(p-q) (p + {\mu^2 \over p})
                - {g^2 \over 4} {\cal FP} \int \ov{r^2} \tilde
                M(p+r,q+r) [dr]
    \eeq
Thus, the equations of motion are quadratically non-linear integral
equations for $\tilde M(p,q)$. A further complication comes from the
quadratic constraint $(\eps + M)^2 = 1$. Thus the large-$N$ limit of
2d QCD is a highly non-linear interacting theory of mesons.

Let us first consider static solutions, given by $[E'(M), \eps + M]
= 0$. The simplest static solutions are given by the extrema of the
energy $E'(M) = 0$. The minimum of energy on each connected
component of the phase space describes the ground-state of Classical
Hadrondynamics for a given baryon number.

On the baryon number zero component of the phase space, the
obvious static solution is $M = 0$, the vacuum. Small oscillations
around the vacuum in the $B = 0$ sector describe an infinite
number of non-interacting mesons. To see this, we linearize the
equations of motion around $M = 0$:
    \beqs
    {i \over 2} {d \tilde M(p,q) \over dt} &=& -\half (K(p)-K(q))
    \tilde M(p,q) - {\tilde g^2 \over 4} {\cal FP} \int [ds]
    \ov{s^2} \bigg[ \cr &&
    \tilde M(p+s,q+s) \sgn(q) - \sgn(p) \tilde M(p+s,q+s)\bigg]
    \eeqs
where $\tilde K(p) = \half (p+ {\mu^2 \over p})$ is the kinetic
energy. Moreover, the constraint $(\eps + M)^2 = 1$ when
linearized becomes $\bigg[ \sgn(p) + \sgn(q) \bigg] \tilde M(p,q)
= 0$ so that one may take $p \geq 0, q \leq 0$. The equation of
motion after eliminating the constraint is
    \beq
    {i \over 2} {d \tilde M(p,q) \over dt} = -\half [\tilde K(p) - \tilde
    K(q)] \tilde M(p,q) + \half \tilde g^2 {\cal FP} \int {\tilde M(p+s,q+s) \over
    s^2} [ds]
    \eeq
Since the vacuum $M = 0$ is translation invariant, the total
momentum $P = p-q$ is a conserved quantity. The remaining
independent variable can be taken as momentum fraction $\xi =
p/P,~ 0 \leq \xi \leq 1$, and we look for small oscillations
$\tilde M(p,q,t) = e^{i \omega t} \chi(\xi)$. The resulting
eigenvalue problem for $\omega$ is
    \beq
    {\cal M}^2 \chi(\xi) = [{\mu^2 \over \xi} + {\mu^2 \over (1-\xi)}
    ] \chi(\xi) - {\tilde g^2 \over \pi} {\cal FP} \int_0^1 {\chi(\eta) \over (\xi - \eta)^2}
    d \eta
    \eeq
where ${\cal M}^2 = 2 \omega P - P^2$ is the square of the meson
mass. This is the linear integral equation for the meson spectrum of
2d QCD in the large-$N$ limit. It was originally obtained by 't
Hooft by summing planar Feynman diagrams \cite{thooft-2d-meson}.

Thus, the 't Hooft linear integral equation for meson masses
corresponds to the linear approximation around the vacuum, to our
non-linear and non-local formulation of large-$N$ 2d QCD. We do not
know of any simple way of deriving this non-linear theory by summing
planar diagrams. It is only in this linearized approximation that
mesons are non-interacting in the large-$N$ limit.

Heuristically, we may regard free mesons as an infinitesimal
departure from the vacuum since they involve only the promotion of
${\cal O}(1)$ quarks from the Dirac sea (where each fully occupied
state contains ${\cal O}(N)$ quarks) to a positive momentum state.
Thus the large-$N$ limit is crucial.

What then is a finite disturbance to the vacuum? It must involve the
addition, removal or rearrangement of ${\cal O}(N)$ quarks. The
simplest example of such a state is a baryon. It is an old idea of
Witten that baryons may be described in a Hartree Fock approximation
in the large-$N$ limit. He outlined this in a non-relativistic
context \cite{wittenN}. In what follows we will develop the fully
relativistic theory of the baryon in two dimensions and see that a
variational approximation to this theory is indeed a kind of
Hartree-Fock approximation to a many body theory. We turn to the
study of the baryon now.

\chapter{Ground-state of the Baryon}
\label{ch-grnd-state-baryon}

\thispagestyle{myheadings}

\markboth{}{Chapter \ref{ch-grnd-state-baryon}. Ground-state of the
Baryon}


The ground-state of the baryon in the large-$N$ limit is the static
solution corresponding to a minimum of energy on the baryon number
$B=1$ component of the phase space. Thus we have a realization of
Skyrme's idea \cite{Skyrme1,Skyrme2,Skyrme3,bal-tasi} that baryons
must arise as solitons of a theory of mesons. The essential novelty
is that our classical bilocal theory of the meson variable $M(x,y)$
is not a low energy effective theory, but is equivalent to the
large-$N$ limit of 2d QCD for all energies. In particular, this
theory has an infinite number of mesons, unlike many low energy
effective theories which contain only a finite number of mesonic
excitations. This chapter is based on our papers
\cite{ipm,bogoliubov-anti-quark,solitonparton}. Other useful
references are
\cite{rajeev-derive-str-fn,rajeev-bedaque,rajeev-2dqhd}.

We first describe a numerical method for determining the minimum
energy configuration, based on a geometric adaptation of steepest
descent to the curved grassmannian phase space.

Then we will present a variational approach. The primary hurdle in
the minimization is the quadratic constraint satisfied by
$M(x,y)$. An ascending family of sub-manifolds of the grassmannian
is found, corresponding to operators of increasing rank. They
allow us to replace the constraint with simpler ones. The minimum
on these sub-manifolds provides a variational approximation to the
true minimum. The dynamics on these reduced phase spaces is shown
to correspond to interacting models of quarks and anti-quarks.
Thus we reconcile two seemingly disparate pictures of the baryon:
the soliton and parton pictures.

The minimum energy configuration $\tilde M(p,q)$ is the form factor
of the baryon. The diagonal elements $\tilde M(p,p)$ are the quark
and anti-quark probability densities in the baryon in the large-$N$
limit. We determine these by a variational approximation. As an
application, we use these to model the quark distributions measured
in deep inelastic scattering (\S \ref{s-appl-to-DIS}).

Small oscillations around the ground-state of the baryon would
correspond to excited states of the baryon, such as $\Delta$ and
$N^*$. We do not discuss these here. One could also study the
minimum of energy on the sectors of higher baryon number; they
correspond to two-dimensional nuclei.

\section{Steepest Descent on the Grassmannian}
\label{s-steepest-descent}


To determine the ground-state of the baryon, we must minimize the
energy $H$ (\ref{e-energy-of-CHD}) on the baryon number one
component of the phase space holding the momentum $P$
(\ref{e-momentum-of-CHD}) fixed. A Lorentz invariant formulation is
to minimize the squared mass ${\cal M}^2(M) = P(2H-P)$ subject to
the Pauli exclusion principle quadratic constraint $(\eps + M)^2 =
1$ and baryon number $B = -\half \tr M = 1$ constraint. The
quadratic constraint complicates matters significantly since we are
minimizing a non-linear function of $M(x,y)$ on a non-linear
manifold, the grassmannian.

The first method we tried to solve this problem with, is steepest
descent. The latter is a way of minimizing a real-valued function,
usually on a linear space. It is an iterative technique: beginning
with a conveniently chosen initial configuration, we move a small
distance in a direction opposite to the gradient of the function.
The procedure is repeated at the new point. After a sufficiently
large number of iterations, we reach a minimum of the function.

Now, the grassmannian is not a linear space, due to the quadratic
constraint. It is good to keep in mind the sphere embedded in
three dimensional euclidean space when thinking of the
grassmannian. Indeed, the sphere is the simplest finite
dimensional grassmannian. A tangent vector $V$ to the grassmannian
at the point $\Phi = \eps + M$ must satisfy $[\Phi,V]_+=0$. This
comes from linearizing the constraint $\Phi^2 = 1$ at the point
$\Phi$. We can think of the anticommutator as the generalization
of the dot product: the condition that a vector be tangential to
the sphere is implemented by requiring that its dot product with
the radius vector at that point vanish. Now, the gradient vector
${\cal M}'(M)$ is in general not tangential to the grassmannian.
So if we move in a direction opposite to it, we will fly off the
grassmannian! Actually, there is more bad news: if we move in a
straight line in the space of {\em all} hermitian operators $M$,
along the tangential projection $T$ of ${\cal M}'(M)$, we will
still leave the grassmannian. This is analogous to moving along a
tangent vector to the sphere. Now $T = \half[\Phi,Y]$ where $Y =
\half[\Phi,{\cal M}'(\Phi)]$. To understand the formula for the
tangential projection, think of the commutator as the analogue of
the cross product. Taking the cross product of an arbitrary vector
${\cal M}'$ with the radius vector $\Phi$ produces a vector $Y$
tangential to the sphere, but rotated by a right angle. We need to
take a second cross product with the radius vector $\Phi$ to get
the tangential projection $T$.

A way to stay on the grassmannian is to move along a curve
$\gamma(\tau) = \Phi(\tau) - \eps$ which starts out tangential to
the tangential projection $T$ of the gradient vector ${\cal M}'$.
The natural choice for a curve is the great circle, i.e. the
geodesic. The great circle through a point $\Phi$ on the sphere
tangential to a vector $T$ is obtained by rotating the radius
vector $\Phi$ about an axis orthogonal to $T$. In our situation,
$Y$ is the necessary vector orthogonal to $T$. Moreover, the
analogue of rotation is the action of the unitary group on the
grassmannian (see \S \ref{s-CHD-grassmannian}). So a rotation
about the axis $Y$ is implemented by the unitary operator $e^{\tau
Y}$. In other words, the geodesic through $\Phi$ tangential to $T$
and parameterized by $\tau$ is
    \beq
        \Phi(\tau) = e^{\tau Y} \Phi e^{-\tau Y}
    \eeq

We are finally ready to formulate the steepest descent algorithm
for the grassmannian by taking small steps along the geodesic that
goes in the direction of maximal reduction of squared mass.
{\begin{enumerate}
    \item Start with an initial configuration $M_1$ and a small
        $\tau$.
    \item At the $k^{\rm th} $ configuration $M_k$ evaluate the
        gradient ${\cal M}'(M_k)$ and the rotation generator
        $Y_k = \half[\eps + M_k,{\cal M}'(M_k)]$.
    \item Let $M_{k+1} = e^{\tau Y_k} [\eps + M_k] e^{-\tau Y_k} -
        \eps$ ($M$ transforms inhomogeneously unlike $\Phi$ (\ref{e-action-of-restr-unitary-on-M}))
    \item Repeat till there is no further reduction in the mass of
    the baryon.
\end{enumerate}    }

We implemented this steepest descent method numerically using
Mathematica. The value of the time step $\tau$ was arrived at
empirically to ensure quick convergence. As for the initial
configuration, we used the solution of the separable ansatz, a
more analytic variational approximation method which we turn to
next. We do not present the numerical results here, some earlier
numerical results are given in the paper \cite{ipm}.

\section{Separable Ansatz: Formulation}
\label{s-sep-ansatz-formulation}

\subsection{Rank-One Configurations}
\label{s-rank-one-configs}

$\tilde M(p,q)$ depends on a pair of variables. In Feynman's valence
parton model \cite{bjorken-paschos,feynman-photon-hadron}, the
valence-quark distribution is a function of a single momentum
variable. Since we know that this gives a reasonable description of
the proton, it must be a good approximation to assume that $\tilde
M(p,q)$ is built out of a single function of one variable, i.e. it
is a separable or rank-one operator:
    \beq
    M_1 = -2 ~\psi \otimes \psi^{\dag} ~~~ {\rm i.e.} ~~~
    \tilde M_1(p,q) = -2~ \tilde \psi(p) \tilde \psi^*(q)
    \eeq
The quadratic constraint $(\eps + M)^2 = 1$ becomes
    \beq
    (\eps - 2 \psi \otimes \psi^\dag)^2 = \eps^2 -
    2 \eps \psi \otimes \psi^\dag - 2 \psi \otimes \psi^\dag \eps
    + 4 \psi \otimes \psi^\dag \psi \otimes \psi^\dag
    \eeq
The constraint is satisfied if $\tilde \psi(p)$ vanishes for $p <
0$ i.e. $\eps \psi = \psi$ and is of unit norm $\psi^\dag \psi =
1$. The former means this configuration contains no anti-quarks.
The latter ensures $M_1$ is a configuration of baryon number one:
$B = - \half \tr M_1 = \int_0^{\infty} \tilde \psi(p) \tilde
\psi^*(p) [dp] = 1$.

Since $M_1$ is unaltered by a change in the phase of $\psi$, the
phase space of the rank-one ansatz is the projective space on
positive momentum wave functions ${\cal P(H_+)}$. The Poisson
brackets of $\tilde M(p,q)$ are satisfied if
    \beqs
    \{\tilde\psi(p),\tilde\psi(p')\} ~= & 0 & =~
        \{\tilde\psi^*(p),\tilde\psi^*(p')\}, \cr
    \{\tilde\psi(p),\tilde\psi^*(p')\} &=& -i~ 2\pi\delta(p-q)
    \eeqs
The hamiltonian when written in terms of $\psi$ is
    \beq
    E_1(\psi) = \int_0^\infty\half[p+{\mu^2 \over p}]
        |\tilde\psi(p)|^2 [dp] + \half \tilde g^2
        \int |\psi(x)|^2|\psi(y)|^2 {|x-y| \over 2}dxdy.
    \eeq
Thus, we have a self contained hamiltonian dynamical system on the
reduced phase space of rank-one configurations.

\subsection{Quantizing the Separable Ansatz: Interacting Valence Quarks}
\label{s-quantize-rank-1}

In order to make precise the relation of the separable ansatz to an
interacting valence-quark model, we quantize the classical theory of
the previous section. We denote the parameter that measures quantum
corrections by $\ov{N}$. The constraint on norm is implemented by
restricting attention to states $|V>$ satisfying
    \beq
    <V|\int_0^{\infty} \hat{\tilde\psi}^{*}(p){\hat{\tilde\psi}}(p)[dp]|V>=1.
    \eeq
The Poisson brackets go over to commutation relations
    \beq
    [\tilde {\hat \psi}(p),\tilde{\hat \psi}(p')]=0=
    [\tilde{\hat \psi}^{\dag}(p),\tilde{\hat \psi}^{\dag}(p')], ~
    [\tilde{\hat\psi}(p),\tilde{\hat\psi}^{\dag}(p')] = {1 \over N}
    2 \pi \delta(p-p').
    \eeq
A representation for these commutation relations is given by the
canonical commutation relations of bosonic creation-annihilation
operators:
    \beq
    [\hat{\tilde{b}}(p),\hat{\tilde{b}}(p')] = 0 = [\hat{\tilde{b}}^{\dag}(p),\hat{\tilde{b}}^{\dag}(p')], \quad
    [\hat{\tilde{b}}(p),\hat{\tilde{b}}^{\dag}(p')] = 2 \pi \delta(p-q).
    \eeq
where $\hat\psi = \ov{\sqrt{N}} \hat b, ~~ \hat\psi^\dag =
\ov{\sqrt{N}} \hat b^\dag$. In terms of these the constraint
becomes
    \beq
    <V| \int_0^\infty \hat b^\dag(p) \hat b(p) [dp] |V> = N
    \eeq
Thus we have a system of $N$ bosons, so $N$ must be an integer.
The hamiltonian operator becomes
    \beq
    N \hat E_1 = \int_0^\infty\half[p+{\mu^2\over p}]
    \tilde b^\dag(p) \tilde b(p) ~[dp] + {\tilde g^2 \over 2 N} \int
    b^\dag(x) b^\dag(y) {|x-y| \over 2} b(y) b(x) ~dxdy.
    \eeq
The baryon must be in a simultaneous eigenstate of the hamiltonian
and momentum operators, whose eigenvalues $E_N$ and $P$ are the
energy and total momentum of the baryon. The momentum operator is
    \beq
    \hat P= \int_0^\infty p ~ b^\dag(p) b(p)~ [dp].
    \eeq
If we work in the momentum basis, an $N$ particle momentum
eigenstate is
    \beq
    |p_1,p_2,\cdots ,p_N> =
    b^\dag(p_1) b^\dag(p_2) \cdots b^\dag(p_N) |0>.
    \eeq
A general state $|\psi>$ containing $N$ particles is given by the
wave function  $\tilde \psi (p_1,\cdots,p_N)$, where
    \beq
    |\psi>= \int_0^\infty [dp_1] \cdots [dp_N] \tilde\psi(p_1,\cdots,p_N)|p_1,\cdots,p_N>.
    \eeq
The expectation value of the hamiltonian in such a state is
    \beqs
    <\psi| N \hat E | \psi>  &=& \int_0^\infty \sum_{i=1}^{\
    N}\half[p_i+{\mu^2\over p_i}]|\tilde\psi(p_1,\cdots,
    p_N)|^2 [dp_1] \cdots [dp_N] \cr & & + {\tilde g^2 \over 2 N}
    \int_0^\infty \sum_{i\neq j} {|x_i-x_j| \over 2}
    |\psi(x_1,\cdots,x_N)|^2 dx_1\cdots dx_N.
    \label{e-expect-val-N-body-egy}
    \eeqs
It is now clear that we have a system of $N$ bosons interacting
through a linear two-body potential in the null coordinates. What
are these bosons? They are just the valence-quarks of the parton
model, whose dependence on color has been factored out. Though
quarks are fermions, the total baryon wave function must be totally
antisymmetric in the color indices so that it is invariant under
$SU(N)$. Thus, the baryon wave function (suppressing spin and flavor
dependence) $\tilde \Psi$ factorizes as
    \beq
    \tilde\Psi(a_1,p_1;a_2,p_2;\cdots;a_{N},p_{N}) = \eps_{a_1,a_2,\cdots a_N}
    \tilde\psi(p_1,p_2,\cdots,p_N).
    \eeq
Here $a_k$ denote color indices. Thus, if we ignore color,
valence-quarks behave like bosons.

In the parton model, the momentum fraction carried by a quark
satisfies $0 \leq \xi = {p \over P} \leq 1$. How do we recover
this in our picture? Null momentum is positive for a quark. Thus,
the momentum of the particles created by $b^\dag(p)$ is always
positive whence the total momentum $\hat P$ is a positive
operator. On an $N$ particle momentum basis state, $\hat P$ is
just the sum of individual momenta, each of which is positive:
    \beq
    \hat P|p_1,p_2\cdots,p_N> = [p_1+p_2+\cdots p_N] |p_1,p_2,\cdots,p_N>.
    \eeq
An eigenstate of $\hat P$ with eigenvalue $P$ must satisfy
    \beq
    \left[p_1+\cdots p_N\right] \tilde\phi(a_1,p_1;\cdots a_N,
    p_N)= P\tilde\phi(a_1,p_1;\cdots a_N, p_N).
    \eeq
Since each of the momenta $p_i$ are positive, it follows that $0
\leq {p_i \over P} \leq 1$. Thus, quantizing the rank-one
approximation to our classical hadron theory gives us an interacting
valence-quark model \cite{ipm,bogoliubov-anti-quark}.

\subsection{Hartree Approximation and the Large-$N$ limit}
\label{s-hartree-approx}

It is an old idea of Witten \cite{wittenN} that for the baryon, the
large-$N$ limit should correspond to a sort of Hartree approximation
to a many body theory. We have a relativistic implementation of this
idea in two dimensions.

We would like to find approximate simultaneous eigenstates of $\hat
E_1$ and $\hat P$. The ground-state of a many boson system can often
be described by mean-field theory: each boson moves in the field
created by all the others. Moreover, all the bosons can be assumed
to occupy the same single particle state in this ground-state. The
naive choice is a wave function $\psi$ which is a product of single
particle wave functions. But such a choice would not be a momentum
eigenstate, so we enforce it by requiring the momenta to add up to
$P$.
    \beq
    \tilde\psi(p_1 ,p_2, \cdots ,p_N)= 2 \pi\delta(\sum_ip_i-P)
    \tilde\psi(p_1)\tilde\psi(p_2)\cdots \tilde\psi(p_N).
    \eeq
In particular, the fraction of momentum carried by each parton is
less than one $\tilde \psi(p)=0 ~{\rm unless}~ 0 \leq p/P \leq 1$.
In terms of the single parton wave function, the energy
(\ref{e-expect-val-N-body-egy}) of the baryon per quark becomes
($N(N-1) \sim N^2$)
    \beq
    E_1(\psi) = \int_0^P \half[p+{\mu^2 \over p}]
    |\tilde\psi(p)|^2 [dp] + {\tilde g^2 \over 2} \int
    |\psi(x)|^2|\psi(y)|^2 {|x-y| \over 2}dxdy.
    \eeq
Here, we are retaining the condition $0 \leq p \leq P$. The
remaining correlations imposed by the factor $\delta(\sum_ip_i-P)$,
are suppressed for large-$N$. As a compromise to the momentum
eigenstate condition we will require the expectation value of $\hat
P$ to equal $P$. This is analogous to using the canonical ensemble
as an approximation to the micro-canonical ensemble. This gives rise
to the momentum sum rule:
    \beq
    N \int_0^P p|\tilde\psi(p)|^2 [dp] ~=~ P.
    \eeq
The wave function also satisfies the normalization condition
$\int_0^P |\tilde\psi(p)|^2 ~[dp] ~=~1 $.

We see that the Hartree approximation to the interacting
valence-quark model has reproduced our classical hadron theory
restricted to the separable submanifold of the phase space, except
for the requirement that the wave function vanish beyond $p = P$.
This subtle difference is a semi-classical (finite $N$) effect. To
recover the true classical limit, we must let $N$ become truly
infinite. Since the total momentum is an extensive variable, $P \to
\infty$ as $N \to \infty$. So for $N = \infty$, the valence-quark
wave function is not required to vanish for any finite value of $p$.

\subsection{Pl\"ucker Embedding and Density Matrix}
\label{s-fock-space-den-matrix-rank-1}

In order to find the Fock space state corresponding to the classical
rank-one ansatz, we use an infinite dimensional version of the
Pl\"ucker embedding \cite{chern,mickelsson,rajeev-2dqhd} of the \gr
in the Fock space ${\cal F}=\sum_{r,s=0}^\infty \Lambda^{s}{\cal
H}_-' \otimes \Lambda^{r}{\cal H}_+$. Here ${\cal H = {\cal
H}_-\oplus {\cal H}_+}$ is the splitting into positive and negative
momentum subspaces.

Given a point $M$, the state in ${\cal F}$ to which it is mapped by
the Plucker embedding, is the wedge product of the occupied states.
These are the eigenstates of the density matrix $\rho = \half(1 - M
- \eps)$ with eigenvalue $1$ or equivalently, the $-1$ eigenstates
of $\Phi = \eps + M$. In the case of the rank-one ansatz $M = -2
\psi \otimes \psi^\dag$, we look for states $\chi$ with
    \beq
    \Phi_1 \chi = (\eps -2 \psi \otimes \psi^{\dag}) \chi = - \chi
    \eeq
Every negative momentum state (i.e. those with $\eps \chi = - \chi$)
is a solution. The only other solution is $\chi = \psi$ since, $\eps
\psi = \psi$ and $\psi^\dag \psi = 1$. These are the filled states.
Thus the Fock space state (denoted $|M_1>$) corresponding to the
rank-one ansatz $M_1$ is a semi-infinite wedge product:
    \beq
    \cdots f^{2'} \wedge f^{1'} \wedge \psi
    \eeq
Here $f^{j '}$ is a dual basis in ${\cal H}_-$. The only
interesting information in this state is its departure from the
vacuum. If we denote by $|\tilde 0>$ the wedge product $\cdots
f^{2'} \wedge f^{1'}$ we see that $|\tilde 0>$  is the filled
negative  energy sea. Then $|M_1>$ is obtained by populating
$\psi$, starting from the vacuum $|\tilde 0>$.
    \beq
    |M_1> = b_\psi^{\dag} |\tilde 0>
    \eeq
Here, the quasi-particle created by $b_\psi^{\dag}$ is not a quark,
but the colorless quasi-particle of the rank-one ansatz. This point
of view is easier for computations and has the advantage of avoiding
the unobservable color quantum number altogether.

The quantized rank-one baryon can also be described in terms of the
creation annihilation operators of colored quarks. Let $|V>$ denote
the second quantized state of the baryon. In the Hartree
approximation, it is determined by the condition that the
expectation value of the density operator in $|V>$ must equal the
classical density matrix of the rank-one ansatz:
    \beq
        <V|\tilde{\hat \rho}(p,q)|V> = \tilde\rho_1(p,q)
    \eeq
For the rank-one baryon $\tilde \rho_1(p,q)~=~\tilde\psi(p)
\tilde\psi^{*}(q) + \half (2\pi \delta(p-q))(1-\sgn{p})$. A short
calculation shows that
    \beq
    |V> = q^{1 \dag}_{\tilde{\psi}} \cdots q^{N \dag}_{\tilde{\psi}} |0>.
    \eeq
satisfies the above requirement. Here, $q^{a \dag}_{\tilde \psi}$
creates a quark with color $a$ in the state $\tilde \psi$. These
operators satisfy canonical anti-commutation relations (CAR): $ \{
q_{i u},q^{j \dag}_{v} \} = \delta_{i}^{j} <u,v> $ with respect to
the Dirac vacuum $|0>$: $q_{\tilde {\psi}_-}^{a^{\dag}} |0> = 0$
and $q_{b \tilde {\psi}_+} |0> = 0$ where $\tilde {\psi}_-(p)$
vanishes for $p \geq 0$ and $\tilde {\psi}_+(p)$ for $p \leq 0$.

This formalism will be useful when we generalize the theory to
include sea and anti-quarks (section \ref{s-rank-three-ansatz}).
Before doing so, we discuss the actual determination of the
ground-state of the baryon within the rank-one approximation.

\section{Separable Ansatz: Solution}
\label{s-sep-ansatz-soln}

The energy of the baryon in the separable ansatz is
    \beq
        E/N = \int_0^P \half(p + {\mu^2 \over p}) |\tilde
        \psi(p)|^2 [dp] + {\tilde g^2 \over 2}
        \int_{-\infty}^{\infty} |\psi(x)|^2|\psi(y)|^2{|x-y|
        \over 2}dxdy
    \eeq
where $\mu^2 = m^2 - {\tilde g^2 \over \pi}$. Its momentum is
    \beq
        \bar P = P/N = \int_0^P p |\tilde \psi(p)|^2 [dp]
    \eeq
A Lorentz invariant formulation is to minimize the squared mass,
which in null coordinates is ${\cal M}^2 = 2EP - P^2$ (\S
\ref{s-large-N-2d-QCD}):
    \beqs
        { {\cal M}^2 \over {N^2 \tilde g^2}} = \bigg[\int_0^P
        p|\tilde\psi(p)|^2 [dp]\bigg] &*& \bigg[\bigg({\msqovgsq} - {1 \over
        \pi} \bigg) \int_0^P {|\tilde\psi(p)|^2 \over p} [dp] \cr
        && + \int_{-\infty}^{\infty} |\psi(x)|^2|\psi(y)|^2{|x-y|
        \over 2}dxdy\bigg].
    \label{e-masssq-rank-1}
    \eeqs
while holding $\int_0^P |\tilde \psi(p)|^2[dp] = 1$. The only
parameter in the theory is the dimensionless ratio $\nu = {m^2
\over \tilde g^2}$. The total baryon momentum $P$ can be scaled
out by expressing momenta in terms of the ratio $\xi  = {p \over
P}$. It is convenient to introduce notation for the kinetic, self
and potential energies:
    \beqs
    KE = \int_0^P {1 \over p} |\tilde\psi(p)|^2 [dp], &&
    SE = -{1 \over \pi} \int_0^P {1 \over p} |\tilde\psi(p)|^2
        [dp] \cr
    PE &=& \int_{-\infty}^{\infty} dxdy |\psi(x)|^2  |\psi(y)|^2 \half
        |x-y|.
    \eeqs
Then,
    \beq
    {{\cal M}^2 \over \tilde g^2 N^2}= \bar P *(\nu KE + SE +
    PE).
    \eeq

\subsection{Potential Energy in Momentum space}
\label{s-pe-in-mom-space}

While the kinetic and self energies are simple in momentum space,
the potential energy is more easily expressed in position space.
However, the integral equation for minimization of energy is
simplest in momentum space. Moreover, the condition that
$\tilde\psi(p)$ vanish for negative momenta is harder to implement
in position space. It implies that $\psi(x)$ is the boundary value
on the real line of a function analytic in the upper half plane.
So it is useful to rewrite the potential energy in momentum space.
However, the kernel of the integral operator is singular, and we
take some care to define it correctly. We write
    \beq
    PE = \int_{-\infty}^{\infty} |\psi(x)|^2 V(x) dx,~ {\rm
    where}~
    V(x) = \int_{-\infty}^{\infty} |\psi(y)|^2 {|x-y| \over 2} dy.
    \eeq
The self-consistent potential $V(x)$ obeys Poisson's equation
$V''(x) = |\psi(x)|^2$ along with a pair of boundary conditions.
    \beq
    V(0) = \int_{-\infty}^{\infty} |\psi(y)|^2 {|y| \over 2} dy ,~~
    V'(0) = \half \int_{-\infty}^{\infty} |\psi(-y)|^2 {\sgn(y)} dy
    \eeq
If we define the momentum space wave function and potential by
    \beq
    \psi(x) = \int_0^P \tilde\psi(p) e^{ipx} [dp] ~
    {\rm and} ~ \tilde V(r) = \int_{-\infty}^{\infty} e^{-irx} V(x) dx
    \eeq
then the potential energy becomes
    \beq
    PE = {\cal FP} \int [dp]
        \int [dr] \tilde\psi(p) \tilde\psi^*(p+r) \tilde V(r).
    \eeq
When the limits of integration are not indicated, they are from
the lower to upper limit of the interval over which the integrand
is supported. Poisson's equation in momentum space becomes
    \beq
    \tilde V(p) = {\cal FP} {-1 \over p^2} \int [dq] \tilde
    \psi(q) \tilde\psi^*(q-p) = {\cal FP} {-1 \over p^2} \tilde W(p)
    \eeq
We see that $\tilde V(p) \sim {-1 \over p^2}$ is singular for
small $p$ and use ${\cal FP}$ to denote an appropriate ``finite
part''. $\tilde W(-p) = W^*(p)$ is supported from $-P$ to $P$. The
rules for integrating this singularity are given by the boundary
conditions to Poisson's equation:
    \beq
    {\cal FP}\int {-1 \over p^2} \W(p) [dp] = \int |\psi(y)|^2
    {|y| \over 2} dy ~~ {\rm and} \eeq \beq -i {\cal FP} \int {1 \over
    p} \W(p) [dp] = \half \int_0^{\infty} \{-|\psi(y)|^2 + |\psi(-y)|^2 \} dy
    \eeq
We will primarily be interested in the ground-state of the baryon.
Assuming that the ground-state wave function is real, $\tilde W(p)$
is even. In Appendix \ref{a-finite-part} we show (for real wave
functions $\tilde\psi(p)$ which vanish at the origin like $p^a, p>0,
a>0$) that $\tilde W'(0) = 0$ and
    \beq
    {\cal FP} \int_{-P}^{P} {\tilde W(p) \over p^2} [dp]
        = \int_{-P}^{P} {\tilde W(p) - \tilde W(0) \over p^2} [dp]
        -  {\tilde W(0) \over \pi P}
    \eeq
Thus we can rewrite the potential energy as
    \beq
    PE = - {\cal FP} \int_{-P}^{P} {|\tilde W(r)|^2 \over
        r^2} [dr] = {1 \over \pi P} + 2 \int_0^P {|\tilde
        W(0)|^2 - |\tilde W(r)|^2 \over r^2} [dr]
    \eeq
Here $\tilde W(0) = 1$ and $|\tilde W(r)|^2 = |\tilde W(-r)|^2$ even
if $\tilde\psi(p)$ is not real. Notice that in the large-$N$ limit,
where $P = \infty$, the term $1 \over \pi P$ may be omitted. We can
now derive the integral equation for the minimization of the baryon
mass.

\subsection{Integral Equation for Minimization of Baryon Mass}

The condition for the minimization of ${\cal M}^2$ is equivalent
to a non-linear integral equation for the wave function
$\tilde\psi(p)$. Recall that
    \beq
    {{\cal M}^2 \over \tilde g^2 N^2}= \bar P *(\nu KE +
        SE + PE).
    \eeq
\noindent We vary this with respect to $\tilde \psi^*(p)$ and
impose the constraint on the norm through a Lagrange multiplier
$\lambda$.  The condition for an extremum is:
    \beq
    \bar P(\nu ~ \delta KE + \delta SE + \delta PE) + \delta \bar P
    (\nu KE + SE + PE) = \lambda \tilde\psi(p).
    \eeq
Here
    \beq
    \delta PE =- {\cal FP} ~ \delta \int_{-P}^P {|W(r)|^2
    \over r^2} [dr] ~~{\rm and}~~ {\delta W(r) \over \delta
    \tilde \psi^*(p)} = \tilde \psi(p+r).
    \eeq
Thus,
    \beqs
    \delta PE &=& - {\cal FP} \int_{-p}^{-p+P} [dr]
    {\tilde \psi(p+r) \tilde W^*(r) \over r^2} - {\cal FP}
    \int_{p-P}^{p} [dr] {\tilde \psi(p-r) \tilde W^(r) \over r^2}
    \cr &=& ~ 2 {\cal FP} \int_{-P+p}^p [dr] \tilde V(r)
    \tilde \psi(p-r)
    \eeqs
This leads to the non-linear integral equation
    \beqs
    \bar P \bigg[{\nu \over p} \tilde \psi(p) &-& {1 \over \pi p}
    \tilde \psi(p) + 2 {\cal FP}\int_{-P+p}^{p} \tilde V(r) \tilde\psi(p-r)
    [dr] \bigg] \cr &+& p \tilde \psi(p) (KE + SE + PE) =
    \lambda \tilde \psi(p).
    \label{e-integ-eqn-for-valence-1}
    \eeqs
We can write $\delta PE$ in a more convenient form, so that the
self energy term $- {1 \over \pi p} \tilde \psi(p)$ is cancelled.
Let us work in the limit $N = \infty$ so that $P = \infty$. Since
$W(-r) = W^*(r)$, we have
    \beq
    \delta PE(p) = -2 {\cal FP}  \int_0^p {[dr] \over r^2}
    [\tilde W(r) \tilde \psi(p-r) + \tilde W(-r) \tilde \psi(p+r)]  +
    2 \int_p^{\infty} V(-r) \tilde \psi(p+r) [dr]
    \eeq
The second integral is finite, but with hindsight we add and
subtract $\tilde \psi(p) \over \pi p$ from it. We may rewrite the
first integral using the definition of the ``finite part''
(Appendix \ref{a-finite-part}). Note that $\tilde W(0) = 1$.
    \beqs
    \delta PE &=& {\tilde \psi(p) \over \pi p} - 2 \int_0^p {[ds] \over s^2}
    [\tilde \psi(p-s) \tilde W(s) + \tilde \psi(p+s) \tilde W(-s) - 2
    \tilde \psi(p)] \cr && - 2 \int_p^{\infty} {[ds] \over s^2} [\tilde \psi(p+s)
    \tilde W(-s) - \tilde \psi(p)].
    \eeqs
Thus $\delta SE$ is cancelled by the first term in $\delta PE$.
The integral equation becomes:
    \beqs
    && {\nu \over p} \tilde \psi(p) - 2 \int_0^p {[ds] \over
    s^2} [\tilde \psi(p-s) \tilde W(s) + \tilde \psi(p+s) \tilde W(-s) - 2
    \tilde \psi(p)] \cr &-& 2\int_p^{\infty} {[ds] \over s^2} [\tilde \psi(p+s)
    \tilde W(-s) - \tilde \psi(p)] + {(\nu KE + SE + PE) \over \bar P}p \tilde \psi(p) =
    {\lambda\over \bar P} \tilde \psi(p)
    \label{e-integ-eqn-for-valence-2}
    \eeqs
The quantities $\bar P, KE, SE, PE$ (defined in \S
\ref{s-sep-ansatz-soln}) depend on the wave function and are to be
determined self-consistently. We now study this integral equation,
to find the ground-state of the baryon. Let us begin with the
behavior of the wave function for small arguments.

\subsection{Small Momentum Behavior of Wave function}
\label{s-frobenius}

To understand the behavior of the valence-quark wave function for
small momenta, we do a Frobenius-type analysis of the integral
equation for the minimization of the squared mass of the baryon:
    \beqs
    \lambda \tilde \psi(p) &=& \bar P \bigg[{\nu \over p} \tilde \psi(p) - {1 \over \pi p}
    \tilde \psi(p) +  2 {\cal FP}\int_{-P+p}^{p} \tilde V(r)
    \tilde\psi(p-r) [dr] \bigg] \cr && + p ~\tilde \psi(p)
    (\nu KE + SE + PE)
    \eeqs
For small $p$, we may ignore the last term on the RHS and the one
on the LHS, since the other terms are more singular in the $p \to
0$ limit. Dividing by $\bar P > 0$ we get the integral equation
    \beq
    (\nu - {1 \over \pi}) {1 \over p} \tilde\psi(p) +
    2 {\cal FP} \int_{p-P}^{p} \tilde{V}(q) \tilde\psi(p-q) [dq] = 0
    \eeq
where the self-consistent potential $\tilde{V}(q) \sim {-1 \over
q^2}$ for small q.  Now we assume a power law behavior
$\tilde\psi(p) \sim p^a$ for small $p > 0$ and derive an equation
for $a$.
    \beq
        (\nu - {1 \over \pi}) =  2 {\cal FP} \int_{1-{P
        \over p}}^1 {(1-y)^a \over y^2} {dy \over 2\pi}
    \eeq
where $y = {q \over p}$. Since $p << P$, we have
    \beq
        \pi \nu -1 = {\cal FP} \int_0^1
        \frac{(1-y)^a +  (1+y)^a}{y^2} dy + \int_1^{\infty}
        \frac{(1+y)^a}{y^2} dy
    \eeq
The first of these integrals is singular and we evaluate it
according to the definition of the finite part (Appendix
\ref{a-finite-part}). The result is a transcendental equation for
$a$:
    \beq
    \pi \nu = h(a) \equiv
    \int_0^1 \frac{(1+y)^a + (1-y)^a -2}{y^2} dy -1 + {{\rm _2 F_1}(1-a,-a,2-a,-1) \over 1-a}
    \eeq
Here $_2 F_1$ is the Hypergeometric function and $\nu = {m^2 \over
\tilde g^2}$. It is easily seen that for $m = 0$, $a = 0$ is a
solution. In the limit of chiral symmetry, $\tilde \psi(p) \sim
p^0$ as $p \to 0$. Calculating $h(a)$ shows that for a positive
current quark mass $m$, the wave function rises like a power
$\tilde\psi(p) \sim p^{a}$, $a > 0$. The plot shows $h(a)$.  The
solution for the exponent $a(m)$ for $m \geq 0$ is the point at
which the horizontal line ${\pi m^2 \over \tilde{g}^2}$ intersects
the curve.
\begin{figure}
\centerline{\epsfxsize=6.truecm\epsfbox{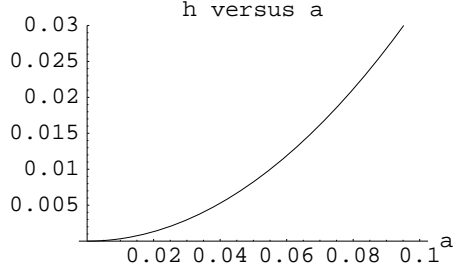}}
\caption{ $h(a)$. The solution for the exponent $p^{a(m)}$ is the
point at which the horizontal line ${\pi m^2 \over \tilde g^2}$
intersects the curve $h(a)$. For $m = 0$, $a = 0$ is a solution: in
the limit of chiral symmetry, the wave function goes to a non-zero
constant at the origin.}
\end{figure}
For sufficiently small $\msqovgsq$, we may solve the
transcendental equation for $a$, $\pi \msqovgsq  = h(a)$
analytically.  Expanding $h(a)$ in a power series we get for small
$a$,  $h(a) \to {\pi^2 \over 3} a^2$. Therefore, $a \to \sqrt{3
\over \pi}  {m \over \tilde g}$.

Our conclusion that in the chiral limit the momentum space wave
function tends to a non-zero constant as $p \to 0$ implies that it
is  discontinuous at $p=0$ since $\tilde \psi(p)=0$ for $p < 0$.
Therefore, the corresponding position space  wave function decays
like $1 \over x$ as $|x| \to \infty$.  This may be expected from
another point of view. The baryon we are studying, is a soliton of
the meson field $\hat M(x,y)$. It should be possible to approximate
the baryon wave function with the meson wave function for large
spatial x. At large distances we are essentially far away from the
baryon.  The meson wave function has been calculated by 't Hooft
\cite{thooft-2d-meson} in  the large-$N$ limit of 2d QCD. In the
chiral limit, the 't Hooft meson wave function has a discontinuity
at $p=0$ and consequently decays like $1 \over x$ as $|x| \to
\infty$. This is a non-trivial consistency check.

\subsection{Exact Solution for $N = \infty, m = 0$: Exponential Ansatz}
\label{s-exp-ansatz}

There is an exact solution to the integral equation for minimization
of baryon squared mass in the chiral and large-$N$ limits: $\tilde
\psi(p) \sim e^{-p}, p > 0$ where $b$ is fixed by the Lorentz
reference frame. Moreover, we find that the mass of the baryon
vanishes for this particular rank-one configuration. Therefore, in
the chiral and large-$N$ limits, the exact minimum of $(mass)^2$
occurs for a configuration with no anti-quarks. While this is to be
expected in the non-relativistic limit ${m^2 \over \tilde{g}^2} \to
\infty$ \cite{gk-ugrad-thesis}, it is rather  surprising in this
ultra-relativistic chiral limit, since conventional wisdom would
suggest that as ${m^2 \over \tilde g^2} \to 0$, the phase space for
production of virtual $q\bar{q}$ pairs is enhanced. We find this is
not true in the two-dimensional theory.

To see this, let us derive the analytic solution. We will use a
variational approach that also allows us to estimate corrections due
to finite $m$ and $1/N$. From the Frobenius analysis (section
\ref{s-frobenius}) we know that the wave function must vanish like a
power $p^a$ as $p \to 0^+$ with $a \to 0^+$ in the chiral limit. For
$N=\infty$ the valence-quark wave function is not required to vanish
for any finite value of $p$. But for finite $N$, it must vanish
beyond $p = P$. If the former is to be a good approximation, it must
fall off rapidly for large momenta.

As the simplest ansatz that satisfies these boundary conditions
and also facilitates explicit calculation, we take the variational
wave function $\tilde \psi(p) = p^a e^{-bp}$. The Lorentz
invariant parameter $a$ is determined by minimizing the
$(mass)^2$. In the null coordinates, a rescaling $p \to \lambda p$
is actually a  Lorentz boost in the longitudinal direction.
Therefore, $b$ is determined by the choice of the reference frame.
For the purposes of determining $a$, let us work in a reference
frame in which $b = 1$. We will fix $b$ subsequently, based on the
momentum sum rule, which serves to pick a reference frame.

In the notation we introduced earlier, ${{\cal M}^2 \over \tilde
g^2 N^2}= \bar P * (\nu KE + SE + PE)$. The normalized wave
function for the exponential ansatz is $\tilde\psi_a(p) =
\sqrt{\pi \over 2^{-2-2a} \Gamma{(1+2a)}} p^{a}e^{-p}$. For
non-zero current quark mass, $KE$ diverges for negative $a$.
Therefore, it is physically reasonable to consider only positive
values of $a$. In position space,
    \beq
    \psi_a(x) = {\sqrt{2^{-2a} \Gamma{(1+2a)}} \over \Gamma{(a +
    \half)}} (1~-~ix)^{-a-1}.
    \eeq
Then
    \beq
    \bar P(a) = {1 + 2a \over 2}, ~~ KE(a) = {1 \over a},
    ~~ SE(a) = -{1 \over \pi a}.
    \eeq
The potential energy $ PE(a) = \int_{-\infty}^{\infty}
|\psi_a(x)|^2 V_a(x) dx$ is determined by solving  $V_a^{''}(x) =
|\psi_a(x)|^2 $ subject to boundary conditions $V_a(0) =
\int_0^{\infty} x |\psi(x)| dx = {\Gamma{(a)} \over 2 \sqrt{(\pi)}
\Gamma{(\half + a)}}$ and $V_a^{'}(0) = 0$. Integrating once,
    \beq
    V_a^{'}(x) = {x \Gamma{(1+a)}  ~{\rm _2F_1} (\half,1+a,{3 \over
    2},-x^2)   \over \sqrt{\pi}\Gamma{(\half + a)}}.
    \eeq
And the second integration gives
    \beq
    V_a(x) = {\Gamma{(a)} [(1+x^2)^{-a} +  2a x^2 ~ {\rm _2F_1}
    (\half,1+a,{3 \over 2},-x^2)] \over  2 \sqrt{(\pi)} \Gamma{(\half
    + a)} }.
    \eeq
We arrive at
    \beq
        PE(a) = {2^{1-2a} \pi^2 \over \Gamma{(\half+2a)}\Gamma{(1-a)}
        (\Gamma{(\half + a)})^3 (-\sin{(\pi a)} + \sin{(3\pi a)}) }.
    \eeq
The Frobenius analysis suggests that the $(mass)^2$ is minimized
if $a = 0$ in the chiral limit. Therefore, we expand $PE(a)$ in a
Laurent  series around $a = 0$:
    \beq
        PE(a) = {1\over \pi a} + {\pi a \over 3} - {12 \zeta{(3)}
        a^2\over \pi} + O(a^3).
    \eeq
The pole at $a =0$ in the potential energy is cancelled by a
similar pole in the self energy.

Alternatively, we can calculate $PE(a)$ in momentum space using
the definition of the ``finite part'' integrals (see \S
\ref{s-pe-in-mom-space}). We use this as a check on our definition
of ${\cal FP}$. Let
    \beq
        \tilde \psi_a(p) = p^a e^{-p} {\sqrt{\pi} \over {\sqrt{(2^{-2-2a}
        \Gamma(1+2a))}}}.
    \eeq
Then
    \beq
        PE(a) = - 2 \int_0^{\infty} {|\W(s)|^2 - 1 \over s^2} [ds],
    \eeq
where
    \beq
        \W_a(s) = \int_0^{\infty} \tilde \psi_a(s+t) \tilde \psi_a^*(t)
        [dt]
        ~ {\rm for}~ s \geq 0, ~ \W(-s) = \W^*(s).
    \eeq
We get
    \beq
        \W_a(s) = { {{(2s)}^{\half + a} \sqrt{\pi}  K_{\half +
        a}(s) \csc{\pi a}} \over \Gamma(-a) \Gamma(1+2a)}
    \eeq
where $K_n(z)$ is the modified Bessel function of the second kind.
This results in the same expression for $PE(a)$ as before.
Collecting the results, the baryon $\masssq$ is:
    \beqs
     {{\cal M}^2 \over \tilde g^2 N^2} = ({1 + 2a \over
    2}) [{\nu \over a} - {1 \over \pi a}  + {2^{1-2a}
    \pi^2 \over \Gamma{(\half+2a)}\Gamma{(1-a)} (\Gamma{(\half +
    a)})^3 (-\sin{(\pi a)} + \sin{(3\pi a)}) }]
    \eeqs
For small $`a'$, we may write
    \beq
        {{\cal M}^2(a) \over \tilde g^2 N^2} = ({1 + 2a \over 2})
        ({\nu \over a} + {\pi a \over 3} - {12 \zeta{(3)} a^2\over
        \pi}    + O(a^3))
    \eeq

\subsubsection{The Chiral Limit}
\label{s-chiral-limit-for-large-N}

In the chiral limit $m =0 \Rightarrow \nu =0$,
    \beq
        {{\cal M}^2(a,m=0) \over \tilde g^2 N^2}
        = ({1 + 2a \over 2}) ({\pi a \over 2} - {12 \zeta{(3)}
        a^2\over \pi}  + O(a^3))
    \eeq
is the product of two factors, both of which are monotonically
increasing functions for small $a$. Therefore, the baryon mass is
minimized for $a = 0$, with a minimum value of $0$. Since the
$\masssq$ is a Lorentz scalar,  it is unchanged if we Lorentz
boost this wave function. Thus, the  $\masssq$ must vanish for
$\tilde \psi(p) = A e^{(-bp)}$ for any $b > 0$.

Assuming the  positivity of $(mass)^2$ on physical grounds, we
conclude that in the large-$N$ and chiral limits, the absolute
minimum of $(mass)^2$ occurs for a configuration containing only
valence-quarks. We emphasize that the exponential solution is not
just the absolute minimum of the baryon mass on the space of
separable configurations but on the {\it entire phase space} of
Hadrondynamics, in the chiral  and large-$N$ limit.

The masslessness of this two-dimensional baryon may be puzzling at
first. But there is a precedent for this. 't Hooft found that in the
chiral limit,  the lowest lying meson was massless in (the large-$N$
limit of)  2d QCD \cite{thooft-2d-meson}. Moreover, numerical
calculations of Hornbostel et. al. \cite{hornbostel} also show that
the baryons and mesons of 2-dimensional QCD are massless in the
chiral limit.

\subsubsection{Momentum Sum Rule and Determination of $`b'$}
\label{s-mom-sum-rule-and-b}

The valence-quark wave function in the chiral and large-$N$ limits
is $\tilde \psi(p) = A e^{-bp}, ~ p \geq 0$. Normalization gives $A
= \sqrt{4\pi b}$. A Lorentz boost is just a rescaling of momenta, so
the momentum sum rule
    \beq
        \int_0^{\infty}{p|\tilde\psi(p)|^2 [dp]} = P/N = \bar P
    \eeq
fixes the reference frame and determines $b = {1/ 2 \bar P}$.
Therefore,
    \beq
        \tilde \psi(p) = \sqrt{2\pi \over  \bar P} \exp{({-p \over 2
        \bar P})}. \label{e-exact-exp-soln}
    \eeq
This is the exact valence-quark wave function for $N=\infty, m=0$.
Thus the exact form factor of the baryon in the large-$N$ and chiral
limits is
    \beq
    \tilde M(p,q) = -\bigg({4\pi \over  \bar P}\bigg) e^{-(p+q) /2
    \bar P }
    \label{e-exact-exp-sol-for-M}
    \eeq
For finite $N$, we work in a reference frame in which the total
baryon momentum is $P = 1$. In the case $N=3$, this corresponds to
$\bar P = P/N ={1 \over 3}$. So to compare this wave function with
that for finite $N$, $\bar P = 1/3$ is the appropriate choice.

\subsection{Effect of Finite Current Quark Mass}
\label{s-eff-finite-curr-qk-mass}

\begin{figure}
\centerline{\epsfxsize=6.truecm\epsfbox{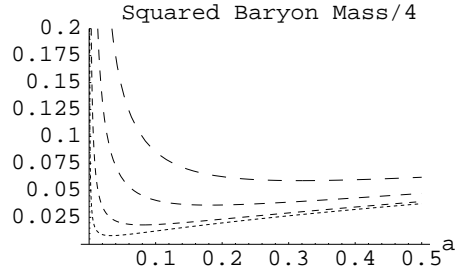}}
\caption{Estimate of ${1 \over 4}$ baryon mass$^2$ in units of
$\tilde g^2 N^2$ as a function of $a$. The curves top to bottom are
for $\msqovgsq = .05, .02, .005, .001$. As $\msqovgsq \to 0^+$, both
the minimum value of $\masssq$ and the value of $a$ that minimizes
it tend to $0^+$, recovering the exact massless exponential solution
in the chiral limit. } \label{f-mass-vs-a}
\end{figure}

\begin{figure}
\centerline{\epsfxsize=8.truecm\epsfbox{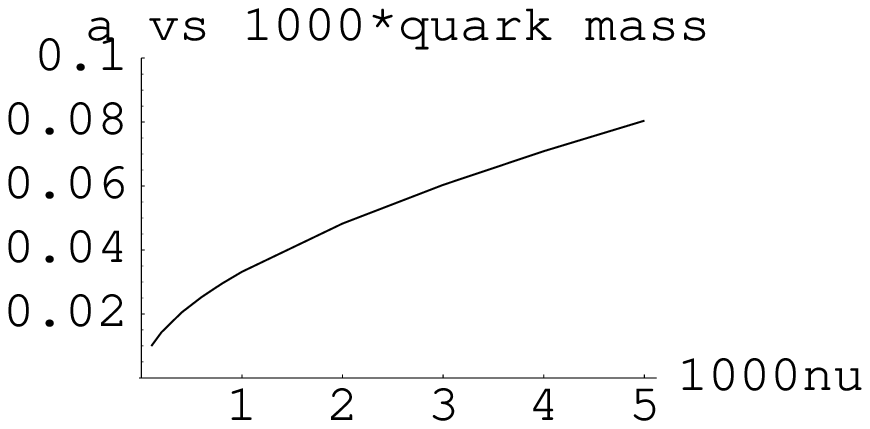}}
\caption{ The power $a$ in the wave function $\tilde \psi(p)= Ap^a
e^{-bp}$ plotted as a function of $1000 \msqovgsq$.  }
\label{f-a-vs-msqovgsq}
\end{figure}

We estimate the corrections due to non-zero $\nu = \msqovgsq$
variationally using the above exponential ansatz. From the
Frobenius analysis, we already expect that $a \to 0^+$ as
$\msqovgsq \to 0^+$. To investigate whether there is a minimum in
$a$, we expand ${{\cal M}^2(a) \over \tilde g^2 N^2}$ in a Laurent
series:
    \beq
    {{\cal M}^2(a) \over \tilde g^2 N^2} =  {\nu \over
    2a} + \nu + {a\pi \over 6} + a^2 ({\pi \over
    3} -  {6 \zeta{(3)} \over \pi} + O(a^3))
    \eeq
We see that for small $a$, the  $O({1 \over a})$ and $O(a)$ terms
dominate and have opposing  tendencies. Thus there is a positive
value of $a$ for which ${\cal M}^2$ is minimized as fig.
\ref{f-mass-vs-a}  shows.
For any given $\msqovgsq$, we can solve the transcendental
equation for the minimization of ${\cal M}^2$ numerically and find
the optimal $a$. We show this in fig. \ref{f-a-vs-msqovgsq}. We
see that $a \to 0^+$ as $\msqovgsq \to 0^+$, recovering the exact
exponential solution in the chiral limit.
However, for small enough current quark mass, we can get the
asymptotic  power law $a(\msqovgsq)$ analytically. Minimizing in
$a$, we find that $a \to \sqrt{{3 \over \pi}} {m \over \tilde g}$
as we found from the Frobenius analysis (section
\ref{s-frobenius}). Moreover, ${{\cal M}^2 \over \tilde g^2 N^2}
\to  {\sqrt{{\pi \over 3}{\msqovgsq}}} + \msqovgsq$ as $\msqovgsq
\to 0$.

In other words, for sufficiently small current quark masses, our
variational estimate for the valence-quark wave function is $\tilde
\psi(p)=Ap^{a}e^{-bp}$, with $a = {m \over \tilde g} \sqrt{3 \over
\pi}$.  The variational upper bound for the baryon mass is ${\cal M}
= {\tilde g} N ({\pi \over 3})^{1 \over 4}  ({m \over \tilde
g})^{\half}$. Though the baryon mass is of order $N$, in the chiral
limit, the proportionality constant vanishes in two dimensions!

\subsection{Variational ground-state for Finite $N$}
\label{s-var-grnd-state-finite-N}

We have seen (section \ref{s-hartree-approx}) that the leading
effect of finite $N$ is to restrict the range of quark momenta to
$0< p < P$. Thus $\tilde \psi(p)$ vanishes for $p
> P$. For finite, but large-$N$ we expect the ground-state wave
function to be well approximated by the exact exponential solution
(\ref{e-exact-exp-soln}) for $N = \infty$, in the chiral limit.
Since
    \beq
        (1-{p \over \bar P N})^{N} \to e^{-p \over \bar P}
    \eeq
the wave function
    \beq
        \tilde \psi(p) = A(1-{p \over P})^b,~0 < p < P
    \eeq
should be a good ansatz for the ground-state. From the above
argument, we expect $b$ to be proportional to $N$. More generally,
the Frobenius analysis (\ref{s-frobenius}) shows that
    \beq
        \tilde \psi(p) = A p^a (1-{p \over P})^b,~0 < p < P
    \eeq
should be a good ansatz for the ground-state even for non-zero
current quark mass. $A$ is fixed by normalization; $a$ is determined
by the minimization of baryon $\masssq$. From the Frobenius analysis
we know that for small $\msqovgsq$, $a \to \sqrt{3 \over \pi}  {m
\over \tilde g}$. $b$ can be eliminated using the momentum sum rule:
    \beq
        N \int_{0}^{P} p |\tilde \psi(p)|^2 [dp] = P
    \eeq
Thus
    \beq
        b = {N \over 2} - 1 + a({N} -1) ~ \to ~ {N \over 2} - 1
        + \sqrt{3 \over \pi} {m \over \tilde g} ({N} -1)
    \eeq
which is proportional to $N$ as we expected. The valence-quark
probability density is $V(\xi) =  {1 \over 2\pi} |\tilde \psi(\xi
P)|^2,~ 0 \le \xi \le 1$,
    \beq
        V(\xi) = {\xi^{2a} (1-\xi)^{2b} \over {\rm Beta}(1+2a,1+2b)}.
    \eeq
In the chiral limit,
    \beq
        V(\xi) = (N - 1) (1-\xi)^{{N} - 2}
    \eeq
which agrees with to the distribution obtained from numerical
discretized light cone calculations by Hornbostel et. al. in
\cite{hornbostel}.

\section{Beyond the Separable Ansatz}
\label{s-beyond-sep-ansatz}

\subsection{Fixed Rank Submanifolds of Grassmannian}
\label{s-rank-r}

We have been studying a reduced dynamical problem on the separable
submanifold of the phase space. By the variational principle, the
minimum of energy on a subset of the phase space provides an upper
bound for the  true minimum. To improve on this estimate, we must
allow for a larger set of configurations. We will now give an
increasing family of submanifolds  of the phase space, whose union
is dense in the \gr. Each of these is the  phase space for a reduced
dynamical system. They turn out to correspond to interacting quark
models, that go beyond the valence-quark approximation.

Let us generalize the rank-one ansatz to a rank $r$ ansatz ($a,b$
here do not stand for color, the meaning should be clear from the
context):
    \beq
        M_r=\sum_{a,b=1}^r\xi_b^a\psi_a\otimes \psi^{\dag b}.
    \eeq
We must pick $\xi_b^a$ and the $\psi_a$ such that the configuration
$M_r$ is admissible: $tr M_r^2 < \infty$ and satisfies the Pauli
exclusion principle:  $M_r^{\dag} = M_r$, $(M_r + \eps)^2 = 1$. The
admissibility condition is automatic since $M_r$ is a finite-rank
operator.  We pick the $\psi_a$ to be a finite number $r$ of
orthonormal  eigenstates of $\eps$, the sign of momentum.
    \beq
        \eps \psi_a = \eps_a \psi_a,~ \eps_a = \pm 1.
    \eeq
Suppose there are $r_+$ positive momentum vectors and $r_- = r -
r_+$ negative momentum vectors.  Let $\tilde \eps =$
Diag$(\eps_a)$ be the diagonal $r \times r$ matrix  representing
the restriction of $\eps$ to the sub-space of  ${\cal H_-} \oplus
{\cal H_+}$ spanned by the $\psi_a$.  This ansatz satisfies the
constraints if the $r \times r$ matrix $\xi$  is hermitian and
satisfies the constraint
    \beq
        (\tilde \eps + \xi)^2  = 1,
    \eeq
a mini-version of the constraint on $M_r$. If we let $\tilde \phi
= \tilde \eps + \xi$, then the constraints on $\tilde\phi$ are
    \beq
        \tilde\phi^{\dag} = \tilde\phi, ~ ~ \tilde\phi^2 = 1
    \eeq
Thus $\tilde \phi$ (and hence $\xi$) lies on a finite dimensional
\gr. The connected component of this finite dimensional \gr is
fixed by Baryon number: $B = -\half tr M_r = -\half tr \xi$.

Specializing to the case of baryon number one, it is easily seen
that the simplest solution to the constraints on $\xi$ is
$r_+=1,r_-=0$, so that $\xi$ is  just a $1 \times 1$ matrix, the
number -2. Then  $M_1 = -2 \psi \otimes \psi^{\dag}$. This is just
the rank-one separable ansatz that gave rise to the valence-quark
model.

To go beyond the valence-quark approximation, we need to find other
non-trivial solutions $r_+, r_-, \xi$ which satisfy these
constraints. If $r = r_+ + r_-  = 2$, the only non-trivial solutions
to the constraints reduces to  the rank-one solution. To see this we
first note that for $r = 2$,  $\tilde \phi$ is a point on the finite
dimensional grassmannian of  ${\bf C}^2$, the set of subspaces of
${\bf C}^2$. This \gr has three  connected components, corresponding
to $\tr \tilde\phi = -2, 0, 2$.  The baryon number constraint
implies that $tr \tilde\phi = r_+ - r_- -2$.  For $\tilde \phi$ to
lie on the finite dimensional \gr, we must have  $r_+ = 1,2$. The
case $r_+ = 1$ gives $\tilde \phi = \pmatrix{-1&0\cr 0&-1}$, and
$\tilde \eps = \pmatrix{1&0\cr 0&-1}$, so that $\xi = \tilde\phi -
\tilde\eps  = \pmatrix{-2&0\cr 0&0}$, which is the same as the
rank-one solution.

The case $r_+ = 2, r_-=0$ may also be reduced to the  rank-one
ansatz. In this case $\tr \tilde\phi = 0, \tilde \eps = {\bf 1}$.
Therefore, $\tilde\phi$ is a point on ${\bf CP}^1$,  which may be
parameterized as $\tilde \phi = {\bf 1} - 2 z z^{\dag}$.  Here
$z~\eps~{\bf C}^2, ||z|| = 1$, though $\phi$ is independent of the
overall phase of $z$. Then $\xi = \tilde\phi - \tilde\eps = -2 z
z^{\dag}$ and $M_2 = -2 (z z^{\dag})^a_b \psi_a \otimes \psi^{b
\dag}$.  However, we notice that $M_2$ has a {\it gauge symmetry},
under the simultaneous action of $U(2)$ on both $z$ and $\psi$. This
symmetry may be used to bring $z$ to the standard form
$\pmatrix{1\cr 0}$. It is then clear that $M_2$ reduces to the
rank-one ansatz.

    Thus, $r = 2$ does not provide any new solutions. The physical
reason for this is clear. Roughly speaking, $r = 2$ corresponds to
a baryon that  contains only valence and sea quarks $(r_+ = 2)$ or
a baryon that contains only valence and anti-quarks $(r_+ = 1)$.
However, we know that a consistent description of the baryon is
possible only if we have all  three components: valence, sea and
anti-quarks. This suggests that we should try a rank three
configuration. Indeed, a new non-trivial rank three solution
exists; we discuss it in the next section.

\subsection{Rank Three Ansatz}
\label{s-rank-three-ansatz}

The next simplest possibility is of rank three, with $r_+ = 2, r_- =
1$ so that $\tilde\eps = \pmatrix{-1&0&0\cr 0&1&0\cr 0&0&1}$. Let us
denote the negative momentum vector $\psi_-$ and the two positive
momentum vectors $\psi, \psi_+$. It will turn out that the former
represents anti-quarks while the latter represent valence and
``sea'' quarks.  The constraints on $\tilde\phi = \tilde\eps + \xi$
are
    \beq
        \tilde\phi^{\dag} = \tilde\phi,~ \tilde\phi^2 = 1,~ tr
        \tilde\phi = -1
    \eeq
Each solution to these constraints determines a one dimensional
subspace of ${\bf C}^3$, the $-1$ eigenspace of $\tilde\phi$;
i.e., a point in ${\bf CP}^2$. We may parameterize ${\bf CP}^2$ by
unit vectors $\zeta$ in ${\bf C}^3$, whose overall phase is
irrelevant:
    \beq
        \tilde\phi = -{\bf 1} + 2 \zeta \otimes \zeta^{\dag},~ ~
        ||\zeta||^2 = 1,~ ~ \zeta  \sim e^{i\alpha} \zeta.
    \eeq
Letting $\zeta = \pmatrix{\zeta_-\cr \zeta_0 \cr \zeta_+}$, we get
    \beq
        \xi = \tilde\phi - \tilde\eps = \pmatrix{ 2|\zeta_-|^2 &  2
        \zeta_-\zeta_0^* & 2 \zeta_-\zeta_+^* \cr 2 \zeta_0\zeta_-^* & -2
        + 2|\zeta_0|^2   & 2 \zeta_0\zeta_+^* \cr 2 \zeta_+\zeta_-^* &  2
        \zeta_+\zeta_0^* & -2 + 2|\zeta_+|^2}
    \eeq
Then $M_3 = \xi_b^a \psi_a \otimes \psi^{b \dag}$ has a
$U(1)\times SU(2)$ a global {\it gauge symmetry}. The phase of the
negative momentum vector $\psi_-$ may be changed and the two
positive momentum vectors $\psi$ and  $\psi_+$ may be rotated into
each other, provided we transform $\zeta$ in the same way.

Let us use the three degrees of freedom in the $SU(2)$ symmetry to
rotate the vector $\pmatrix{\zeta_- \cr \zeta_+}$ to the standard
form $\pmatrix{0 \cr \zeta_+}$, with $\zeta_+$ real and positive.
Since $||\zeta|| = 1$, $\zeta_+ = \sqrt{1 - |\zeta_-|^2}$. Then
$M_3$ has the form
    \beq
        M_3 = -2\psi\otimes \psi^\dag +  2 |\zeta_-|^2 [
        \psi_-\otimes\psi_-^\dag -\psi_+\otimes\psi_+^\dag] + 2
        [\zeta_-\zeta_+\psi_-\otimes\psi_+^\dag +  \zeta_+\zeta_-^*
        \psi_+\otimes\psi_-^\dag].
    \eeq
We see that $M_3$ is still invariant under
    \beq
        \psi \to e^{i \alpha} \psi,~ \psi_+ \to e^{i \alpha_+} \psi_+,~ \psi_- \to e^{i
        \alpha_-} \psi_-,~ \zeta_- \to e^{i (\alpha_+ - \alpha_-)}
        \zeta_-
    \eeq
In the case of the separable ansatz, the phase space of rank-one
configurations was ${\cal P}({\cal H}_+)$. The analogous space of
rank three configurations is more involved. $\psi$ and $\psi_+$
determine a point in the flag ${\cal F}l_{2,1}({\cal H}_+)$.
$\psi_-$ determines a point in the projective space $P({\cal H}_-)$.
The total space of rank three configurations is then a bundle over
${\cal F}l_{2,1}({\cal H}_+) \times  P({\cal H}_-)$, with fibre the
unit interval $[0,1]$, parameterized by the absolute value of
$\zeta_-$.

We may use the $\zeta_- \to e^{i (\alpha_+ - \alpha_-)} \zeta_-$
freedom to make $\zeta_-$ real. Thus, the rank three
configurations  which satisfy the constraints are given by
    \beq
        M_3 = -2\psi\otimes \psi^\dag +  2 \zeta_-^2
        [\psi_-\otimes\psi_-^\dag -\psi_+\otimes\psi_+^\dag] + 2
        \zeta_-\zeta_+ [\psi_-\otimes\psi_+^\dag  + \psi_+\otimes\psi_-^\dag].
    \eeq
Baryon number may be expressed as
    \beq
        B= -\half tr M_3 =  \int_0^\infty\left\{|\tilde\psi(p)|^2+
        \zeta_-^2\left[|\tilde\psi_+(p)|^2-|\tilde\psi_-(-p)|^2\right]\right\}
        [dp].
    \eeq
The total momentum $P = -\half N \int p \tilde M(p,p) [dp]$
becomes
    \beq
        \bar P = {P \over N} = \int_0^\infty p\big\{|\tilde\psi(p)|^2|
        +\zeta_-^2[|\tilde\psi_{-}(-p)|^2+|\tilde\psi_{+}(p)|^2]\big\}[dp]
    \eeq
The variational quantities $\psi,~ \psi_-,~ \psi_+,~ \zeta_-$ are
determined by minimizing the energy of the baryon. The kinetic
energy is simple in momentum space
    \beq
        K=\int_0^\infty \half[p+{\mu^2\over p}]
        \big\{|\tilde\psi(p)|^2|
        +\zeta_-^2[|\tilde\psi_-(-p)|^2+|\tilde\psi_{+}(p)|^2]\big\}
        [dp].
    \eeq
The expression for potential energy is simpler in position space,
but is lengthy since it is quadratic in $M(x,y)$. Unlike in the
rank-one case, our wave functions also depend on a discrete index. A
discrete analog of the Fourier transform, which diagonalizes $\xi$
makes the potential energy simpler. Let,
    \beq
        \psi_1={1\over \surd2}\bigg\{\surd[1-\zeta_-]\psi_-  -
        \surd[1+\zeta_-]\psi_+\bigg\} \eeq \beq \psi_2={1\over \surd
        2}\bigg\{\surd[1+\zeta_-]\psi_-  + \surd[1-\zeta_-]\psi_+ \bigg\}
    \eeq
In this basis $\xi = \pmatrix{2\zeta_- & 0 & 0 \cr 0 & -2 & 0 \cr
0 & 0 & -2 \zeta_-}$ is diagonal. However, the new wave functions
$\psi_{1,2}$ are no longer eigenstates of $\eps$. $M_3$ becomes
($\psi_0 := \psi$):
    \beq
        -\half M_3 = \psi_0\otimes \psi_0^\dag +
        \zeta_-\big\{\psi_1\otimes\psi_1^\dag-\psi_2\otimes\psi_2^\dag\big\}
    \eeq
And the potential energy is given by
    \beqs
        U&=&{\tilde g^2 \over 2} \int \bigg\{|\psi_{0}(x)|^2
        V_{00}(x)+ \zeta_-^2|\psi_{1}(x)|^2
        V_{11}(x)+\zeta_-^2|\psi_{2}(x)|^2 V_{22}(x)\bigg\}dx\cr & &
        +{\tilde g^2\over 2}2\;{\rm Re}\; \int   \bigg\{\zeta_-
        \psi_{0}(x)\psi_{1}^*(x) V_{10}(x)-\cr & &
        \zeta_-\psi_{0}(x)\psi_{2}^*(x)
        V_{20}(x)-\zeta_-^2\psi_{1}(x)\psi_{2}^*(x) V_{21}(x)\bigg\}dx.
    \eeqs
The mean-fields $V_{\alpha \beta}$ are defined as
    \beq
        V_{\alpha\beta}(x) = \half \int dy \psi_{\alpha}(y) |x-y|
        \psi_{\beta}^*(y)
    \eeq
and satisfy the hermiticity condition $V_{\alpha\beta}^*(x) =
V_{\beta\alpha}(x)$. They are the solutions to the differential
equations
    \beq
        V_{\alpha\beta}''(x)=\psi_{\alpha}(x)\psi_{\beta}^*(x).
    \eeq
with the boundary conditions \beq V_{\alpha\beta}(0) = \half \int
dy \psi_{\alpha}(y) |y| \psi_{\beta}^*(y) \eeq and \beq
V_{\alpha\beta}'(x)\to \half\delta_{\alpha\beta},\;{\rm for}\;
|x|\to \infty.  \eeq

Thus we have defined the phase space and hamiltonian of the rank
three ansatz.  As in the rank-one case, (section
\ref{s-quantize-rank-1}), we can work out the Poisson bracket
relations among $\psi,~\psi_{\pm}$ implied by  the Poisson brackets
of the $\tilde M(p,q)$. We can then quantize this classical
dynamical system to obtain an interacting quantum system of quarks
and anti-quarks, whose color is not being counted explicitly. These
particles would be the rank three analogs of the $N$ bosons we
encountered in the rank-one ansatz. However, we will follow a
different strategy this time, which clarifies the connection of the
rank three ansatz to the quark model.

\subsection{Fock Space Description}
\label{s-fock-space-rank-3}

In what follows, we will give two Fock space descriptions of the
rank three configurations. In the first case, we will generalize the
rank-one ($N = \infty$) formula $|M_1> = b^{\dag}_{\psi} |\tilde 0>$
to the case $M = M_3$. This will give us a description of the baryon
in terms of quasi-particles which are certain colorless combinations
of the true quarks and anti-quarks. Next we will generalize the
(finite $N$) formula $|V> = q^{1 \dag}_{\tilde{\psi}} \cdots q^{N
\dag}_{\tilde{\psi}} |0>$ in order to get a description in terms of
colored particles.

In order to find the Fock space state corresponding to $M_3$, we use
an analog of the Pl\"ucker embedding \cite{chern,mickelsson} of the
\gr in the Fock space, just as we did in the rank-one case in \S
\ref{s-fock-space-den-matrix-rank-1}. The filled states $\chi$ are
$-1$ eigenstates of $\Phi_3$
    \beq
        \Phi_3 ~\chi = (\eps + \xi_b^a \psi_a \psi^{b \dag}) ~\chi =
        ~ - \chi
    \eeq
There are two types of solutions to this eigenvalue problem. First
let us look in the orthogonal complement of the subspace spanned
by $\{\psi_a\}_{a = 1}^r$:
    \beq
        \eps \chi = - \chi~ {\rm and}~ <\psi_b,\chi> = 0,~ b=1,\cdots r.
    \eeq
This means $\chi$ is a negative momentum state, orthogonal to
$\psi, \psi_{\pm}$. Thus, all negative momentum states, orthogonal
to $\psi_-$ are filled.

For the other type of solution, we consider those $\chi$ that  lie
in the span of the $\psi_a$.  Let $\chi = \sum_{a=1}^r
\tilde\chi_a \psi_a$.
    \beq
        \eps_a \tilde\chi_a + \xi_a^b \tilde \chi_b = -\tilde\chi_a.
    \eeq
Thus, we are looking for $-1$ eigenvectors of $\tilde\eps + \xi =
\pmatrix{-1+2|\zeta_-|^2 & 0 & 2 \zeta_- \zeta_+ \cr 0 & -1 & 0
\cr 2 \zeta_+ \zeta_- & 0 & 1 - 2|\zeta_-|^2}$.  $\tilde
\chi^{(1)} = \pmatrix{0 \cr 1 \cr 0}$ and $\tilde\chi^{(2)} =
\pmatrix{-\zeta_+ \cr 0 \cr \zeta_-}$ are the solutions.  Thus,
the states
    \beq
        \chi^{(1)} = \psi ~ {\rm and} ~ \chi^{(2)} = -\zeta_+\psi_- +
        \zeta_- \psi_+
    \eeq
are also filled. Thus the Fock space state corresponding to the
rank three ansatz $M_3$ is a semi-infinite wedge product:
    \beq
        \cdots f^{2'} \wedge f^{1'} \wedge \psi \wedge (-\zeta_+\psi_- + \zeta_- \psi_+)
    \eeq
Here $f^{j '}$ is a dual basis in the orthogonal complement of
$\psi_-$ in ${\cal H}_-$. As before, we focus on the deviation
from the vacuum:
    \beq
        |\tilde 0> = \cdots f^{2'} \wedge f^{1'} \wedge \psi_{-}^{'}
    \eeq
Then $|M_3>$ is obtained by populating $\psi$ and $(-\zeta_+\psi_-
+ \zeta_- \psi_+)$ and depopulating $\psi_-$, starting from the
vacuum $|\tilde 0>$.
    \beq
        |M_3> = (-\zeta_+ b_{\psi_-}^{\dag} + \zeta_-
        b_{\psi_+}^{\dag}) b_{\psi_-} b_\psi^{\dag} |\tilde 0>
    \eeq
However, the operators $b_{\psi_\pm}^{\dag},b_\psi^{\dag},
b_{\psi_\pm},b_\psi$ are not the creation-annihilation operators
of quarks. They create and annihilate the quasi-particles of the
rank three ansatz. This point of view is simpler for computations,
and avoids the unobservable color quantum number.

However, we are more accustomed to thinking in terms of the Fock
space of quarks and anti-quarks, which carry the color quantum
number. We now  determine the state consisting of ``valence'',
``sea'' and anti-quarks, $|VSA>$  that corresponds to the rank
three ansatz. We will see that the advantage of  this point of
view is that it has a dual description in terms of an embellished
quark model, which we discuss in the next section.

Recall that the Fock space state $|V>$ corresponding to the rank-one
ansatz is
    \beq
        |V> = q^{1 \dag}_{\tilde{\psi}} \cdots q^{N \dag}_{\tilde{\psi}} |0>
    \eeq
where, $q^{a \dag}_{\tilde \psi}$ creates a quark with color $a$ in
the state $\tilde \psi$. These operators satisfy canonical
anti-commutation relations. As before, the state $|VSA>$ in the
Hartree approximation is determined by the condition that the form
factor of the quark density matrix n the solitonic state $|VSA>$
should be the classical density matrix in the large-$N$ limit
    \beq
        <VSA|\hat \rho(p,q)|VSA> = \tilde\rho_3(p,q).
    \eeq
Here $\hat \rho(p,q) = {1 \over N} q^{a \dag}(p)~q_a(q)$ and
$\tilde\rho_3(p,q) = \half (2\pi \delta(p-q) - \tilde M_3(p,q) -
\tilde \eps(p,q))$. While it was easy to guess the form of $|V>$,
we will use an indirect method to find $|VSA>$. We know that $|V>$
satisfies the condition $ <V|\hat \rho(p,q)|V> = \tilde\rho_1(p,q)
$. We proceed by finding the operator that transforms
$\tilde\rho_1$ into $\tilde\rho_3$. Then we represent this
operator on the Fock space and use it to transform $|V>$ into
$|VSA>$. This is feasible since the \gr carries a transitive
action of the infinite dimensional restricted unitary group
(section \ref{s-CHD-grassmannian}). We may obtain the rank three
configuration $\Phi_3 = M_3 + \eps$ from the separable ansatz
$\Phi_1 = M_1 + \eps$ by a unitary transformation:
    \beq
        U^{\dag}(\eps - 2 \psi \otimes \psi^{\dag})U = \eps +
        \xi_b^a\psi_a\otimes \psi^{\dag b}.
    \eeq
(We work with $\Phi = M + \eps$ or $\rho$ since they transform
homogeneously under unitary transformations, while due to normal
ordering,  $M$ does not.) The infinite dimensional restricted
unitary group is disconnected, unlike  its finite dimensional
counterparts. However, since both $M_1$ and $M_3$ are
configurations of baryon number one, $U$ lies in the connected
component  of the identity, and is of the form $U = e^{iA}$, for
hermitian $A$.  From the expressions for $M_1$ and $M_3$,
    \beqs
        M_1 &=& -2~ \psi\otimes \psi^\dag \cr
        M_3 &=& -2~ \psi\otimes \psi^\dag +  2 \zeta_-^2 [
            \psi_-\otimes\psi_-^\dag -\psi_+\otimes\psi_+^\dag ] \cr &&
            + 2 ~ \zeta_-\zeta_+ [\psi_-\otimes\psi_+^\dag +
            \psi_+\otimes\psi_-^\dag].
    \eeqs
We see that $U$ must be the identity except on the span of
$\psi_-$ and $\psi_+$. The restriction of $\Phi_1$ to this
subspace is $-\sigma_3 = \pmatrix{-1 & 0 \cr 0 & 1}$, while the
restriction of $\Phi_3$ is $s \sigma_1 + (r-1) \sigma_3$, where $r
= 2 \zeta_-^2$ and $s = 2 \zeta_-\zeta_+$. Here $\sigma_i$ are the
Pauli matrices. Thus, the restriction of $A$ to this subspace
satisfies
    \beq
        e^{-iA} (-\sigma_3) e^{iA} = s \sigma_1 + (r-1) \sigma_3.
    \eeq
Therefore $A$ is a $2 \times 2$ traceless hermitian matrix ${\bf
\sigma} . {\bf w}$. Here {\bf w} is the vector in {\bf R}$^3$
about which $(0,0,-1)$ must be rotated by an angle $2|{\bf w}|$ to
reach $(s,0,r-1))$.  We get
    \beq
        A(p,q) = {i \arcsin(\zeta_-)} (\psi_-(p)\psi_+^*(q) -
        \psi_+(p)\psi_-^*(q)) ~ ~ {\rm and} ~ ~ \tilde U(p,q) = e^{i A(p,q)}.
    \eeq
Let us denote $\theta = \arcsin(\zeta_-)$. Therefore we have
($\rho = \half (1 - \Phi)$),
    \beq
        \int e^{-iA(p,r)} \tilde \rho_1(r,s) e^{iA(s,q)} [dr] [ds] =
        \tilde \rho_3(p,q).
    \eeq
Now we look for a representation of $\tilde U(p,q)$ as a unitary
operator $\hat U$ acting on the quark creation annihilation
operators. Moreover, suppose we can show that
    \beqs
        \hat U^\dag \hat q(q) \hat U = \int [ds] \hat q(s) \tilde
        U^\dag(s,q) \cr \hat U^\dag \hat q^\dag (q) \hat U = \int
        [dr]
        \tilde U(p,r) \hat q^\dag (r).
    \eeqs
Then
    \beq
        |VSA> = \hat U ~ |V>.
    \eeq
since for this choice, the expectation value condition is
satisfied:
    \beqs
        <VSA| {1 \over N} \hat q^\dag(p)~ \hat q(q) |VSA> ~ &=& ~ <V| \hat
        U^\dag \hat q^\dag(p) \hat U \hat U^\dag \hat q(q) \hat U|V> \cr &=& \int
        [dr][ds]<V|\tilde U(p,r) \hat q^\dag(r) \hat q(s) \tilde U^\dag
        (s,q)|V> \cr &=& \int [dr] [ds] \tilde U(p,r) \tilde \rho_1(r,s)
        \tilde U^\dag(s,q) \cr &=& \tilde \rho_3(p,q).
    \eeqs
It only remains to find the second quantized representative $\hat
U$ and show that the quark creation and annihilation operators
transform in the desired manner under it. $\hat U$ may be obtained
by promoting the wave functions that appear in $U = e^{i A(p,q)}$
to operators:
    \beq
        \hat U = e^{i \hat A} = e^{- \arcsin(\zeta_-) (\hat q_{a
        \psi_-}  \hat q^{a \dag}_{\psi_+} - \hat q_{a \psi_+} \hat
        q^{a  \dag}_{\psi_-})}
    \eeq
We sum over colors to produce a color-singlet. We can check that
$\hat q(p), ~ \hat q^\dag(q)$ have the desired transformation
properties under $\hat U$. For example, we show that $\hat U^\dag
\hat q^\dag (p) \hat U = \int [dr] \tilde U(p,r) \hat q^\dag (r)$.
For this we will need the preliminary result
    \beq
        [\hat A,~ \hat q^\dag(p) ] = \tilde \psi_+(p) \hat q^\dag_{\tilde \psi_-}
        - \tilde \psi_-(p) \hat q^\dag_{\tilde \psi_+}
    \eeq
which is easily verified. Then
    \beqs
        \hat U^\dag \hat q^\dag(p) \hat U &=& e^{-i\hat A} \hat
        q^\dag(p) e^{i\hat A} \cr &=& (1-i\hat A +O(\theta)^2) ~ \hat
        q^\dag(p) ~ (1+i\hat A +O(\theta)^2) \cr &=& \hat q^\dag(p) -
        i[\hat A,~\hat q^\dag(p) ] + O(\theta)^2 \cr &=& \hat q^\dag(p) -
        i(\tilde \psi_+(p) \int [dr] \tilde \psi_-^*(r) \hat q^\dag(r) - \tilde \psi_-(p)
        \int [dr] \tilde \psi_+^*(r) \hat q^\dag(r)) + O(\theta)^2 \cr &=& \hat
        q^\dag(p) - i \int [dr] \hat q^\dag(r) (\tilde \psi_+(p) \tilde \psi_-^*(r) -
        \tilde \psi_-(p)\tilde \psi_+^*(r)) + O(\theta)^2 \cr &=& \int [dr] [ 2\pi
        \tilde \delta(p,q) + i \tilde A(p,q) + O(\theta)^2 ] \hat
        q^\dag(r) \cr &=& \int e^{i \tilde A(p,r)} \hat q^\dag(r) [dr] \cr
        \hat U^\dag \hat q^\dag(p) \hat U &=& \int [dr] \tilde U(p,r) \hat
        q^\dag(r)
    \eeqs
as desired. We conclude that
    \beq
        |VSA> = e^{-i \hat A} |V> = e^{\arcsin{\zeta_-} [q_{a\psi_-}
        q^{a\dag}_{\psi_+}-{\rm h.c.}]}|V> .
    \eeq
$|VSA>$ is the second quantized Fock space state corresponding to
the rank three ansatz $M_3$. Here the $q^{\dag a}, q_a$ create and
annihilate colored quarks. We see that $|VSA>$ may be obtained from
the valence-quark state by a rotation. In the next section we will
show that the same result could have been obtained by a Bogoliubov
transformation, applied to the valence-quark model.

        \subsection{Valence, Sea and Antiquarks: Bogoliubov Transformation}
        \label{s-bogoliubov}

The interacting valence-quark model we derived earlier (section
\ref{s-quantize-rank-1}) ignored the anti and sea-quark degrees of
freedom. Now, we show  how they can be included by performing a
unitary transformation on the  Fermionic Fock space. This is like a
Bogoliubov transformation that mixes  positive and negative momentum
states.

In general, the baryon can be in a superposition of states
containing $N+\eta$ quarks and $\eta$ anti-quarks for $\eta =
0,1,\cdots,\infty$.  This is because only the difference, baryon
number, is a  conserved quantum number in the full theory. The
valence-quark approximation amounts to ignoring all states for $\eta
> 0$ and in the Hartree approximation, corresponds to the state
$|V> = q^{1 \dag}_{\tilde{\psi}} \cdots q^{N \dag}_{\tilde{\psi}}
|0>$.

Let us continue to work within a factorized Hartree approximation,
ignoring correlations except when required by the Pauli exclusion
principle or color-invariance.  The next simplest possibility beyond
the valence approximation is to consider only states with $\eta \leq
N$. In the {\it ground-state} of the baryon, we  find that this is a
good approximation.

The quarks cannot all occupy the positive momentum state
$\tilde\psi(p)$ as in the valence approximation, because they would
violate the Pauli exclusion principle: there is no completely
anti-symmetric tensor in more than $N$ indices transforming under
$SU(N)$.  In the ground-state, it is energetically  favorable to
introduce just one more positive momentum state  $\tilde\psi_+(p)$
to accommodate the quarks. As for the anti-quarks,  they occupy the
lowest available negative momentum state $\tilde\psi_-(p)$.

To obtain a state containing valence, ``sea'' and anti-quarks (say
$|VSA>$), we apply a unitary transformation to the valence-quark
state $|V>$. It will be determined by the variational wave functions
$\psi, \psi_+$ and  $\psi_-$ in additional to an ``angle of
rotation'' $\theta$. These are to be determined by minimizing the
$\masssq$ of the baryon.  In order to produce a state $|VSA>$ of
norm 1, we require the transformation to be unitary. In analogy with
a Bogoliubov transformation, the operator that creates a quark in
state $\psi_+$ and an anti-quark in state $\psi_-$ is
$a_{b\psi_-}a^{b\dag}_{\psi_+}$. We have summed over colors to
produce a color-singlet state. Our variational ansatz for the state
$|VSA>$ is then $e^{\theta[a_{b\psi_-}a^{b\dag}_{\psi_+}-{\rm
h.c.}]}|V>$.

We see that this is identical to the state $|VSA>$ we obtained
from  the rank three ansatz, provided we identify $\zeta_-$ with
$\sin\theta$ where $\theta$ is the angle of the Bogoliubov
rotation.

This point of view confirms the interpretation of $\psi_-$ as the
wave function describing anti-quarks. $\psi$ and $\psi_+$ may be
interpreted as ``valence'' and ``sea'' quark wave functions. Their
orthogonality can be understood as a consequence of the Pauli
exclusion principle.  $0 \leq \zeta_- \leq 1$ measures the departure
from the valence-quark approximation. The probability of finding an
anti-quark in the baryon is ${\zeta_-^2 \over {1 + 2 \zeta_-^2}}$.

While the physical meaning of these quantities are best understood
in the context of the quark model, actual computations are most
easily performed in the context of Hadrondynamics. We turn next to
an estimation of the anti-quark content of the two-dimensional
baryon.

\subsection{Variational Estimate of Antiquark Content}
\label{s-anti-content-for-large-N}

In \S \ref{s-exp-ansatz} we found that the absolute minimum of
energy (baryon mass) in the limit of chiral symmetry occurs for a
rank-one configuration, in  large-$N$ limit. In this limit, the
anti-quark content of the baryon vanishes. We find that for small
values of current quark mass (${m \over {\tilde g}} << 1$), the
baryon has a small but non-vanishing  anti-quark content. We already
saw that the leading effect of finite $N$ was to make the wave
functions $\tilde \psi_a(p)$ vanish beyond  the total baryon
momentum $P$.

    To determine the rank three configuration, we can derive
integral  equations for the minimization of ${\cal M}$. However,
from our exact solution, we expect that the anti-quark content is
small for small current quark masses.  To estimate
$\zeta_-(\msqovgsq)$, we do a variational minimization of  ${\cal
M}$, assuming some simple functional forms for $\psi_\pm, \psi$,
which are suggested by the form of the exact solution. Moreover,
we work  in the limit $N = \infty$. As before, we pick
    \beq
        \tilde\psi(p)  = C ({p \over {\bar P}})^a  e^{{- b p} \over
        {\bar P}},~  p \geq 0
    \eeq
Here $\bar P = {P \over N}$ is the mean baryon momentum per color,
$C$ is fixed by normalization and $b$ is fixed by the choice of
Lorentz reference  frame. $\psi_+$ must be orthogonal to $\psi$,
so it must have a node:
    \beq
        \tilde\psi_+(p) = C_+ ({p \over {\bar P}})^a ({p \over {\bar
        P}} - C_1)  e^{{- b p} \over {\bar P}}, ~ p \geq 0
    \eeq
Here $C_1$ is fixed by orthogonality. In the ground-state the
anti-quark wave function has no nodes (besides at $p = 0$), and for
ease of computation we pick it to be the same function (but for $p
\leq 0$) as the valence-quark wave function. (Note: this is {\it
not} required by any symmetry.)
    \beq
        \tilde\psi_-(p) = \tilde\psi(-p) , ~ p \leq 0
    \eeq
Since  a Lorentz boost in the longitudinal direction corresponds
to  a rescaling $p \to \lambda p$ in null coordinates, we pick a
reference frame in which $b = 1$, and minimize the Lorentz
invariant ${\cal M}^2$ in that frame. This determines the Lorentz
invariant quantities $\zeta_-({\msqovgsq})$ and $a(\msqovgsq)$.

\begin{figure}
\centerline{\epsfxsize=6.truecm\epsfbox{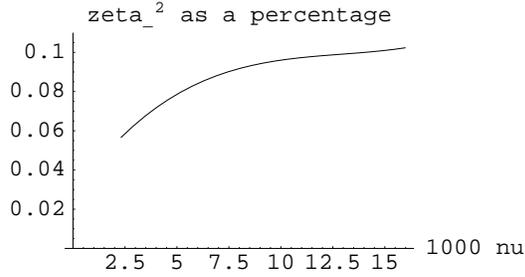}} \caption{
$\z^2$ measuring departure from valence-quark approximation plotted
as a percent as a function of $1000 \msqovgsq$. The probability of
finding an anti-quark in the baryon is $\z^2 \over {1 + 2 \z^2}$.
The valence-quark approximation becomes exact in the chiral limit.
} \label{f-zeta-minus}
\end{figure}

\begin{figure}
\centerline{\epsfxsize=6.truecm\epsfbox{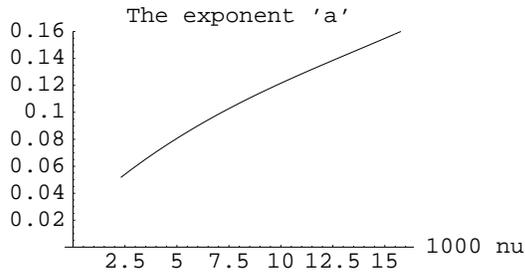}} \caption{ The power
$p^a$ governing behavior of the wave functions near the origin,
plotted as a function of  $1000 \nu = 1000 \msqovgsq$. The wave
functions tend to a constant as $p \to 0$ in the chiral limit.}
\label{f-power-a-rank-3}
\end{figure}

The $\masssq$ of the baryon is ${\cal M}^2 = P(2E-P)$
    \beqs
        {{\cal M}^2 \over \tilde g^2 N^2} &=&  \bigg[-\half \int
        p\tilde M(p,p)[dp] \bigg] \bigg[-\nu \int {{\tilde M(p,p)} \over 2p} [dp]
        + {1 \over \pi} \int {{\tilde M(p,p)} \over 2p}  [dp] \cr && + {1
        \over 4}\int dxdy |M(x,y)|^2\half|x-y|\bigg]
    \eeqs
where ($\nu = {m^2 \over \tilde g^2}$). We abbreviate
    \beq
        {{\cal M}^2 \over \tilde g^2 N^2} =
        \bar P * (\nu KE + SE + PE).
    \eeq
Using the rank three ansatz,
    \beq
        M_3=-2\psi\otimes \psi^\dag +  2 \zeta_-^2 [
        \psi_-\otimes\psi_-^\dag -\psi_+\otimes\psi_+^\dag ] + 2
        \zeta_- \sqrt{1 - \z^2} [\psi_-\otimes\psi_+^\dag +
        \psi_+\otimes\psi_-^\dag]
    \eeq
for the above choices of wave functions we find:
    \beq
        \bar P(\zeta_-,a) = \half + 2 \zeta_-^2 + a(1 +
        2 \zeta_-^2), ~ KE(\zeta_-,a) = {{1 + 2 \zeta_-^2} \over a},~
        SE(\zeta_-,a) = - {{1 + 2 \zeta_-^2} \over \pi a}
    \eeq
    \beq PE(\zeta_-,a) = (U_0(a) + \zeta_-
        \sqrt{1 - \zeta_-^2} U_1 (a) +  \z^2 U_2(a) + \z^3 \sqrt{1 - \z^2}
        U_3(a) + \z^4 U_4(a))
    \eeq
The potential energies $U_i(a)$ can be calculated analytically for
the most part, and we use the symbolic package Mathematica for
this purpose. We must minimize ${\cal M}^2$ in the strip $0 \leq
\zeta_- \leq 1, a \geq 0$.  We find that for small but
non-vanishing $\msqovgsq$, ${\cal M}^2$ has a minimum  for small
but non-zero values of both $\z$ and $a$, indicating a  small
anti-quark content. To see that such a non-trivial minimum exists,
it suffices to consider the factor $(\nu KE + SE + PE)$, since
$\bar P$ is monotonic in $\z$ and $a$. Moreover, since $\z$ turns
out to be small, it is enough to consider only terms in $PE(\z,a)$
which dominate for small $\z$
    \beq
        PE(\zeta_- \to 0,a) \to (U_0(a) + \zeta_-  \sqrt{1 -
        \zeta_-^2} U_1 (a) + \z^2 U_2(a))
    \eeq
$U_0(a)$ is the same as in the valence-quark approximation and was
obtained in \S \ref{s-exp-ansatz}. For small $a$,
    \beq
        U_0(a) = {1 \over {\pi a}}    +   {\pi a \over 3} + O(a^2)
    \eeq
We also find that $U_1(a)$ is negative and goes to zero as $a \to
0$.  As for $U_2(a)$, we find,
    \beq
        U_2(a) \to {2 \over \pi a } + {\rm constant} . a + O(a^2)
    \eeq
Thus, the terms proportional to $1 \over a$ in PE and SE cancel
out exactly, and we get
    \beq
        (\nu KE + SE + PE)(\z,a) = d_1 a +  \nu + d_2 a \z^2
        + 2 \nu {\z^2 \over a} - d_3 a^{d_4} \z \sqrt{1 - \z^2}
    \eeq
for positive constants  $d_i, ~ i = 1, \cdots, 4$. Isolating the
$a$ and $\z$ dependence below we see that for $\nu > 0$, $(\nu KE
+ SE + PE)(\z,a)$, and hence ${\cal M}^2$  must have a non-trivial
minimum for  some $\z > 0$, $a > 0$:
    \beqs
        (\nu KE + SE + PE)(\z) &=& f_0 - f_1 \z \sqrt{1 - \z^2}  + f_2 \z^2, \cr
        (\nu KE + SE + PE)(a) &=& {h_{-1} \over a} - h_0 a^{h_2}  + h_1 a.
    \eeqs
Here the $f_i(a), h_j(\z)$  are positive. Thus we estimate that the
minimum of baryon $\masssq$ on the rank three sub-manifold occurs
for a small but non-trivial anti-quark content, when the parameter
$\msqovgsq$  is small. However, as the plots (fig.
\ref{f-power-a-rank-3} \& \ref{f-zeta-minus}) show, in the limit of
chiral symmetry, $\zeta_- \to 0, ~a \to 0$, recovering the exact
rank-one solution.
\begin{figure}
\centerline{\epsfxsize=6.truecm\epsfbox{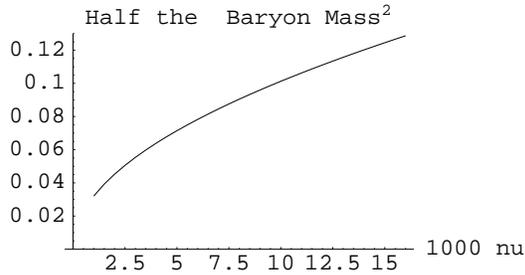}} \caption{
Variational upper bound for $\half$ mass$^2$ of baryon in the rank
three approximation plotted as a function of $1000 \msqovgsq$. The
baryon is massless in the chiral limit.  }
\end{figure}
\begin{figure}
\centerline{\epsfxsize=6.truecm\epsfbox{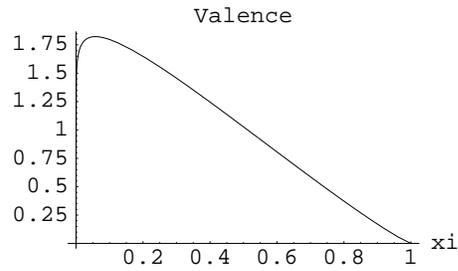}} \caption{
Variational approximation for valence-quark density $V(\xi) = 2\pi
|\psi(\xi)|^2$ in the rank three ansatz. It is normalized to 1.}
\end{figure}
\begin{figure}
\centerline{\epsfxsize=6.truecm\epsfbox{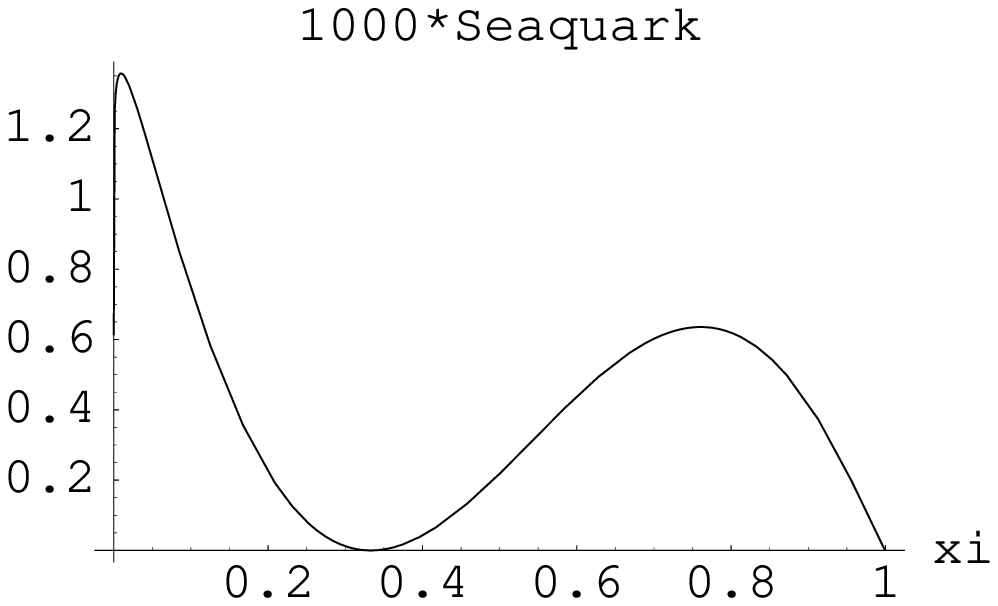}} \caption{
Variational approximation for sea quark distribution $2\pi \z^2
|\psi_+(\xi)|^2$ in the rank three ansatz. It is normalized to
$\z^2$. }
\end{figure}
\begin{figure}
\centerline{\epsfxsize=6.truecm\epsfbox{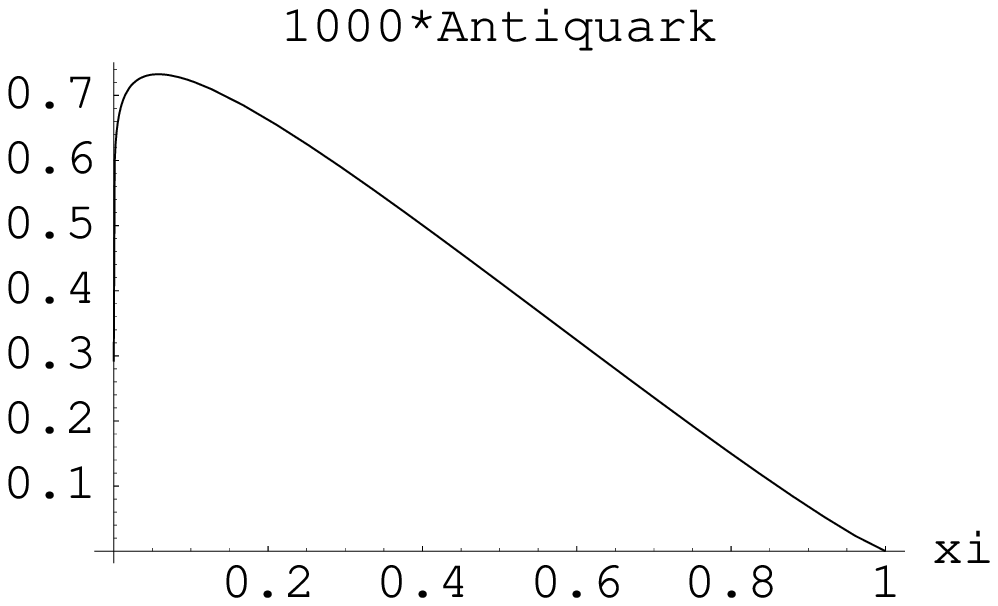}}
\caption{Variational approximation for antiquark quark distribution
$2\pi \z^2 |\psi(\xi)|^2$ in the rank three ansatz. It is normalized
to $\z^2$.}
\end{figure}
\begin{figure}
\centerline{\epsfxsize=6.truecm\epsfbox{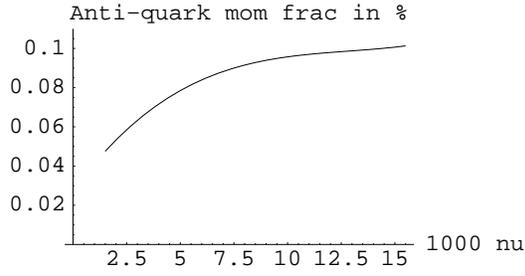}}
\caption{Fraction of baryon momentum carried by anti-quarks as a
function of $\nu = {m^2 \over \tilde g^2}$.}
\end{figure}
To include the leading $1 \over N$ corrections the wave functions
must vanish beyond $p=P$. Setting $\xi = p/P$, we pick
    \beqs
        \psi(\xi) = C \xi^a (1 - \xi)^b, && \psi_+(\xi) = C_+ \xi^a (\xi-C_1)
        (1-\xi)^b ~{\rm for}~ 1 \geq \xi \geq 0 \cr  {\rm and} ~~
        \tilde\psi_-(\xi) &=& \tilde\psi(-\xi)~ {\rm for} ~ -1 \leq \xi \leq 0
    \eeqs
The behavior of the wave functions for $p \to 0$, is insensitive to
the boundary condition for large $p$, so we may use the value of $a$
determined in the large-$N$ limit. Similarly, we ignore the $1 \over
N$ correction to the small parameter $\z$ (See Fig
\ref{f-power-a-rank-3} \& \ref{f-zeta-minus}). $b$ is determined in
terms of $a$ and $\z$ by the momentum sum rule:
    \beq
        {P \over N} = \int_0^P p\big\{|\tilde\psi(p)|^2|
        +\zeta_-^2[|\tilde\psi_{-}(-p)|^2+|\tilde\psi_{+}(p)|^2]\big\}[dp].
    \eeq
Ignoring the small difference between the momentum carried by sea
and anti-quarks (both of which are suppressed by factors of $\z^2$
compared to the valence-quark contribution) we find
    \beq
        b \sim -1 + N(\half + \z^2) + a(-1 + N + 2 N \z^2)
    \eeq
We plot the valence, sea and anti-quark  densities as functions of
$\xi = {p \over P}$ for $\msqovgsq = 10^{-3}$, $N = 3$. For these
parameters we estimate  $\z^2 = 4 \times 10^{-4}$, $a = 3.5 \times
10^{-2}$ and $b = 0.57$.

\section{Application to Deep Inelastic Scattering}
\label{s-appl-to-DIS}

The structure of nucleons (proton and neutron) is measured in Deep
Inelastic Scattering (DIS) experiments. Electrons (or muons,
neutrinos) are scattered off nucleons by the exchange of a
space-like ($q^2 < 0$) virtual photon (or weak gauge boson). Thus,
the electro-weak force is used as a probe of the strong force
binding hadrons, just as Rutherford used alpha particles to study
atomic structure. The DIS experiments of the late 1960s and early
1970s \cite{friedman-kendall} discovered scaling. The structure of
the proton was roughly independent of the wave length of the
virtual photon used as a probe: the measured structure functions
were independent of $q^2$. The quark-parton model of Feynman
\cite{bjorken-paschos,feynman-photon-hadron} explained this
phenomenon in terms of point like quarks, which had to be weakly
interacting at high energies. Soon after, it was discovered that
the only renormalizable four dimensional quantum field theory of
quarks with this property of asymptotic freedom at high momentum
transfers is a non-abelian gauge theory of quarks coupled to
gluons, which we now refer to as QCD. Perturbative QCD predicts
small logarithmic violations of scaling. The evolution of
structure functions as $Q^2 (\equiv -q^2)$ is increased is given
by the DGLAP equations \cite{dglap}. This $Q^2$ dependence has
been experimentally verified. However, the measured structure
functions also contain information on how quarks and gluons bind
to form a relativistic bound state. This is encoded in the
momentum fraction $\xi = p/P$ or Bjorken variable $x_B = {Q^2
\over 2 P.q}$ dependence of structure functions, or parton
distribution functions. Like atomic wave functions, they tell us
the probability of finding a quark with a given momentum $p$ in
the proton whose momentum is $P$. It has not been possible to
describe the formation of hadronic bound states by perturbing
around the limit of non-interacting quarks. What is more, unlike
the electrons in an atom, quarks appear to be permanently confined
inside hadrons, and have never been isolated. Understanding the
bound state structure of hadrons and the phenomenon of confinement
remains a challenge in $3+1$ dimensions.

However, in Deep Inelastic Scattering, the transverse momenta of
partons are small compared to their longitudinal momenta.
Moreover, a perturbative treatment of tansverse momenta with an
upper cut off $Q$ leads to the same leading $Q^2$ dependence for
structure functions as predicted by the operator product expansion
in perturbative QCD. We have determined the bound state structure
of quarks in a baryon in $1+1$ dimensions in preceding sections.
This leads us to propose a model for the $x_B$ dependence of
structure functions at an initial low value of $Q^2 = Q_0^2$ where
transverse momenta are ignored. Our quark distribution functions
are to be used as initial conditions for $Q^2$ evolution and
compared with experimental data at higher values of $Q^2$.

It is an experimental fact that at low $Q^2$, about half the
momentum of the proton is carried by gluons \cite{cteq}. However,
two-dimensional QCD does not contain any dynamical gluon degrees of
freedom, which are necessary to find the gluon distribution. Rather
than two-dimensional QCD, it is more appropriate to study the
dimensional reduction of QCD from four to two dimensions, which
includes the transverse polarization states of the gluon field. But
understanding the non-perturbative effects of the gluon degrees of
freedom is substantially harder. Indeed, the second part of this
thesis is motivated by a desire to do just that. For the present, we
account for the omission of gluons by merely assuming that the
quarks carry only a fraction $f$ of the baryon momentum. Moreover,
we found that anti-quarks and sea quarks carry a very small portion
of the baryon momentum in two dimensions (see \S
\ref{s-anti-content-for-large-N}). Phenomenological fits
\cite{cteq,mrst} to experiment show that anti and sea quarks carry
very small fractions of baryon momentum at low $Q^2$, the bulk of it
is carried by valence-quarks and gluons. Thus we make the following
ansatz: at an initial low value $Q_0^2$, the valence-quarks carry a
fraction $f$ of baryon momentum and their probability distribution
$V(\xi) = {P \over 2\pi} |\tilde \psi(\xi P)|^2$ is determined by
our approximate solution of two-dimensional QCD.

Now, under $Q^2$ evolution, the quark, anti-quark and gluon
distributions mix, in general. However, the difference between quark
and anti-quark distributions $q^V(\xi,Q^2) = q(\xi,Q^2) - \bar
q(\xi,Q^2)$ evolves independently of the gluon distribution. This
difference is referred to as the ``valence'' quark distribution in
the perturbative QCD literature. Moreover, the experimentally
measured structure function $F_3(x_B,Q^2) \simeq q^V(\xi,Q^2)$ where
the Bjorken variable $x_B$ is approximately equal to the momentum
fraction $\xi$ up to corrections suppressed by the ratio of the
proton target mass to $Q^2$: $x_B \approx \xi \bigg(1 - {\xi^2 M_p^2
\over Q^2} \bigg)$ \cite{target-mass}. Here we consider the isospin
averaged valence-quark distribution, which corresponds to $F_3$
averaged over neutrino and anti-neutrino DIS on isoscalar targets.
Finally, the relation between the valence-quark distribution
function and the valence-quark probability density is $q^V(\xi,Q^2)
= \nu(Q^2) V(\xi)$ where $\nu(Q^2)$ is a normalization constant, the
perturbatively renormalized number of valence-quarks, given by the
GLS sum rule  \cite{gls}. It is determined to high order since it is
obtained by integrating the DGLAP equation from $Q^2 = \infty$ down
to $Q_0^2$
    \beq
    \int_0^1 q^V(\xi,t) d\xi = \nu(t) = 3 \bigg[1 -
    {\alpha_s(t) \over \pi} + {\alpha_s(t) \over \pi}^2 (-{55 \over 12} +
    {N_f \over 3}) + \cdots \bigg]
    \eeq
where $t = \log{\bigg({Q^2 \over \Lambda_{QCD}^2} \bigg)}$, the
number of flavors is $N_f = 2$ (up and down) and $\alpha_s(t)$ is
the strong coupling constant.

Within the rank-one approximation, the valence-quark wave function
is determined in the large-$N$ limit, by minimizing the baryon
mass$^2$ (\ref{e-masssq-rank-1}) holding $||\psi||^2 = 1$ subject to
the modified momentum sum rule
    \beq
    N \int_0^P p |\tilde \psi(p)|^2 [dp]  = fP
    \eeq
where $f$ is the fraction of baryon momentum carried by
valence-quarks. Our variational solution is $\tilde \psi(p) = C ({p
\over P})^a (1- {p \over P})^b$ where in the chiral limit, $a = 0$
and $b = {N \over 2 f} -1$. We work in the chiral limit of massless
current quarks since the up and down quarks are essentially massless
($5-7$ MeV) compared to the scale of strong interactions
($\Lambda_{QCD} = 200$ MeV). The valence-quark probability density
is then $V(\xi) = ({N \over f} -1) (1-\xi)^{{N \over f} -2}$ where
we set the number of colors to its physical value $N = 3$. According
to our ansatz, the valence distribution at $Q_0$ is $q^V(\xi,Q_0^2)
= \nu(Q_0^2) V(\xi)$.

The two parameters of our model, $Q_0^2$ and $f$ are to be
determined by fitting to data. We expect $Q_0$ to be low, since we
predict the anti-quark distribution to be small at $Q_0$. And we
expect $f$ to be roughly a half since gluons and valence-quarks
share baryon momentum roughly equally at low $Q_0$. To compare with
experimental data we evolve the valence-quark distribution from
$t_0$ to $t > t_0$ according to the leading order DGLAP \cite{dglap}
equation
    \beq
    {dq^V(\xi,t) \over dt} = {\alpha_s(t) \over 2\pi} C_2
    \bigg[ \int_\xi^1  {q^V({\xi \over \zeta},t) (1+\zeta^2)
    - 2\zeta q^V(\xi,t) \over \zeta (1-\zeta)} d\zeta
    + {3\over 2} q^V(\xi,t)\bigg]
    \eeq
We use the leading order DGLAP equation since we are evolving only
over a small range of $Q_0^2 \leq Q^2 \leq 13 GeV^2$ Here $C_2 =
{N^2 -1 \over 2 N}$. The evolution is done numerically by
discretizing $\xi$ and $t$ with increments of $\delta \xi= .01$
and $\delta t = .025$ respectively. Since the DGLAP equation is
linear, we may normalize $q^V(\xi,t)$ to $\nu(t)$ at any value of
$t$. But in practice we normalize it at $t_0$ to reduce
discretization errors in the region $0 \leq \xi \leq \delta \xi$
which contributes significantly to $\nu(t)$ since $q^V(\xi,t)$
diverges logarithmically for small $\xi$ for $t > t_0$. The loss
of knowledge of $q^V(\xi,t)$ for $0 \leq \xi \leq \delta \xi$ for
$t > t_0$ does not affect the evolution for $\xi > \delta \xi$
because information flows from higher to lower values of $\xi$ as
is evident from the evolution equation.

$F_3$ has been experimentally measured by the CCFR and CDHS
collaborations \cite{cdhsccfr} at a variety of values of $Q^2$ and
$x_B$. The parameters $f, Q_0^2$ should be determined by a best
fit to experimental data. For now, we assign to them the values $f
= \half$ and $Q_0^2 = 0.4$ GeV$^2$ suggested by phenomenological
fits to data \cite{cteq,mrst}. We may now compare our prediction
with experimental measurements of $F_3$ at higher values of $Q^2$.
\begin{figure}
\centerline{\epsfxsize=6.truecm\epsfbox{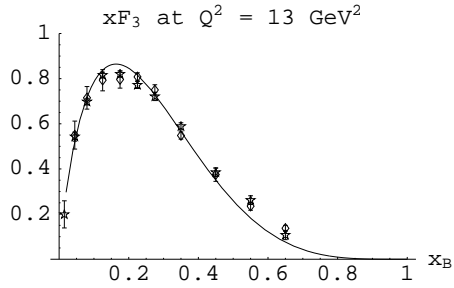}}
\caption{Comparison of predicted $xF_3$ at $Q^2=13$ GeV$^2$ (solid
curve) with measurements by CCFR (star) at $12.6$ GeV$^2$ and CDHS
(diamond) at $12.05 \leq Q^2 \leq 14.3$ GeV$^2$. $Q_0^2 = 0.4$
GeV$^2$ and $f = \half$.} \label{f-F3-str-fn-comparison}
\end{figure}
In Fig.\ref{f-F3-str-fn-comparison}, we show a comparison with
$xF_3$ measurements by the CDHS and CCFR collaborations
\cite{cdhsccfr} at $Q^2 \sim 13$ GeV$^2$. We have ignored the
difference between the Bjorken variable $x_B$ and the momentum
fraction $\xi$ which is small at $Q^2 \sim 13$ GeV$^2$. The plot
shows that our model gives a zeroth order explanation of the data.

Note: Too much should not be read into the values $f = \half$ and
$Q_0^2 = 0.4$ GeV$^2$  we have picked for these parameters. In a
more detailed comparison, and to further constrain the fit, one
should substitute $x_B F_3$ which has not been measured for $x_B >
0.7$ with the structure function $F_2$. The two are approximately
equal for large momentum fractions, where the contributions of
anti-quarks and gluons are small. It appears that our choice for $f$
and $Q_0$ produces a prediction that underestimates $F_3$ for large
momentum fractions. Increasing $f$ makes the valence-quark
distribution vanish slower as $\xi \to 1$, since the exponent is
$(1-\xi)^{{N \over f} - 2}$ at $Q^2 = Q_0^2$. Merely changing $f$
from 50\% to 60\% would change the exponent from $4$ to $3$ at $Q^2
= Q_0^2$. Increasing $Q_0$ also has a similar effect: smaller the
range of $Q^2$ over which the distributions are evolved, smaller is
the momentum that flows from large to small values of $\xi$. We do
not discuss this any further. Better still, it should be possible to
eliminate these fits to data and predict the values of both $Q_0$
and $f$. This requires a much deeper understanding of
renormalization and gluon dynamics respectively. We now turn to the
problem of understanding matrix-valued dynamical gluons, which we
have ignored so far. We do this within the regularized context of
matrix models.

\part{Large-$N$ Matrix Models as Classical systems}
\label{part-large-N-mat-mod-class-systems}

\markboth{}{Part \ref{part-large-N-mat-mod-class-systems}. Large-$N$
Matrix Models}


\fl {\bf \Large Large-$N$ Matrix Models as Classical systems}

\medskip

We now turn to $N \times N$ hermitian multi-matrix models, which are
to gluons what $N$-component vector models are to quarks. Matrix
field theories are much harder to deal with since there are of order
$N^2$ rather than $N$ degrees of freedom at each point in
space-time. Non-perturbative techniques and the necessary
mathematics to understand them are only just being developed.
Understanding even a finite number of matrix degrees of freedom is a
challenge, and there are many conceptually new phenomena to contend
with.

We will follow a strategy similar to that in Part
\ref{part-baryon-large-N}, and study the large-$N$ limit, which
turns out to be a classical limit. The analogs of the color-singlet
meson variable $M \sim \chi^{\dag i}(x) \chi_i(y)$, which are
unitary invariants of a collection of vectors, are the gluon Green's
functions, the traces of products of matrices: cyclically symmetric
tensors $G_I = G_{i_1 i_2 \cdots i_n}$. We find that the classical
configuration space is the space of non-commutative probability
distributions, whose coordinates are the gluon correlations.
Non-commutative probability theory (introduced in \S
\ref{ch-free-alg-non-comm-prob}) is the probability theory of
operator-valued random variables, and is an outgrowth of the study
of von Neumann algebras, especially due to D. Voiculescu
\cite{Voiculescu}. The entropy from confinement of color degrees of
freedom manifests itself in the large volume of matrices that
correspond to the same set of correlations, a non-commutative
entropy.

In the euclidean formalism, matrix models are defined by an action
$S$, which is usually the trace of a polynomial in the matrices.
Classical equations of motion for the gluon correlation tensors are
the factorized Schwinger-Dyson equations, also known as the loop
equations. The solution of these classical equations is a point on
the configuration space, a particular non-commutative probability
distribution. We discover a classical action principle for these
loop equations: maximize non-commutative entropy holding the
correlations conjugate to the coupling tensors fixed. The main
subtlety is that there is a cohomological obstruction to expressing
the entropy in terms of the gluon correlations. To circumvent this,
we express the configuration space as a coset space, taking
inspiration from the Wess-Zumino-Witten model \cite{wzw-model}. The
automorphism group of the tensor algebra acts as changes of variable
on the space of probability distributions. Indeed, the latter is a
coset space of the former, by the isotropy subgroup fixing a given
reference distribution. We express the entropy as a function on the
automorphism-group invariant under the action of the isotropy
subgroup. This allows us to get an explicit formula for the
large-$N$ classical action of euclidean matrix models. In other
words, we express the correlations of a given matrix model in terms
of those of a reference model and the change of variables that
relates the two. This also leads to a variational approach to
estimating the large-$N$ correlations and partition function. We
parameterize the change of variable in terms of variational
parameters, and use the extremization of the classical action to
find their optimal values. In other words, given a submanifold of
the configuration space parameterized by variational parameters, our
action principle finds the point on this submanifold that best
approximates the true correlations of a given matrix model. The
variational estimate for the logarithm of the partition function is
then the maximum value of entropy.

Part \ref{part-large-N-mat-mod-class-systems} of this thesis, is
based on our paper \cite{entropy}. The reader is also referred to
the related paper \cite{abhi-rajeev-loop-ym-wzw} and
\cite{information}, where we study the large-$N$ limit of
hamiltonian matrix models as classical systems. Chapter
\ref{ch-free-alg-non-comm-prob} deals with some mathematical
background on tensor algebras (free algebras) and non-commutative
probability. Chapter \ref{ch-eucl-mat-mod} deals with euclidean
large-$N$ matrix models as classical systems. Appendix
\ref{a-one-mat-mod} discusses the case of a single random matrix
(one-matrix model). Appendix \ref{a-wigner-distr} summarizes some
facts about the simplest of matrix models, the gaussian.

\chapter{Free Algebras and Free Probability Theory}
\label{ch-free-alg-non-comm-prob}

\thispagestyle{myheadings}

\markboth{}{Chapter \ref{ch-free-alg-non-comm-prob}. Free Algebras
\& Free Probability}


In this chapter we will cover some mathematical background on tensor
algebras, operator-valued random variables and the non-commutative
version of probability theory that is relevant to their study. We
will follow a point of view that is a synthesis of our own work,
previous work in the physics literature
\cite{mehta-book,brezin-et-al,Cvitanovic,Cvitanovic-et-al,
Rajeev-turgut,gopakumar-gross,migdal-makeenko,thooft-coll-papers,
sakita-book,Yaffe,cwh-lee-rajeev,Douglas-master-fld} and that of
Voiculescu \cite{Voiculescu} in the mathematics literature.

A pair of real-valued random variables are statistically independent
if their joint probability distribution factorizes as a product of
individual probability distributions. Quantum mechanics can be
considered as the ordinary (commutative) probability theory of
real-valued random variables such as energy, momentum, position and
spin; their expectation values being given by averaging with respect
to a weight specified by the wave function. If a pair of quantum
systems are independent, then the total wave function is just the
tensor product of the individual wave functions. Thus, the notion of
statistical independence of classical probability theory is based on
the tensor product. When the tensor product is replaced by the free
product, the notion of statistical independence is replaced by the
mathematical notion of freeness. (Note: freeness must not be
confused with free field theory in physics!) The resulting
probability theory is called free probability theory (developed by
Voiculescu \cite{Voiculescu}). Of course, not all random variables
in free probability theory need be free of each other, just as not
all real-valued random variables are statistically independent.
Real-valued random variables commute and are elements of commutative
algebras such as the algebra of measurable functions on ${\mathrm
R}^M$. The random variables in free probability theory are no longer
real-valued, they do not commute in general and are elements of
non-abelian algebras of ``measurable operators'', also known as
$W^*$ algebras or von Neumann algebras. Thus the study of von
Neumann algebras is just a non-abelian extension of measure theory.
Free probability theory is also referred to as non-commutative
probability theory. Somewhat counter intuitively, the simplest
examples of non-commutative probability theory occur when the
algebra of random variables is maximally non-abelian, i.e. the basic
operators (generators) satisfy no relations. Then the operator
algebra is just a free product of the operator algebra for each of
the generators. Examples are (1) the free algebra on hermitian
generators (full tensor algebra) which is associated to hermitian
large-$N$ matrix models; (2) the group algebra of the free group
(the algebra of convolution operators on the free group) which has
to do with large-$N$ unitary matrix models; and (3) the Cuntz and
Toeplitz algebras of creation-annihilation operators satisfying the
Cuntz relations having to do with quantum matrix models in the
canonical formalism \cite{cwh-lee-rajeev} or the master field
formalism for matrix models
\cite{gopakumar-gross,Douglas-master-fld,Voiculescu}.

The three examples are related. For example, the information in
the connected moments (cumulants) of the tensors in the free
algebra can also be packaged as the vacuum expectation values of
so-called master fields which are thought of as elements of the
Cuntz algebra. Similarly, the probability theory on (1) and (2)
are related by a kind of exponentiation. We will follow an
approach based on example (1), since it is most closely tied to
the hermitian matrix models studied in this thesis.

\section{Tensor Algebras and Operator-valued Random Variables}
\label{s-tensor-alg-op-val-rnd-var}

In the usual commutative probability theory of $M$ real-valued
random variables $x_1 \cdots x_M$ we are concerned with determining
the expectation values of observables which are functions of the
random variables: $\langle f(x_1, \cdots, x_M) \rangle$. These
expectation values are determined by averaging with respect to a
joint probability distribution function (pdf or measure)
$\rho(x_1,\cdots , x_M)$ on ${\mathrm R}^M$. The expectation values
of functions that can be written as power series in the $x_i$ can be
built up from the expectation values of monomials in the random
variables: the moments $G_{i_1 \cdots i_n} = \langle x_{i_1}\cdots
x_{i_n} \rangle$ which are symmetric tensors since real-valued
random variables $x_i$ commute. Thus the algebra of observables of
commutative probability theory can be taken to be the commutative
algebra of polynomials (more generally formal power series) in $M$
indeterminates. In solving the classical moment problem, nineteenth
century mathematicians showed that it is possible to reconstruct the
joint pdf from a knowledge of the moments, under mild hypotheses
\cite{akhiezer}. Indeed, one can formulate almost the entire theory
in terms of moments. (But moments do not tell the whole story. They
are not adequate to compute expectation values of functions
non-analytic in the random variables, for instance. In fact, one of
the most important quantities in probability theory, the
Boltzmann-Shannon entropy of a probability distribution $- \int
\rho({\bf x}) \log \rho({\bf x}) d{\bf x}$ cannot be expressed in
terms of the moments alone.) For instance, the condition on the
positivity of a probability measure $\rho({\bf x})\geq 0$ translates
into the moment inequalities: for $M =1$, they are the conditions
that the expectation value of the square of any polynomial $f(x) =
\sum_n f_n x^n$ must be positive, $\sum_{m,n} f_m f_n G_{m+n} \geq
0$. This is the same as the positivity of the Hankel matrix $G_{m,n}
:= G_{m+n}$. Similarly, the normalization constraint $\langle 1
\rangle= 1$ becomes $G_{\emptyset} = 1$ where $\emptyset$ denotes
the absence of any indices.

Now suppose $\xi_1, \cdots, \xi_M$ are $M$ operator-valued random
variables. We take them to be hermitian. We can think of the $\xi_i$
as hermitian matrices. What is the analogue of a joint probability
distribution? If there was just a single random hermitian operator,
the distribution of its eigenvalues may be regarded as the
probability distribution. But when we have two or more operators
which do not commute, they cannot be simultaneously diagonalized. In
this situation, the moments come to the rescue and serve to define
the joint probability distribution.

First, what is the algebra of observables? We shall consider
random variables which are polynomials (or more generally formal
power series) in the basic ones (the generators) $\xi_i$ :
    \beq
        f(\xi) = f^{i_1 \cdots i_n} \xi_{i_1} \cdots \xi_{i_n} =
        f^I \xi_I
    \eeq
where $\xi_I := \xi_{i_1 \cdots i_n} := \xi_{i_1} \cdots \xi_{i_n}$
and repeated indices are summed. Thus random variables are
determined by their complex-valued coefficient tensors $f^I$. Just
like functions, these formal power series may be multiplied by
complex scalars, added and multiplied ``point-wise''
    \beq
        [\alpha f]^I = \alpha f^I; ~~ [f + g]^I = f^I + g^I, ~~
        [f g]^I = \delta^I_{I_1 I_2} f^{I_1} g^{I_2}
    \eeq
where $\delta^I_{I_1 I_2} = 1$ if $I_1 I_2 = I$ and vanishes
otherwise. The multiplication is the associative direct product of
tensors. If we think of the string $I$ as a path in the space of
indices, then this is the product induced by concatenation of paths.
The simplest situation is the one in which the $\xi_i$ satisfy no
relations, this can be modelled by $N \times N$ hermitian matrices
in the limit of large-$N$, where the constraints imposed by the
vanishing of characteristic polynomials may be ignored. In this
case, the algebra of observables is the full tensor algebra or the
free algebra ${\cal T}_M$ on $M$ generators. Note that the tensors
are not required to satisfy any symmetry property. ${\cal T}_M$ is
the free product of $M$ copies of ${\cal T}_1$
    \beq
        {\cal T}_M = {\cal T}_1 * \cdots * {\cal T}_1
    \eeq
The free product of a pair of algebras ${\cal A}, {\cal B}$ is the
linear span of all products of the form $a_1 b_1 a_2 b_2 \cdots
a_n b_n$ where $a_i \in {\cal A}$ and $b_i \in {\cal B}$ are
arbitrary. The multiplication is the concatenation product given
above. Other associative algebras on hermitian generators can be
obtained from this free algebra by quotienting by the ideal
generated by the relations satisfied by the generators.

\section{Non-Commutative Probability Distributions}
\label{s-non-comm-prob-distr}

Now, a non-commutative probability distribution is a rule to
calculate the expectation values of the random variables in the
algebra of observables. They are determined by the moment tensors
    \beq
        G_{i_1 \cdots i_n} = \langle \xi_{i_1} \cdots \xi_{i_n}\rangle
    \eeq
By analogy with commutative probability theory, we define the
non-commutative joint probability distribution to be a sequence of
moment tensors $G_{\emptyset},G_{i_1}, G_{i_1 i_2}, \cdots$
satisfying the normalization, hermiticity and positivity
conditions
    \beq
        G_{\emptyset} = 1; ~~ G_{I}^* = G_{\bar I}; ~~ G_{IJ} f^{\bar
        I*} f^J \geq 0
    \eeq
for all polynomials $f(\xi) = f^I \xi_I$. The positivity condition
can be regarded as the positivity of the Hankel matrix $G_{I;J} :=
G_{IJ}$. If $I = i_1 \cdots i_n$, $\bar I = i_n i_{n-1} \cdots
i_1$. The expectation value of an arbitrary polynomial is then
    \beq
    \langle f(\xi) \rangle = f^I G_I
    \eeq
In the case of relevance to matrix models (pure gluondynamics or
closed string theory), the moments are cyclically symmetric
tensors since they are modelled by the expectation values of
traces of products of matrices. The space of such moment tensors
is the space of non-commutative probability distributions in $M$
generators, ${\cal P}_M$. More generally, if we were to include
the quark degrees of freedom along with the gluons, we would have
to deal with open strings of gluons with quarks at the ends
(\ref{e-meson-observables}). The corresponding moments would not
be cyclically symmetric.

Hermitian large-$N$ matrix models provide examples of
non-commutative probability distributions. The moments are given by
integrals over large hermitian matrices (see \S
\ref{s-loop-eq-class-eq-mot})
    \beq
        G_I = \lim_{N \to \infty} \ov{Z} \int dA e^{-N \tr S(A)}
            \Ntr A_I
    \eeq

More generally, Voiculescu defines a non-commutative probability
space as an associative algebra ${\cal A}$ over the complex
numbers with unit element $1$ and linear functional (or ``state''
in quantum statistical mechanics \cite{Haag}) $\tau$ such that
$\tau(1) = 1$. The elements of the algebra are the random
variables. He adds probabilistic content to this algebraic
definition by requiring the algebra to be a weakly closed operator
algebra with the adjoint operation $a \mapsto a^*, (a^*)^* = a$
($W^*$ algebra or von Neumann algebra). The state must be real
$\tau(a^*) = \tau^*(a)$ and positive elements must have positive
expectation values $\tau(a^* a) \geq 0$. The case of cyclically
symmetric moment tensors above corresponds to a ``tracial'' state
$\tau(ab) = \tau(ba)$. Other examples of operator algebras that
admit a normalizable tracial state $\tau$ include Type I algebras:
finite dimensional matrix algebras and finite type II factor
algebras such as the von Neumann algebra of operators in the left
regular representation of a group with infinite conjugacy classes
(e.g. group algebras of free groups).

The precise definition of freeness can be given within the
algebraic context. A family of subalgebras ${\cal A}_i$ of a
non-commutative probability space are a free family if
    \beq
        \tau(a_1 \cdots a_n) = 0
    \eeq
whenever $a_j \in {\cal A}_{i_{j}}$ with $i_j \neq i_{j+1}$ for $1
\leq j < n$ and $\tau(a_j) = 0$ for $1 \leq j \leq n$. In other
words, a family of subalgebras are free of each other if the
expectation value of a product of random variables vanishes if
they each have zero mean and adjacent random variables are from
distinct subalgebras. Random variables are free of each other if
they are from subalgebras that are free of each other.

For example, consider a matrix model with action $S(A_1,A_2) =
S_1(A_1) + S_2(A_2)$. Then random variables $A_1$ and $A_2$ are free
of each other in the large-$N$ limit. In commutative probability
theory, the joint pdfs of a pair of statistically independent random
variables is determined by their individual pdfs. Similarly, the
freeness property can be used to find all the mixed moments such as
$G_{1122}$ once the pure moments $G_{11}, G_{22}$ etc are known:
    \beqs
        0 &=& \langle (\xi_1 \xi_1 - \langle\xi_1 \xi_1\rangle)
        (\xi_2 \xi_2 - \langle\xi_2 \xi_2\rangle)\rangle \cr
        \Rightarrow && G_{1122} - G_{11}G_{22} - G_{11}G_{22} + G_{11}G_{22} = 0
        \cr
        \Rightarrow && G_{1122} = G_{11} G_{22}
    \eeqs
The most famous of non-commutative probability distributions is
the Wigner distribution. We discuss it in Appendix
\ref{a-wigner-distr}.

\section{Automorphisms and Derivations of Tensor Algebras}
\label{s-der-automor-tensor-alg}

We will be interested in how one non-commutative probability
distribution can be transformed into another. We can use the group
of effective transformations as a way to parameterize the space of
probability distributions. This group plays a key role in the
theory to be developed in \S \ref{ch-eucl-mat-mod}.

Let us consider transformations $\phi$ of the algebra of
observables that preserve its structure $\phi(fg) = \phi(f)
\phi(g)$. These are the automorphisms of the tensor algebra ${\cal
T}_M$
    \beq
        \xi_i \mapsto \phi(\xi)_i = \phi_i^I \xi_I
    \eeq
provided they are invertible. The composition of a pair of
automorphisms is
    \beq
        [(\psi \circ \phi)]_i^J = \delta_{K_1 \cdots K_n}^J
        \psi_i^{k_1 \cdots k_n} \phi_{k_1}^{K_1} \cdots
        \phi_{k_n}^{K_n}
    \eeq
The inverse, $(\phi^{-1})_i(\xi) \equiv {\psi}_i(\xi)$, is
determined by the conditions
    \beq
    [({\psi} \circ \phi)_i]^K = \delta^K_{P_1 \cdots P_n}
    {\psi}_i^{j_1 \cdots j_n} \phi_{j_1}^{P_1} \cdots \phi_{j_n}^{P_n} =
    \delta_i^K.
    \eeq
They can be solved recursively for ${\psi}^i_J$:
    \beqs
    {\psi}^i_j &=& (\phi^{-1})^i_j \cr
    {\psi}^i_{j_1 j_2} &=& - {\psi}^{k_1}_{j_1} {\psi}^{k_2}_{j_2}
    {\psi}^i_{l_1} \phi^{l_1}_{k_1 k_2} \cr
    \cdots \cr
    {\psi}^i_{j_1 \cdots j_n} &=& - \sum_{m<n} \delta^{P_1 \cdots
    P_m}_{k_1 \cdots k_n} {\psi}^{k_1}_{j_1} \cdots {\psi}^{k_n}_{j_n}
    {\psi}^i_{l_1 \cdots l_m} \phi^{l_1}_{P_1} \cdots
    \phi^{l_m}_{P_m}.
    \eeqs
A necessary condition for an automorphism to be invertible is that
its linear part $\phi^i_j$ be invertible: $\det{\phi^i_j} \neq 0$.
Even if the linear part is invertible, the above recursive procedure
may not converge in a finite number of steps. Indeed, we would not
be able to find inverses of polynomials within the world of
polynomials. This forces us to work with the ``completed'' tensor
algebra of formal power series in the generators $\xi_i$, rather
than just non-commutative polynomials. This is why we allowed our
random variables to be formal power series in the basic ones. (Note
that the composition rule given above makes sense for two formal
power series since the operations involve only finite sums.)
However, we pay a price for using formal power series instead of
polynomials for our algebra of observables. It may not be possible
to evaluate the expectation value of an arbitrary formal power
series in the generators. However, it turns out that in the
large-$N$ limit, many of the interesting probability distributions
are in a sense compactly supported so that moments do not grow too
fast (see for example the Wigner semi-circular distribution given in
Appendix \ref{a-wigner-distr}; an even stronger restriction on the
growth of moments is the Wilson area law conjecture (Section \S
\ref{s-qcd-wilson-loop-gluon-correlations})) and it is possible to
evaluate the expectation value of many interesting observables.

We will further restrict to the ``orientation preserving''
automorphisms, $\det{\phi^i_j} > 0$. Thus, the automorphisms of the
tensor algebra form a group ${\cal G}_M = {\rm Aut}~ {\cal T}_M$.
For $M = 1$ it is the diffeomorphism group of the real line, where
the real line is thought of in terms of the commutative algebra of
formal power series ${\cal T}_1$ on it.
    \beq
        {\cal G}_1 = \bigg\{\phi(x) = \sum_{n=1}^\infty
            \phi_n x^n \bigg| \phi_1 > 0 \bigg\}
    \eeq
with the composition law
    \beq
        [\psi \circ \phi]_n = \sum_{k=1}^n \psi_k \sum_{l_1 +
            \cdots + l_k = n} \phi_{l_1} \cdots \phi_{l_k}
    \eeq
and inverses given by Lagrange's recursion relations
    \beq
        \phi^{-1} = \psi;~~~ \psi_1 \phi_1 = 1;~~~ \psi_n = -\ov{\psi_1^n}
        \sum_{k=1}^{n-1} \psi_k \sum_{l_1 + \cdots + l_k = n}
        \phi_{l_1} \cdots \phi_{l_k}.
    \eeq
For $M > 1$ the free algebra ${\cal T}_M$ is non-abelian and can
at best be considered as the algebra of functions on a
non-commutative manifold. Then ${\cal G}_M$ is a non-commutative
analogue of the diffeomorphism group.

Finally, the action of the automorphism group on probability
distributions is given by the transformation rule for moments. If
$\Gamma_I = \langle \xi_I \rangle$ and $\xi_i \mapsto
\phi(\xi)_i$, then
    \beq
        G_I = \langle \phi(\xi)_{i_1} \cdots \phi(\xi)_{i_n}
        \rangle= [\phi_* \Gamma]_{i_1 \cdots i_n} = \phi_{i_1}^{J_1} \cdots
        \phi_{i_n}^{J_n} \Gamma_{J_1 \cdots J_n}
    \eeq
However, only a subgroup of the automorphism group preserves the
positivity of probability distributions
    \beq
        {\tilde{\cal G}}_M = \{{\phi \in \cal G}_M | [\phi_* \Gamma]_{IJ}
        f^{\bar I*} f^J \geq 0 ~\forall {\rm ~polynomials ~} f(\xi) \}
    \eeq
Let us now characterize the infinitesimal version of ${\cal G}_M$,
the Lie algebra $\underline{{\cal G}_M}$.

An automorphism $\phi(A)_i = A_i + v^I_i A_I$ that deviates
infinitesimally from the identity is a vector field, a derivation
of the tensor algebra. Such a derivation can be expressed as a
linear combination of the basis vector fields $L_v = v_i^I L_I^i$
where the action of the basis vector fields is
    \beq
        [L^i_I \xi]_j = \delta^i_j \xi_I
    \eeq
and by the chain rule,
    \beq
        L^i_I \xi_J = \delta_J^{J_1 i J_2} \xi_{J_1 I J_2}
    \eeq
Taking expectation values, we get the action on moments
    \beqs
        [L_I^i G]_J &=& \delta_J^{J_1 i J_2} \xi_{J_1 I J_2} \cr
        L_I^i &=& G_{J_1 I J_2} \dd{}{G_{J_1 i J_2}}
    \eeqs
The Lie algebra $\underline{{\cal G}_M}$ of these vector fields
can be worked out
    \beq
        [L_I^i,L_J^j] = \delta_J^{J_1 i J_2} L^j_{J_1 I J_2}
            - \delta_I^{I_1 j I_2} L^i_{I_1 J I_2}
    \eeq
For $M = 1$ it reduces to the Lie algebra of polynomial vector
fields on the real line. Consider the infinitesimal changes of
variable $\phi(x) = x + \eps x^{k+1},~ k = 0,1,2\ldots$.
(translations, corresponding to $k=-1$ are not allowed since they
are not invertible.) Thus a basis for the Lie algebra
$\underline{{\cal G}_1}$ is
    \beq
     L_k = x^{k+1} {\partial \over \partial x} \equiv x^{k} D,~
     k = 1,2 \ldots
    \eeq
and they satisfy the commutation relations of the Virasoro algebra
$[L_p,L_q] = (q-p) L_{p+q}$ (see also Appendix
\ref{a-one-mat-mod-eqns-motion-action}). So far, we have been
thinking of the Lie algebra as acting on the tensor algebra. We
can also think of the Lie algebra as acting on the group itself.
By taking either $\psi$ or $\phi$ to be infinitesimal ($x + \eps
x^{k+1}$) in the composition law $\psi \circ \phi$ for a pair of
group elements, we get the left and right actions of the Lie
algebra on the group ${\cal G}_1$. The left action is $({\cal L}^k
\phi)(x) = \phi^{k}(x)$:
    \beq
     {\cal L}^k \phi_n = \sum_{l_1 + \cdots +l_{k} =n} \phi_{l_1}
     \cdots \phi_{l_{k}} \label{e-left-action-one-variable}.
    \eeq
For a physical application of this see
(\ref{e-pf-lk-chi-equals-eta-k}). The right action is: ${\cal R}^k
\phi(x) = x^{k}D \phi(x)$. They both satisfy the same commutation
relations as the $L_k$.

The reason we studied how probability distributions transform under
a change of variable is that it gives us a novel way of
understanding the entropy of non-commutative probability theory. It
is also essential to understand the classical action principle of
large-$N$ matrix models (see \S \ref{s-classical-action-ppl}).

\section{Free Entropy of Non-Commutative Probability Theory}
\label{s-free-entropy}

In \S \ref{s-chi-as-entropy} entropy of operator-valued random
variables is discussed from the point of view of a variational
principle for matrix models and the cohomology of the automorphism
group of the free algebra. Here, we present Voiculescu's precise
definition \cite{Voiculescu} in terms of matricial microstates.
The free entropy $\chi$ of a collection of self adjoint random
variables $X_1, \cdots, X_n$ in a $W^*$ probability space $({\cal
A},\tau)$ is the logarithm of volume of a set of hermitian $N
\times N$ matrices $A_i$, the matricial microstates. The set of
matricial approximants are restricted by requiring all their $N
\to \infty$ moments to equal those of the random variables $X_i$.
The volume measure on the space of hermitian matrices is the
Lebesgue measure $\lambda$ corresponding to the Hilbert-Schmidt
(HS) norm $\sqrt{\tr(A_1^2 + \cdots + A_n^2)}$. Let
$\Gamma_R(X_1,\cdots,x_n;m,N,\eps)$ be the set of $n$-tuples
$(A_1,\cdots,A_n)$ with
    \beqs
    ||A_j||_{HS} &\leq& R \cr
    |\Ntr A_I - \tau(X_I)| &<& \eps~~~ \forall ~|I| \leq m
    \eeqs
Successively, the regulators are removed $N \to \infty, m \to
\infty, \eps \to 0, R \to \infty$
    \beqs
    \chi_R(X_1,\cdots,X_n;m,N,\eps) &=&
         \log \lambda(\Gamma_R(X_1,\cdots,x_n;m,N,\eps)) \cr
    \chi_R(X_1,\cdots,X_n;m,\eps) &=& \limsup_{N \to \infty}
        (\Nsq  \chi_R(X_1,\cdots,X_n;m,N,\eps)
        + {n \over 2} \log N) \cr
    \chi_R(X_1,\cdots,X_n) &=& \inf\{\chi_R(X_1,\cdots,X_n;m,\eps)|
        m\in {\cal N}, \eps > 0\} \cr
    \chi(X_1,\cdots,X_n) &=& \sup_{R > 0} \chi_R(X_1,\cdots,X_n)
    \eeqs
The cut off $R$ plays a minor role, $\chi_R$ is independent of $R$
once $R$ exceeds the HS norms of all the $X_i$. Some properties of
free entropy $\chi$ that Voiculescu establishes are listed below.
See \cite{Voiculescu} for additional results.

\begin{enumerate}

\item The entropy of a single semi-circular variable $S$ with
$\tau (S^2) = 1$ is $    \chi(S)  = \half \log{2\pi}$.

\item $\chi(X_1,\cdots,X_n) \leq {n \over 2} \log{({2\pi C^2 \over
n})}$ where $C^2 = \tau(\sum_{i=1}^n X_i^2)$. The RHS is the
entropy of the standard Wigner distribution.

\item {\bf n = 1:} If $X$ has distribution $\mu$, then $\chi(X)$
is the logarithmic energy
    \beq
\chi(X) = \int \log|s-t| d\mu(s) d\mu(t) + {3 \over 4} + \half
\log 2\pi
    \eeq

\item {\bf Semi-circular maximum:} If $\tau(X_i^2) = 1 ~\forall~
i$ then $\chi(X_1,\cdots,X_n)$ is maximal iff the $X_i$ are
semi-circular and freely independent. Thus the semi-circular
distribution plays the role of the gaussian of ordinary
probability theory.

\item {\bf Additivity under Freeness:} If $X_i$ are pairwise
freely independent,
    \beq
\chi(X_1,\cdots,X_n) = \sum_{i=1}^n \chi(X_i)
    \eeq
This is the analog of the additivity of entropy of statistically
independent random variables in usual probability theory.
Statistical independence is replaced by freeness here. This
explains the choice of the logarithm of the volume rather than any
other function of volume in defining entropy.
\end{enumerate}

\chapter{Euclidean Large-$N$ Matrix Models}
\label{ch-eucl-mat-mod}

\thispagestyle{myheadings}

\markboth{}{Chapter \ref{ch-eucl-mat-mod}. Euclidean Large-$N$
Matrix Models}


We will consider matrix models, whose random variables are a
collection of $M$ $N \times N$ hermitian matrices, $A_i,~ i = 1,
\ldots, M$. (Appendix \ref{a-one-mat-mod} contains a rapid summary
of some facts about the $M = 1$ matrix model.) We can think of
them as the gluon field at a finite number of points in space-time
labelled by $i$. The $i$ could also label links of a lattice as in
lattice gauge theory, in which case it is more convenient to
consider unitary matrices. The matrix elements are distributed
according to
    \beq
        e^{-N \tr S(A)} ~{dA \over Z(S)}
    \eeq
$dA$ stands for the usual Lebesgue measure on the matrix elements
of the $M$ hermitian matrices. The theory is defined by the action
$S(A)$, which we take to be a polynomial
    \beq
        S(A) = \sum_n S^{i_1 \cdots i_n} A_{i_1} \cdots A_{i_n} \equiv S^I A_I
    \eeq
with cyclically symmetric tensors $S^I$, the coupling constants.
(Note: We will use capital letters to denote strings of indices
$A_{i_1} \cdots A_{i_n} \equiv A_{i_1 \cdots i_n} \equiv A_I$ and
sum repeated indices.) The partition function is
    \beq
        Z(S) = \int e^{-N \tr S(A)} ~dA
    \eeq
and the free energy is defined as
    \beq
        F(S) = -\ov{N^2} \log Z(S)
    \eeq
The expectation value of any function of the matrices is
    \beq
        \langle f(A) \rangle = \ov{Z(S)} \int f(A) e^{-N \tr S(A)}
        dA
    \eeq
The observables we are interested in are those that are invariant
under the simultaneous action of $U(N)$, $A_i \to U A_i U^\dag, ~ i
= 1, \ldots , M$. We can think of these as the global gauge
transformations that remain after local gauge fixing. The simplest
examples of invariants are $\Phi_I = \Ntr A_I$ known as loop
variables. The factor of $\N$ is inserted with hindsight to ensure a
good large-$N$ limit. Also of interest, are the Wilson loop-like
observables $\Ntr e^{A_I}$, though these can be built out of the
loop variables by expanding the exponential in a power series.

\section{Loop Equations, Classical Configuration Space}
\label{s-loop-eq-class-eq-mot}

The loop variables satisfy Schwinger-Dyson equations
    \beq
        S^{J_1 i J_2} \langle \Phi_{J_1 I J_2}\rangle = \delta^{I_1 i
        I_2}_I
        \langle \Phi_{I_1} \Phi_{I_2} \rangle
        \label{e-sq-eq-multi-mat}
    \eeq
which are the conditions for the invariance of the partition
function under an infinitesimal but nonlinear change of
integration variable $A_i  \mapsto  A_i + v_i^I A_I$. In other
words, under the action of the vector field $L_v = v^I_i L^i_I$.
The basis vector fields act as $L^i_I A_j = \delta^i_j A_I$ and by
the Leibnitz rule, $L^i_I A_J = \delta_J^{J_1 i J_2} A_{J_1 I
J_2}$. There are contributions from both the change in the action
and the change in the measure.
    \beqs
        Z(S) &=& \int dA \bigg|\det\bigg({\pdr(A_i + v_i^I A_I)^a_b \over \pdr
            (A_j)^p_q} \bigg)\bigg| e^{-N \tr S(A)} \bigg[1 - N v^I_i \tr L^i_I S(A)
            \bigg] \cr
            &=& \int dA \bigg(1 + v^I_i \delta^{I_1 i I_2}_I \tr A_{I_1} \tr A_{I_2}
            \bigg) \bigg(1 - N v^I_i S^{J_1 i J_2} \tr A_{J_1 I J_2}
            \bigg) e^{-N \tr S(A)}
    \eeqs
where we have used $\det{(1+v T) \simeq 1 + v \tr T}$. Comparing
coefficients of $v^I_i$ and dividing by $Z(S) N^2$ gives the
advertised Schwinger Dyson equations. The latter are not a closed
system of equations for the expectation values of the loop
variables. They are related to those of a product of two loop
variables. Migdal and Makeenko \cite{migdal-makeenko} noticed that
in the large-$N$ limit of 't Hooft \cite{thooft-planar} , holding
the coupling constants $S^I$ fixed, the loop variables satisfy a
closed system of equations, known as the loop equations or
factorized Schwinger-Dyson equations. This is because, in the
large-$N$ limit, the loop variables do not have any fluctuations,
and their expectation values factorize
    \beq
        \langle \Ntr A_{I_1} \Ntr A_{I_2} \rangle =
        \langle \Ntr A_{I_1} \rangle \langle\Ntr A_{I_2}\rangle + O(\ov{N^2})
    \eeq
This property can be proven perturbatively. Thus the large-$N$ limit
is a classical limit and it suffices to restrict attention to the
gluon correlations or moments
    \beq
        G_I = \lim_{N \to \infty} \langle \N \tr A_I \rangle
    \eeq
which are cyclically symmetric tensors. In terms of them, the
factorized Schwinger-Dyson equations are
    \beq
        S^{J_1 i J_2} G_{J_1 I J_2} = \eta^i_I \equiv \delta^{I_1 i I_2}_I G_{I_1}
            G_{I_2}.
        \label{e-fSD}
    \eeq

What is the classical configuration space? It is the space of
allowed moments. From the definition of $Z(S)$ we see that
$G_{\emptyset} = <\Ntr 1> = 1$. Hermiticity of $A_i$ implies
$G_{I}^* = G_{\bar I}$, where $\bar I = i_n \cdots i_1$ if $I = i_1
\cdots i_n$. Moreover, if $f(A) = f^I A_I$ is a polynomial, then the
expectation value of its absolute square should be positive,
$\langle \Ntr f^\dag(A) f(A) \rangle \geq 0$ provided the integrals
converge. In other words, $G_{IJ} f^{\bar I *} f^J \geq 0$ which is
the same as the positivity of the Hankel matrix $G_{I;J} \equiv
G_{IJ}$. This imposes certain inequalities on the moments, which are
known as the moment inequalities. We see that the configuration
space of large-$N$ matrix models is the space of non-commutative
probability distributions in $M$ generators ${\cal P}_M$, as defined
in \S \ref{s-non-comm-prob-distr}. The moments $G_I$ are coordinates
on this configuration manifold. Thus the moments of large-$N$ matrix
models provide an example of non-commutative probability
distributions. If we associate with each matrix $A_i$ a generator of
the free algebra $\xi_i$, then $G_I = <\xi_I>$.

In commutative probability theory of real-valued random variables,
the moments $G_k = \int x^k \rho(x) dx$ determine the probability
distribution $\rho(x)$ up to technicalities \cite{akhiezer}. But
there could still be observables whose expectation values are not
easily expressed in terms of the moments; an example is entropy
$\int \rho(x) \log{\rho(x) } dx$. In matrix models, the large-$N$
limit of the free energy $F = \lim_{N \to \infty} -\ov{N^2} \log Z$
is of much physical interest. Though in principle it is determined
by the moments, we will see that there is no simple formula for it
in terms of the moments. Indeed, this subtlety is quite central to
our entire discussion and we will return to it.

The loop equations (\ref{e-fSD}) are quadratically non-linear in the
moments and in general couple all the moments to each other. The
LHS, $S^{J_1 i J_2} G_{J_1 I J_2}$ comes from the expectation value
of the variation of the action $S(A)$. The RHS $\eta^i_I$ is an
anomaly in the sense that it comes from the change in integration
measure \cite{fujikawa}. It is a universal term since it is
independent of the coupling constants $S^I$. We can regard these as
the classical equations of motion. For a single matrix ($M=1$) these
equations are recursion relations and can be solved (see Appendix
\ref{a-one-mat-mod}). There are a few special choices of action,
such as the gaussian, where $S$ is a quadratic polynomial (see
Appendix \ref{a-wigner-distr} and Ref.
\cite{kazakovABAB,mehtaAB,Staudacher-two-mat,Douglas-and-li}) for
which the partition function or some special classes of moments can
be obtained exactly. But in general, even after the simplifications
of the large-$N$ limit, the loop equations of a multi-matrix model
have not been solved exactly. The situation is not out of the
ordinary, since the generic classical system is not exactly
solvable, it is understood by approximation methods such as
perturbation theory, mean-field theory, variational principles etc.
Perturbation theory around the gaussian leads to the expansion in
terms of planar Feynman diagrams \cite{thooft-planar}. In the case
of finite matrix models, the perturbative series is expected to have
a finite radius of convergence in the coupling constant and one
might hope for a purely perturbative construction of the theory
\cite{thooft-convergence}. But there are limitations to this
approach due to the renormalon singularities upon passage to
full-fledged field theories \cite{renormalon}. Moreover, as we saw
in two-dimensional QCD, there are many interesting phenomena that
are either inaccessible to perturbation theory or more easily
understood by other methods. Understanding the large-$N$ limit of
matrix models as bona-fide classical mechanical systems will permit
us to adapt the methods of classical mechanics to this new class of
dynamical systems.

\section{Classical Action Principle}
\label{s-classical-action-ppl}

\subsection{Anomaly as a Cohomology Class}
\label{s-anomaly-as-cohomology}

One peculiar feature of our derivation of the classical equations
of motion is that we got them sans any knowledge of the classical
action! The original action $S(A)$ will not do since its variation
will not give the anomalous term $\eta^i_I$ which is independent
of coupling constants and came from a change in the measure. We
look for a classical action $\Omega(G)$, the conditions for whose
stationarity under the vector fields $L^i_I$ are the loop
equations
    \beq
        L^i_I \Omega(G) = - S^{J_1 i J_2} G_{J_1 I J_2} +
        \delta_I^{I_1 i I_2} G_{I_1} G_{I_2}
    \eeq
The action of these vector fields on moments is given by $[L^i_I
G]_J = \delta_J^{J_1 i J_2} G_{J_1 I J_2}$. The Lie algebra (which
we denote $\underline{{\cal G}_M}$) of these vector fields is (see
\S \ref{ch-free-alg-non-comm-prob})
    \beq
        [L_I^i,L_J^j] = \delta_J^{J_1 i J_2} L^j_{J_1 I J_2} -
        \delta_I^{I_1 j I_2} L^i_{I_1 J I_2}
    \eeq
Equivalently, we can think of them as first order differential
operators on the configuration space
    \beq
        L_I^i = G_{J_1 I J_2} {\partial \over \partial G_{J_1 i J_2}}
    \eeq
The action dependent term is clearly the variation of ($\N$ times)
the expectation value of the original action $S(A)$
    \beq
        L^i_I (S^J G_J) = S^{J_1 i J_2} G_{J_1 I J_2}.
    \eeq
So let $\Omega(G) = - S^J G_J + \chi(G)$ where
    \beq
        L^i_I \chi(G) = \eta^i_I = \delta_I^{I_1 i I_2} G_{I_1} G_{I_2}
        \label{e-d-chi-equals-eta}
    \eeq
This is a system of inhomogeneous first order linear PDEs on the
configuration space. Necessary integrability conditions (though
not sufficient) for a solution $\chi(G)$ are:
    \beq
        L_I^i \eta_J^j - L_J^j \eta_I^i - \delta_J^{J_1 i J_2} \eta_{J_1 I J_2}^j
        +  \delta_I^{I_1 j I_2} \eta_{I_1 J I_2}^i = 0.
        \label{e-integ-cond-multi-mat}
    \eeq
These complicated looking conditions simply state that $\eta^i_I$
are the components of a closed one-form. They look complicated
because the $L_I^i$ are a non-commuting basis for vector fields
unlike the coordinate basis $\dd{}{G_{iI}}$. Define the $1$-form
$\eta$ such that evaluated on the vector field $L^i_I$ it gives
$\eta(L_I^i) = \eta_I^i$. Then the condition for $\eta$ to be closed
($d\eta =0$) is
    \beq
        (d\eta)(L_I^i,L_J^j) = L_I^i \eta(L_J^j) - L_J^j
        \eta(L_I^i)- \eta([L_I^i,L_J^j]) = 0.
    \eeq
When written in components we get (\ref{e-integ-cond-multi-mat}).
We show in Appendix \ref{a-pf-integ-condition} that the
integrability conditions are satisfied. So $\eta$ is indeed a
closed $1$-form!

In the same vein, (\ref{e-d-chi-equals-eta}) becomes $\eta
\stackrel{?}{=} d \chi$. Is $\eta$ an exact one-form? The answer to
a question on exactness depends on the class of functions on ${\cal
P}_M$ from which $\chi(G)$ can be chosen. We could not find any
formal power series in the moments whose exterior derivative is
$\eta$. Formal power series are natural since the other term in the
classical action ($S^I G_I$) is such a function. To see why
$\chi(G)$ is not a power series in the moments, it suffices to
consider the one-matrix ($M = 1$) case where $\chi = \int
\log{|x-y|} \rho(x) \rho(y) dx dy$ cannot be expressed as a series
in the moments $G_k = \int x^k \rho(x)~ dx$ on account of the
non-analyticity of the logarithm at the origin (see Appendix
\ref{a-one-mat-mod}). Unfortunately, there is no straightforward
generalization of this one-matrix formula for $\chi$, to $M > 1$.
The notion of a probability density $\rho(x)$ makes sense as the
density of eigenvalues of a single matrix. But when we have several
matrices that do not commute, there is no meaning to their
eigenvalues since there is no common basis in which they are all
diagonal. The basis independent notion of a joint spectrum is then
the collection of moments $G_I$.

We conclude that the anomaly $\eta$ is a closed but not exact
one-form, in other words an element of the first cohomology of the
Lie algebra of vector fields $\underline {{\cal G}_M}$ valued in the
space of power series in the moments. But $\eta$ is just the
infinitesimal version of $\chi(G)$ ($\eta = d \chi$). It is
therefore natural to expect $\chi(G)$ to be an element of the first
cohomology of the group ${\cal G}$ whose Lie algebra is $\underline
{{\cal G}_M}$. Now $\underline {{\cal G}_M}$ is spanned by the
vector fields $L^i_I$ which are just infinitesimal automorphisms of
the tensor algebra on $M$ generators (or free algebra) ${\cal T}_M$
(see \S \ref{ch-free-alg-non-comm-prob}). Therefore ${\cal G}_M$
must be the group of automorphisms of the tensor algebra Aut${\cal
T}_M$.

\subsection{Configuration Space ${\cal P}$ as a Coset Space ${\cal G}/{\cal SG}$}
\label{s-config-space-coset-space}

More concretely, to find a formula for $\chi(G)$, we must find
some other way of describing functions on the configuration space
${\cal P}_M$, since power series in $G_I$ were inadequate. To do
this it helped us recall that $\eta$ was the change in the matrix
integration measure under an infinitesimal change of variable.
This suggests another way of parameterizing ${\cal P}_M$ -- in
terms of the finite (invertible) change of variables $\phi$ that
maps the moments of a reference probability distribution
$\Gamma_I$ to those of the probability distribution of interest
$G_I = [\phi_* \Gamma]_I$. Moreover, $\chi(G)$ should be the
change in measure under a finite change of variables.

But such a change of variables is nothing but an automorphism of the
free algebra ${\cal T}_M$! We see that the configuration space
${\cal P}_M$ carries an action of ${\cal G}_M = ~{\rm Aut}~ {\cal
T}_M$ and this action is transitive if we restrict to those
distributions which can be obtained by such a change of variable
starting from $\Gamma$. However, there may be more than one
automorphism that maps $\Gamma$ to $G$. Suppose ${\cal SG}_M$ is the
isotropy subgroup for this action, the subgroup of
measure-preserving automorphisms which map $\Gamma$ to itself. Then
the space of probability distributions ${\cal P}_M$ is the coset
space ${\cal G}_M / {\cal SG}_M$.

Returning to $\chi(G)$, how does this coset space description of
the configuration space get around the cohomological obstruction
to expressing $\chi(G)$ as a power series in the moments? We have
a new way of thinking of functions on ${\cal P}_M$: functions on
${\cal G}_M$ invariant under the action of ${\cal SG}_M$! In
particular, power series in the $G_I$ can always be expressed in
terms of $\phi_i^I$ by substituting $G_I = [\phi_* \Gamma]_I$. But
there are power series in the $\phi_i^I$ (functions on ${\cal
G}_M$) invariant under ${\cal SG}_M$ which cannot be expressed as
power series in the $G_I$. Indeed, we find that $\chi(G)$ is one
such function!

Let us explain our implementation of the rough ideas outlined in
the last few paragraphs. Suppose $\phi$ is a transformation of the
matrices
    \beq
        [A_i]^a_b \mapsto \phi_i(A) = \phi_i^{i_1 \cdots i_n} [A_{i_1}]^a_{a_1}
        [A_{i_2}]^{a_1}_{a_2} \cdots [A_{i_n}]^{a_{n-1}}_b
    \label{e-finite-change-of-var}
    \eeq
which is the finite version of the infinitesimal transformation
$A_i \mapsto A_i + v^I_i A_I$ associated with the vector fields
$L^I_i$ (see \S \ref{s-loop-eq-class-eq-mot}). $\phi$ defines an
automorphism of the free algebra
    \beq
        \xi_i \mapsto \phi_i^I \xi_I
    \eeq
provided its linear part $\phi_i^j$ is an invertible $M \times M$
matrix. Let $\Gamma_I$ be the moments for the standard action
$S_0(A)$ and $G_I$ those for action $S(A)$. Suppose they are
related by the change of variables $S_0(A) = S(\phi(A))$. Under
such an automorphism, the moments and free energy $F(S) = -\lim_{N
\to \infty} \ov{N^2} \log Z(S)$ transform as
    \beqs
        G_{i_1 \cdots i_n} &=&  [\phi_* \Gamma]_{i_1 \cdots i_n} =
            \phi_{i_1}^{J_1} \cdots
            \phi_{i_n}^{J_n} \Gamma_{J_1 \cdots J_n} \cr
        F(S) &=& F(S_0) + \lim_{N \to \infty} <-\ov{N^2} \tr \log
            J>_{S_0}
        \label{e-transf-of-moments-and-free-egy}
    \eeqs
where $J = \dd{\phi(A)}{A}$ is the jacobian matrix for the
transformation and $<>_{S_0}$ denotes expectation value with
respect to the action $S_0$. To see this, consider
    \beq
        G_I = \lim_{N \to \infty} \bigg\{ \int dA e^{-N \tr S(A)} \Ntr A_I \bigg\} /
            \bigg\{ \int dA e^{-N \tr S(A)} \bigg\}
    \eeq
Make the change of variables $A_i = \phi_i(\tilde A)$
    \beqs
        G_I &=& \lim_{N \to \infty} {\int d \tilde A \det(J) e^{-N \tr S_0(\tilde A)}
            \Ntr \phi_{i_1}(\tilde A) \cdots \phi_{i_n}(\tilde A)
            \over \int d\tilde A \det(J) e^{-N \tr S_0(\tilde A)}}
             \cr
            &=& \lim_{N \to \infty} {\int d \tilde A e^{-N \tr S_0(\tilde A)}
            \Ntr \phi_{i_1}(\tilde A) \cdots \phi_{i_n}(\tilde A)
            \over \int d\tilde A e^{-N \tr S_0(\tilde A)}} \cr
            &=& \phi_{i_1}^{J_1} \cdots \phi_{i_n}^{J_n}
            \Gamma_{J_1}\cdots \Gamma_{J_n}
    \eeqs
where we have used the factorization of correlations in the
large-$N$ limit to cancel the jacobian determinants. Similarly,
    \beqs
        Z(S) = \int dA e^{-N \tr S(A)} &=& \int d \tilde A \det(J)
            e^{-N \tr S_0(\tilde A)} \cr
        \ov{N^2} \log \{Z(S) / Z(S_0)\} &=& \ov{N^2}\log
            <e^{\tr \log (J)}>_{S_0}  \cr
        F(S) - F(S_0) &=& -\lim_{N \to \infty} <\ov{N^2} \tr \log
        J>_{S_0}
    \eeqs
where in the last step we have used large-$N$ factorization to
reverse the order of the logarithm and expectation value.

However, not all automorphisms preserve the positivity of the
Hankel matrix $\Gamma_{I;J} = \Gamma_{IJ}$. The subset that do is
$\tilde {\cal G}_M =\bigg\{ \phi \in {\cal G}_M \mid [\phi_*
\Gamma]_{I_1 I_2} {\rm ~a~ positive~ matrix} \bigg\}$ We will
largely ignore the somewhat technical difference between $\tilde
{\cal G}$ and ${\cal G}$. (Does ${\cal G}_M$ have the same Lie
algebra as $\tilde {\cal G}_M$?)

Now we will think of any probability distribution of interest,
$G_I$ as being obtained from a fixed reference distribution
$\Gamma_I$ by the action of a change of variable $\phi \in \tilde
{\cal G}_M$. Can every probability distribution be obtained in
such a manner? Here we show that any distribution in the
infinitesimal neighborhood of a given $\Gamma_I$ can be obtained
by an infinitesimal automorphism, a derivation $v = v^I_i L^i_I$.
To see this, let $G_I = \Gamma_I + \delta\Gamma_I$ be a
distribution that differs infinitesimally from our reference
$\Gamma_I$. Suppose $\Gamma$ satisfies the strict moment
inequalities, i.e. $\Gamma_{I;J} = \Gamma_{IJ}$ is a positive
Hankel matrix so that it is invertible $\Gamma^{I;J} \Gamma_{J;K}
= \delta^I_K$. $\delta\Gamma_I$ are cyclically symmetric tensors
(preserving the moment inequalities). The question is, can we find
a vector field $L_v = v^I_i L^i_I$ such that $[L_v \Gamma]_I =
\delta \Gamma_I$? Let $I = i_1 \cdots i_n$. Then
    \beqs
    [L_v \Gamma]_I &=& v^J_j L^j_J \Gamma_I = v^J_j \delta_I^{I_1 j I_2}
        \Gamma_{I_1 J I_2} \cr
        &=& v_{i_i}^J \Gamma_{J i_2 \cdots i_n} + v_{i_2}^J
            \Gamma_{i_1 J i_3 \cdots i_n} + \cdots v_{i_n}^J
            \Gamma_{i_1 \cdots i_{n-1} J} \cr
        &=& v_{i_1}^J \Gamma_{J i_2 \cdots i_n} + v_{i_2}^J
        \Gamma_{J i_3 \cdots i_n i_1} + \cdots + v^J_{i_n}
        \Gamma_{J i_1 \cdots i_{n-1}}
    \eeqs
Given $\delta \Gamma$, pick any tensor $w$ whose cyclic part is
$\delta \Gamma$,
    \beq
        w_{i_1 \cdots i_n} + w_{i_2 \cdots i_n i_1} + \cdots w_{i_n i_1 \cdots
        i_{n-1}} = \delta \Gamma_{i_1 \cdots i_n}
    \eeq
and put $w_{I} = v_{i}^J \Gamma_{J I}$. Since $\Gamma_{I;J}$ is
positive, we can invert it and solve for $v_{i}^J$
    \beqs
        w_{I} \Gamma^{I;K} &=& v_i^J \Gamma_{JI} \Gamma^{I;K} =
            v_i^J \delta^K_J \cr
        \Rightarrow v_i^K &=& w_I \Gamma^{I;K}
    \eeqs
It follows that $[L_v \Gamma]_I = \delta \Gamma_I$. Thus, we have
found an infinitesimal change of variable $v$ which allows us to
transform from our reference $\Gamma$ to any given distribution in
its infinitesimal neighborhood. In other words $\underline {{\cal
G}_M}$ acts transitively on the tangent space to the classical
configuration space at $\Gamma$.

We have not investigated whether this action of the Lie algebra
can be exponentiated to a transitive action of the automorphism
group on ${\cal P}_M$. Though it seems likely, the necessary
change of variable may turn out to be singular if the probability
distribution $G$ is qualitatively very different from $\Gamma$,
this may be the sign of a ``phase transition''. In that case we
will need one reference distribution for each ``universality
class'' of matrix models.

What is more, for $M > 1$ there are many changes of variable that
transform $\Gamma$ into $G$. They differ by those automorphisms
that fix $\Gamma$, the isotropy subgroup of {\em measure
preserving} automorphisms ${\cal {SG}}$. Infinitesimally, the
isotropy Lie algebra consists of derivations $v$ that leave
$\Gamma$ invariant
    \beqs
        [L_v \Gamma]_I &=& 0 {\rm ~ for ~ all~} I \cr
        \Rightarrow ~~~~~ 0 &=& v_{i_1}^J \Gamma_{J i_2 \cdots i_n} + v_{i_2}^J
        \Gamma_{J i_3 \cdots i_n i_1} + \cdots + v^J_{i_n}
        \Gamma_{J i_1 \cdots i_{n-1}}
    \eeqs
A somewhat more explicit characterization of this isotropy Lie
algebra when the reference $\Gamma$ is the multi-variate Wigner
distribution is given in Appendix \ref{a-wigner-distr}.

The upshot is, we can regard the configuration space of large-$N$
matrix models as a coset space of the automorphism group of the free
algebra: ${\cal P}_M = \tilde{\cal G}_M / {\cal SG}_M$.

\subsection{Finite Change of Variable: Formula for $\chi$ on ${\cal G} / {\cal SG}$}
\label{s-finite-change-of-var-formula-for-chi}

We now use the coset space characterization of the configuration
space to solve for $\chi(G)$ such that $L^i_I \chi(G) = \eta^i_I$.
Suppose $\Gamma$ is a reference probability distribution and
$\phi$ is a change of variable that maps $\Gamma$ to the
distribution $G$. We shall express
    \beq
        \chi(G) - \chi(\Gamma) = c(\phi,\Gamma)
    \eeq
where $c(\phi,\Gamma)$ is a function on the group ${\cal G}$. But
the LHS is independent of the ``path'' $\phi$ that connects
$\Gamma$ to $G$. So to make this a sensible equation, we show that
$c(\phi,\Gamma)$ is invariant under the action of the isotropy
subgroup ${\cal SG}$. Then it will really be a function on the
quotient ${\cal P} = {\cal G}/{\cal SG}$, which is the same as the
space of moments.

Recall (\S \ref{s-anomaly-as-cohomology}) that $\eta$ is an
element of the first cohomology of the Lie algebra $\underline
{\cal G}$ valued in the space of power series in the moments. By
naturality, $\chi(G) - \chi(\Gamma)$ must be in the first
cohomology of the group ${\cal G}$, valued in the space of formal
power series in the moments (See Appendix
\ref{a-gp-lie-alg-cohomology} on group and Lie algebra
cohomology). In other words $c(\phi,\Gamma)$ must be a non-trivial
one cocycle. The analogue of $d \eta = 0$ is the cocycle condition
in group cohomology. Let $\phi$ and $\psi$ be two automorphisms.
The cocycle condition is
    \beq
        c(\phi \psi,\Gamma) =
        c(\phi,\psi_{*}(\Gamma))+ c(\psi,\Gamma)
        \label{e-cocycle-condition}
    \eeq
Our answer for $c(\phi,\Gamma)$ must satisfy this transformation
rule.

Recall that $\eta$ is the expectation value of the jacobian for an
infinitesimal change of variable $\phi_i(\xi) = \xi_i + v_i^I
\xi_I$. Since $\eta = L^i_I (\chi(G) - \chi(\Gamma))$ is the
infinitesimal version of $\chi$ we guess that $c(\phi,\Gamma)$ is
the expectation value of the logarithm of the jacobian determinant
for a finite change of variable $\phi$. Why logarithm? For, the
addition rule for the logarithm of a product will translate into
the cocycle condition.

The jacobian matrix for a change of variables $\phi_i^I A_I$
(\ref{e-finite-change-of-var}), is $J(\phi,A) = \dd{\phi(A)}{A}$.
Under two successive changes of variable, the jacobians multiply
    \beq
        J(\phi \psi,A) = J(\phi,\psi(A)) J(\psi,A)
    \eeq
Then $\sigma(\phi,A) = \ov{N^2} \log \det J(\phi,A)~$ satisfies
$\sigma(\phi \psi,A) = \sigma(\phi,\psi(A)) + \sigma(\psi,A)$.
Taking the expectation value with respect to the reference
distribution $\Gamma$ we define
    \beq
        c(\phi,\Gamma) = <\sigma(\phi,A)> = <\ov{N^2} \log \det J>
    \eeq
and the cocycle condition (\ref{e-cocycle-condition}) is
automatically satisfied.

We will get a formula for $c(\phi,\Gamma)$ as a power series in
$\phi_i^I$ and $\Gamma_J$. The simplest possibility is a linear
change of variable, $[\phi(A)]_i = [\phi_1]_i^j A_j$. In this
case, $\sigma(\phi,J) = \log\det \phi_1$. Since $\phi$ is
invertible, this term is always non-vanishing, and we factor it
out
    \beq
        [\phi(A)]_i = [\phi_1]_i^j ~ [\tilde \phi(A)]_j, ~~~
        \tilde \phi(A)_i = A_i + \sum_{|I| \geq 2} \tilde
        \phi_i^I A_I, ~~~ \tilde \phi_i^I = [\phi_1^{-1}]^j_i
        \phi_j^I
    \eeq
We show in Appendix \ref{a-formula-for-entropy} that
$\sigma(\phi,A)$ can be written in terms of the loop variables
$\Phi_I = \Ntr A_I$
    \beq
        \sigma(\phi,A)= \log\det\phi_1 + \sum_{n=1}^{\infty}{(-1)^{n+1}\over n}
            \tilde\phi_{i_1}^{K_1i_2L_1}\tilde\phi_{i_2}^{K_2i_3L_2}\cdots
            \tilde\phi_{i_n}^{K_ni_1L_n}\Phi_{K_1\cdots K_n}\Phi_{L_n\cdots L_1}.
    \eeq
Taking expectation values and using factorization in the large-$N$
limit
    \beq
        c(\phi,\Gamma) = \log \det \phi_1 + \sum_{n=1}^\infty {(-1)^{n+1}\over
            n}\tilde\phi_{i_1}^{K_1 i_2 L_1} \tilde\phi_{i_2}^{K_2 i_3 L_2}\cdots
            \tilde\phi_{i_n}^{K_n i_1 L_n} \Gamma_{K_1\cdots K_n} \Gamma_{L_n \cdots L_1}.
    \eeq
Now let $\phi$ be an infinitesimal automorphism, $\phi_i(A) = A_i
+ v_i^I A_I$ where $v_i^I$ are small. Then $[\phi_1]_i^j =
\delta_i^j + v_i^j$, $~\tilde \phi_i^J = v_i^J$ and only the
$n=0,1$ terms in the infinite series for $c$ survive
    \beq
        c(A_i + v_i^I A_I, \Gamma) \simeq v_i^i + \sum_{|J|+|K| >0} \phi_{i}^{J i K}
        \Gamma_{J} \Gamma_{K} = \sum_{|I| > 0} v_i^I \delta_I^{I_1 i I_2} \Gamma_{I_1}
        \Gamma_{I_2} = v_i^I \eta_I^i
    \eeq
So $c(\phi,\Gamma)$ reduces to $\eta$ for infinitesimal $\phi$, so
if we define $\chi(G) = \chi(\Gamma) + c(\phi,\Gamma)$, we are
guaranteed that $L_I^i \chi = \eta^i_I$ which is what we wanted
all along!

To complete our construction of a classical action, we need to
show that $c(\phi,\Gamma)$ is a function of $\Gamma$ and $G =
\phi_* \Gamma$ alone and is independent of the choice of $\phi$
that maps one to the other. In other words, that $c(\phi,\Gamma)$
is actually a function on ${\cal P} = {\cal G} / {\cal SG}$. Is
$c(\phi,\Gamma)$ invariant under the right action of the isotropy
subgroup? Suppose $\psi \in {\cal SG}$, i.e. $\psi_* \Gamma =
\Gamma$ Then we need to show
    \beq
        c(\phi \psi,\Gamma) \stackrel{?}{=} c(\phi,\Gamma)
    \eeq
But this follows from the cocycle condition
(\ref{e-cocycle-condition}) if we can show that
    \beq
        c(\psi,\Gamma) = 0 {\rm ~~if~~} \psi_* \Gamma = \Gamma
    \eeq
We establish this at the level of Lie algebras. In other words,
suppose $\psi_i(A) = A_i + v_i^I A_I$ is infinitesimal, and fixes
the distribution $\Gamma$. Then $v_i^I L^i_I \Gamma_J = 0 ~~
\forall J$. On the other hand,
    \beqs
        c(\psi,\Gamma) &=& v_i^I \eta^i_I \cr
            &=& v_i^I L_I^i (S^J_0 \Gamma_J) \cr
            &=& S_0^J (v_i^I L^i_I \Gamma_J) \cr
            &=& S_0^J (0) = 0
    \eeqs
In the second equality we have used the factorized Schwinger Dyson
(\ref{e-fSD}) satisfied by the reference action $S_0$ and
reference moments $\Gamma$. Thus $c(\phi,\Gamma)$ is invariant
under the infinitesimal action of ${\cal SG}$. It should be
possible to exponentiate this to an invariance under the full
isotropy subgroup though we do not address this issue here. Thus
we regard $c(\phi,\Gamma)$ as a function on the coset space ${\cal
P} = {\cal G} / {\cal SG}$ and
    \beq
        \chi(G) = \chi(\Gamma) + \log \det \phi_1 + \sum_{n=1}^\infty {(-1)^{n+1}\over
            n}\tilde\phi_{i_1}^{K_1 i_2 L_1} \tilde\phi_{i_2}^{K_2 i_3 L_2}\cdots
            \tilde\phi_{i_n}^{K_n i_1 L_n} \Gamma_{K_1\cdots K_n} \Gamma_{L_n \cdots L_1}.
        \label{e-entropy-multi-mat}
    \eeq

We have finally solved our problem of finding a classical action for
large-$N$ multi-matrix models! Suppose the distribution $G$ is
obtained by a change of variables $\phi$ applied to the reference
distribution $\Gamma$. Then extremization of $\Omega(G) = -S^I G_I +
\chi(G)$:
    \beqs
        \Omega(\phi,\Gamma) &=& - \sum_{n=1}^\infty S^{i_1\cdots i_n}{\phi}_{i_1}^{J_1}\cdots
        {\phi}_{i_n}^{J_n}\Gamma_{J_1\cdots J_n} + \chi(\Gamma)
        + \log\det \phi^i_{1j} \cr && +\sum_{n=1}^\infty {(-1)^{n+1}\over n}
        {\tilde \phi}^{K_1i_1L_1}_{i_2}{\tilde
        \phi}^{K_2i_2L_2}_{i_3}\cdots {\tilde
        \phi}^{K_ni_nL_n}_{i_1}\Gamma_{K_n\cdots K_1}\Gamma_{L_1\cdots L_n}
    \label{e-action-multi-mat}
    \eeqs
gives the factorized Schwinger Dyson equations (\ref{e-fSD}). This
variational problem is equivalent to computing the partial Legendre
transform of $\chi(G)$: extremize $\chi$ holding fixed the moments
$G_I$ which are conjugate to the couplings $S^I$ that appear in the
action. The coupling constants $S^I$ play the role of Lagrange
multipliers. Furthermore, $\Omega(\phi,\Gamma) = \Omega(G)$ actually
lives on the quotient ${\cal P = G / SG}$. On account of $\eta$ not
being an exact one-form, we could not find a power series in the
moments $G_I$ for $\chi$. But we have found auxiliary variables
$(\phi,\Gamma)$ that allow us to accomplish the task of finding a
classical action.

In principle, the problem of solving a large-$N$ matrix model, given
the coupling constants $S^I$ in its action, reduces to the
extremization of $\Omega$ with respect to the change of variable
$\phi$ for fixed reference moments $\Gamma_I$. Once the optimal
$\phi$ is determined, the moments are given by
    \beq
        G_{i_1 \cdots i_n} = {\phi}_{i_1}^{J_1}\cdots {\phi}_{i_n}^{J_n}
            \Gamma_{J_1\cdots J_n}
    \eeq
Furthermore, recalling the transformation rule for the free energy
under an automorphism (\ref{e-transf-of-moments-and-free-egy}), the
large-$N$ free energy in terms of the optimal $\phi$ is
    \beqs
        F(S) - F(S_0) &=& - ~c(\phi_{optimal},\Gamma) \cr
        \Rightarrow F(S) &=& -~ {\rm constrained ~extremum~of}~ \chi(G)
    \eeqs
where the constrained extremum is with respect to all moments
except the ones conjugate to the couplings $S^I$. Also, $S_0(A)$
is the reference action with moments $\Gamma_I$. If $\tr S(A)$ is
an action that grows as a polynomial as $\tr A^2 \to \infty$ so
that the matrix integrals converge, then the extremum of $\Omega$
is actually a maximum and the free energy is a minimum.

A byproduct of this is a variational approximation method for
large-$N$ matrix models. Rather than extremizing $\Omega$ with
respect to all possible $\phi \in \tilde {\cal G}$, if we choose an
ansatz for $\phi$ with a few variational parameters and maximize
$\Omega$, we will get a variational upper bound on the free energy
and a variational estimate for the moments! As an application we use
this method to solve a two matrix model approximately in \S
\ref{s-var-approx-mat-mod}. But before that, we explain the physical
meaning of $\chi$. It is the entropy that comes from our lack of
knowledge of matrix elements when only the invariants $\Phi_I = \Ntr
A_I$ are observable.

\section{$\chi$ as an Entropy}
\label{s-chi-as-entropy}

The appearance of the Legendre transform, the logarithm of the
jacobian for a change in volume measure and the restriction of
observables to $U(N)$ invariants should all remind us of entropy in
thermodynamics. The non-exactness of $\eta$ is reminiscent of the
heat one-form in thermodynamics \cite{chandrasekhar}.

In thermodynamics, entropy is a measure of the number of
microstates that have a given value for the macroscopic
observables such as energy, pressure or volume. Entropy arises
when we restrict the set of allowed observables of a physical
system. We should expect there to be an entropy due to color
confinement in the strong interactions: quarks and gluons are
confined and the only observable particles are their hadronic
bound states.

In a matrix model, which is a watered-down version of gluon
dynamics, we have restricted the class of observables from matrix
elements $[A_i]^a_b$ to $U(N)$ invariants $\Phi_I$. There should be
an associated entropy. Indeed it is the term $\chi$ in the classical
action we have just obtained. It is the average of the logarithm of
the volume of the space of matrices that have a given set of
moments. It can also be regarded as the expectation value of a
matrix model analogue of the Fadeev-Popov determinant of gauge
theories. Despite the analogies with statistical mechanics, there
are a few differences. The concept of thermal equilibrium is not
relevant to matrix models. What is more, we have an infinite number
of {\em macroscopic} observables $G_I$ while there were only a
finite number such as energy, pressure, density, volume, and
chemical potential in thermodynamics. Furthermore, the microscopic
dynamical variables, the matrices, do not commute as opposed to the
real-valued positions of particles in a gas which commute. As a
consequence, the entropy in matrix models is that of non-commutative
probability theory rather than the usual Boltzmann-Shannon entropy
of commutative probability theory. Nevertheless, the Boltzmann
notion that the entropy should be the logarithm of the volume of
microstates corresponding to a given macrostate continues to makes
sense.

The interpretation of $\chi$ as the entropy is most transparent
for the case of a single matrix. The integral over the matrix
elements of a hermitian matrix $A = U D U^\dag$ can be factored
into an integral over its eigenvalues $D = {\rm diag}(\lambda_1,
\cdots, \lambda_N)$ and the unitary transformation $U$ that
diagonalizes it. The jacobian determinant for this change of
variables is the volume of the space of hermitian matrices which
share a common spectrum $\lambda_1 < \lambda_2 < \cdots \lambda_N$
where the eigenvalues are generically distinct. The well known
result \cite{mehta-book} is
    \beq
        dA = \prod_a dA_a^a \prod_{a < b} d {\rm ~Re} A^a_b~ d {\rm ~Im}
        A^a_b = Vol(U(N))~ \Delta^2 ~\prod_a d \lambda_a
    \eeq
where $Vol(U(N))$ is the volume of the unitary group and $\Delta$
is the Van der Monde determinant $\Delta = \prod_{a<b} (\lambda_a
- \lambda_b)$. Then the logarithm of volume of matrix elements
corresponding to a given spectrum is $\chi = \log{\Delta^2}$, up
to the additive constant $\log{Vol(U(N))}$ which we ignore since
it is independent of the eigenvalues. If $\rho(x) = \ov{N} \sum_a
\delta(x - \lambda_a)$ is the eigenvalue density, then the entropy
is
    \beqs
        \chi &=& 2 \sum_{a < b} \log{|\lambda_a - \lambda_b|} \cr
        &=& {\cal P} \int \rho(x) \rho(y) \log{|x-y|} ~ dx ~ dy
    \eeqs
where we have used the principal value in anticipation of the
passage to the large-$N$ limit where we have a continuous
distribution of eigenvalues. Notice that this formula for entropy is
different from the Boltzmann entropy $- \int ~\rho(x)~ \log \rho(x)
~dx$ of a real-valued random variable with density $\rho$.

Now suppose $G_k = \int \rho_G(x)~ x^k~ dx$ are the moments of the
probability density $\rho_G$ and $\Gamma_k$ are the moments of a
reference distribution $\rho_\Gamma$. Furthermore, let $\phi$ be
the change of variable that relates the two,
    \beqs
        \rho_\Gamma(x) &=& \rho_G(\phi(x)) \phi'(x) \cr
        G_k &=& \int \rho_\Gamma(y) \phi^k(y) dy
    \eeqs
then the above formula for entropy becomes
    \beqs
        \chi(G) &=& {\cal P} \int~ dx dy~ \rho_G(x) \rho_G(y)
            \log{|x-y|} \cr
        \chi(G) &=& \chi(\Gamma) + {\cal P} \int~ dx dy~ \rho_\Gamma(x)
            \rho_\Gamma(y) \log{\bigg|{\phi(x) - \phi(y)\over x-y}
            \bigg|}
        \label{e-entropy-1-mat-rho}
    \eeqs
where we have multiplied and divided the argument of the logarithm
by $|x-y|$ so that we may isolate $\chi(\Gamma)$. The second term
on the right has the physical meaning of the relative entropy of
$G$ with respect to $\Gamma$. Moreover, if we assume the change of
variable is an invertible power series
    \beq
        \phi(x) = \phi_1[x + \tilde \phi(x)]; ~~~ \phi_1 > 0; ~~~ \tilde \phi(x) =
        \sum_{k=2}^\infty \tilde \phi_k x^k
    \eeq
the entropy becomes
    \beq
        \chi(G) = \chi(\Gamma) + \log \phi_1 + \sum_{n=1}^\infty
            {(-1)^{n+1} \over n} \sum_{k_i + l_i > 0} \tilde
            \phi_{k_1 + 1+ l_1} \cdots \tilde \phi_{k_n + 1 + l_n}
            \Gamma_{k_1 + \cdots k_n} \Gamma_{l_1 + \cdots + l_n}
    \eeq
We see that this is just the $M=1$ case of the formula for $\chi$ we
obtained earlier (\ref{e-entropy-multi-mat}) in looking for a
classical action for large-$N$ matrix models. Formula
(\ref{e-entropy-1-mat-rho}) for entropy in terms of an eigenvalue
density has no simple generalization to several non-commuting
matrices. There is no common basis in which the matrices are
diagonal. However, the joint spectrum of non-commuting matrices
makes sense if we think in terms of the basis independent moments.
The series in Taylor coefficients of $\phi$ for entropy continues to
make sense. This is what we obtained in (\ref{e-entropy-multi-mat}).
Thus $c(\phi,\Gamma) = \chi(G) - \chi(\Gamma)$ has the physical
meaning of relative entropy of $G$ with respect to $\Gamma$. For
more on entropy of non-commutative random variables, see \S
\ref{s-free-entropy}.

In light of this, our variational principle from ~\S
\ref{s-finite-change-of-var-formula-for-chi} has a simple physical
interpretation. The moments of the large-$N$ classical limit of a
matrix model are determined by maximizing entropy holding the
moments $G_I$ conjugate to the coupling tensors $S^I$ fixed using
Lagrange multipliers $S^I$. In looking for a variational principle
for matrix models we have naturally been lead to the cohomology of
automorphisms of the free algebra and to the entropy of
non-commuting random variables! Let us now return to our original
problem of determining the moments and free energy of large-$N$
matrix models.

\section{Variational Approximations}
\label{s-var-approx-mat-mod}

We next discuss the use of our variational principle to obtain
approximate solutions of large-$N$ matrix models. (For other
approximation methods in the literature, see for example
\cite{nishimura-gauss-expan-mat-mod}). A linear automorphism with
the wignerian reference leads to an analogue of mean-field theory
for matrix models. We closely follow our presentation in
\cite{entropy,gk-monte-carlo-ym-mat-mod}. Consider a quartic
multi-matrix model with action
    \beq
    S(M) = \tr [\half K^{ij} A_{ij} + {1 \over 4} g^{ijkl} A_{ijkl}].
    \eeq
Let us use the standard multivariate Wigner distribution (Appendix
\ref{a-wigner-distr}) with action $S_0 = \half \delta^{ij} A_i
A_j$ as the reference distribution. The reference moments are
$\Gamma_{ij} = \delta_{ij}, \Gamma_{ijkl} = \delta_{ij}\delta_{kl}
+ \delta_{li}\delta_{jk}~$ etc. Let $E = F(S) - F(S_0)$ denote the
free energy of $S$ relative to $S_0$. For a linear change of
variables $\phi_i(A) = \phi^j_i A_j$. We must maximize
    \beqs
        \Omega[\phi] &=& \tr\log[\phi^j_i] -
        \half K^{i j} G_{i j} - {1 \over 4} g^{ijkl} G_{ijkl}. \cr
        G_{ij} &=& \phi^k_i \phi^k_j; ~~
        G_{ijkl} = G_{ij} G_{kl} + G_{il}G_{jk}
    \eeqs
Thus $G_{ij}$ may be regarded as variational parameters and the
condition for an extremum is
    \beq
        \half K^{pq} + {1 \over 4} [g^{pqkl}G_{kl} +
        g^{ijpq}G_{ij} + g^{pjqk}G_{jk} + g^{ipql}G_{il}] = \half
        [G^{-1}]^{pq}
    \eeq
This non-linear equation for the variational matrix $G$, is
reminiscent of the self consistent equation for a mean-field. We
will use the terms mean-field theory, gaussian ansatz and wignerian
ansatz interchangeably for this variational approximation. To test
it, we consider a two matrix model for which some exact results are
known from the work of Mehta \cite{mehtaAB}.

\subsection{Two Matrix Model with Interaction ${g \over 4}(A^4 + B^4) -2c AB$}
\label{s-mehta-model}

We will study the two matrix model whose action is
    \beqs
    S(A,B) &=& -\tr [\half (A^2 + B^2 - c AB - c BA) + {g \over 4}(A^4 +
    B^4)].
    \eeqs
Kazakov relates this model to the Ising model on the collection of
all planar lattices with coordination number four
\cite{Kazakov-rand-ising}. The partition function of this model has
been calculated exactly by Mehta \cite{mehtaAB}. This is a special
case of the chain matrix models $S(A_1, \cdots A_n) = \sum_{k=1}^n
\tr V(A_k) - \sum_{k=1}^{n-1} c_k \tr A_k A_{k+1}$ which also share
this integrable property. Indeed, as $k \to \infty$, we can replace
it by a continuous index. When the continuous index is thought of as
time, we get the quantum mechanics of a single matrix  whose
ground-state was obtained by Brezin et. al. \cite{brezin-et-al} by
mapping it onto a system of free fermions.

We restrict to $|c| < 1$, where $K^{ij}$ is a positive matrix.
Since $S(A,B)=S(B,A)$ and $G_{AB} = G_{BA}^*$ we may take
    \beq
    G_{ij} = \pmatrix{\alpha & \beta \cr \beta & \alpha}
    \eeq
with $\alpha, \beta$ real. For $g > 0$, $\Omega$ is bounded above
if $G_{ij}$ is positive. Its maximum occurs at $(\alpha, \beta)$
determined by $\beta = {c \alpha \over 1+2g \alpha}$ and
    \beq
    4 g^2 \alpha^3 + 4 g \alpha^2 + (1-c^2-2g) \alpha -1 = 0.
    \eeq
We must pick the real root $\alpha(g,c)$ that lies in the physical
region $\alpha \geq 0$. Thus, the gaussian ansatz determines the
vacuum energy ($E(g,c) = -\half \log{(\alpha^2 - \beta^2)}$) and
$\it all$ the Green's functions (e.g. $G_{AA} = \alpha, ~ G_{AB} =
\beta, ~ G_{A^4} = 2 \alpha^2$ e.t.c) approximately.

By contrast, only a few observables of this model have been
calculated exactly. Mehta \cite{mehtaAB} obtains the exact vacuum
energy $E^{ex}(g,c)$ implicitly, as the solution of a quintic
equation. $G^{ex}_{AB}$ and $G^{ex}_{A^4}$ may be obtained by
differentiation. (Generating series for some other special classes
of moments are also accessible (see
\cite{Staudacher-two-mat,Douglas-and-li}).) As an illustration, we
compare with Mehta's results in the weak and strong coupling
regions. For small $g$ and $c = \half$:
    \beqs
    E^{ex}(g,\half) = -.144 + 1.78 g - 8.74 g^2 + \cdots &&
    E^{var}(g,\half) = -.144 + 3.56 g - 23.7 g^2 + \cdots \cr
    G_{AB}^{ex}(g,\half) = {2 \over 3} - 4.74 g + 53.33 g^2 + \cdots
    &&
    G_{AB}^{var}(g,\half) = {2 \over 3} - 4.74 g + 48.46 g^2 + \cdots \cr
    G_{AAAA}^{ex}(g,\half) = {32 \over 9} - 34.96 g + \cdots
    &&
    G_{AAAA}^{var}(g,\half) = {32 \over 9} - 31.61 g + 368.02 g^2 +
    \cdots \cr
{\rm For ~strong ~coupling~ and ~arbitrary~} c: &&\cr
    E^{ex}(g,c) = \half \log g + \half \log{3}-{3 \over 4}
                + \cdots &&
    E^{var}(g,c) = \half \log{g} + \half \log{2} + {1 \over \sqrt{8g}}
            + {\cal O}({1 \over g}) \cr
    G_{AB}^{ex}(g,c) \to 0 {\rm ~as~} g \to \infty &&
    G_{AB}^{var}(g,c) = {c \over 2g} - {c \over (2g)^{3 \over 2}} +
    {\cal O}({1 \over g^2}) \cr
    G_{AAAA}^{ex}(g,c) = {1 \over g} + \cdots &&
    G_{AAAA}^{var}(g,c) = {1 \over g} - {2 \over (2g)^{3 \over 2}} + {\cal O}({1 \over
    g^2})
    \eeqs
The linear change of variable from a wignerian reference gives a
crude first approximation for both weak and strong coupling. It
fails near singularities of the free energy (phase transitions). As
$|c|\to 1^-$, the energy $E^{ex}$ diverges; this is not captured
well  by the gaussian ansatz. This reinforces  our view that the
gaussian variational ansatz is the analogue of  mean-field theory.

\subsection{Two Matrix Model with $A_1^2 + A_2^2 - [A_1,A_2]^2$ Interaction}
\label{s-ym-2-mat}

Next, consider the two matrix model with Yang-Mills $+$ gaussian
type of action ($m^2 > 0$)
    \beq
    S(A_1,A_2) = \bigg[ {m^2 \over 2} (A_1^2 + A_2^2) - {g^2 \over 4} [A_1,
    A_2]^2 \bigg]
    \eeq
Certain observables in the large-$N$ limit of the gaussian +
commutator squared model (such as the partition function, though not
$G_{ABAB}$ -- this can be seen via the planar diagram expansion) are
the same as in the gaussian + anti-commutator squared two matrix
model. Its partition function has been studied analytically in
\cite{eynard-kristjansen,kostov-3-color-problem} in relation to the
three color problem on a random lattice. The partition function of
this model also arises in the study of the ${\cal N} = 1^*$
supersymmetric gauge theory, the mass deformation of the ${\cal N} =
4$ supersymmetric gauge theory upon integrating out one of the three
adjoint chiral matter fields \cite{dijkgraaf-vafa}. We can also
regard this matrix model as a simple model for Yang-Mills theory.

It has been shown \cite{austing-conv-ym-integ} that the integrals
over matrices for the pure commutator squared action for a two
matrix model are not convergent. To see this, consider the partition
function, and go to the basis in which $A_1$ is diagonal. In this
basis, the integrand is independent of the diagonal elements of
$A_2$, and therefore diverges. The divergence is even worse if we
consider the expectation value of the trace of a polynomial
involving $A_2$, since then the integrand grows for large values of
the diagonal elements of $A_2$. Moreover, from the work of
\cite{eynard-kristjansen,kostov-3-color-problem}, it is expected
that the free energy develops a singularity for sufficiently large
$g^2$ (or small $m^2$). Since there is only one independent coupling
constant we will set $g =1$.

In \cite{gk-monte-carlo-ym-mat-mod} we did a small-scale
Monte-Carlo simulation to test our wignerian variational ansatz
for this model. This ansatz does a good job of estimating $2$ and
$4$ point correlations for $m^2
> 1$ but not in the strongly-coupled region of small $m^2$.
It does not capture the divergence in free energy for small $m^2$,
as we would expect of a mean-field type approximation. Our numerical
simulation is also not expected to be accurate for small $m^2$. We
quote here the results from \cite{gk-monte-carlo-ym-mat-mod}.

\subsubsection{Polynomial Moments}
\label{s-polynomial-moments}

\begin{figure}
\centerline{\epsfxsize=5.truecm\epsfbox{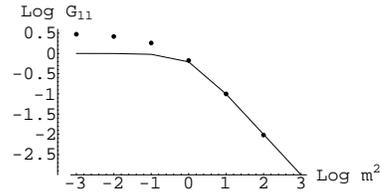}}
\caption{$\log_{10}{[G_{11}]}$ versus $\log_{10}{[m^2]}$ for $g =
1$. Solid line is variational estimate, dots are Monte-Carlo
measurements. The approximation becomes poor for small values of
$m^2$} \label{f-G11}
\end{figure}

\begin{figure}
\centerline{\epsfxsize=5.truecm\epsfbox{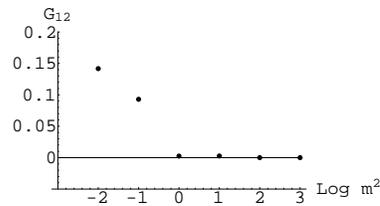}}
\caption{$ G_{12}$ versus $\log_{10}{[m^2]}$ for $g = 1$. Solid line
is variational estimate, dots are Monte-Carlo measurements.}
\label{f-G12}
\end{figure}

For a linear automorphism with standard wignerian reference $S_0 =
\half \delta^{ij} A_i A_j $
    \beq
    \Omega[G] = \half \log \det[G_{ij}] - {m^2 \over 2} (G_{11} + G_{22})
    + {1 \over 2} (G_{1212} - G_{1221})
    \eeq
As before the variational matrix of second moments is
    \beq
    G_{ij} = \left(\begin{array}{cc}
      \alpha & \beta \\ \beta & \alpha \\
    \end{array}
    \right)
    \eeq
Since $<\Ntr A_1^2>~ \geq 0$ and $<\Ntr (A_1-A_2)^2> ~\geq 0$, we
must maximize
    \beq
    \Omega(\alpha,\beta) = \half \log(\alpha^2 - \beta^2) - m^2 \alpha
     + {1 \over 2} (\beta^2 - \alpha^2)
    \eeq
in the region $\alpha \geq 0$ and $\alpha \geq \beta$. We get
    \beq
    G_{11} = G_{22} = \alpha = \sqrt{1+ {m^4 \over 4}} - {m^2 \over 2},~~
         G_{12} = G_{21} = \beta = 0 \label{def-of-alpha}
    \eeq
Figures \ref{f-G11} and \ref{f-G12} compare the variational two
point correlations with Monte-Carlo measurements for a range of
values of $m^2$.
All other correlations can be expressed in terms of these. For
example, the 4-point correlations are (the rest are determined by
cyclic symmetry and $A_1 \leftrightarrow A_2$ exchange symmetry) $
G_{1111} = 2 \alpha^2;    G_{1212} = 0;    G_{1221} = \alpha^2;
G_{1112} = 0$.
\begin{figure}
\centerline{\epsfxsize=5.truecm\epsfbox{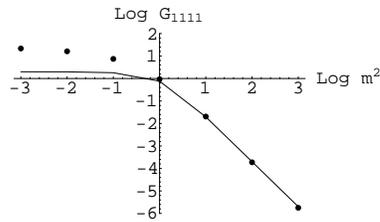}}
\caption{$ \log_{10}{[G_{1111}]}$ versus $\log_{10}{[m^2]}$ for $g =
1$. Solid is line variational estimate, dots are Monte-Carlo
measurements. } \label{f-G1111}
\end{figure}
\begin{figure}
\centerline{\epsfxsize=5.truecm\epsfbox{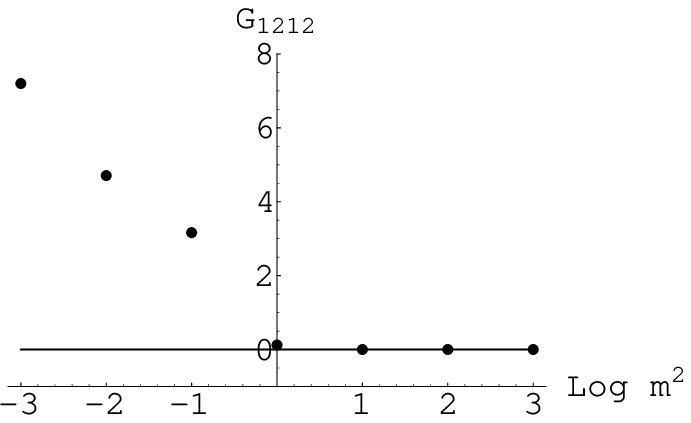}}
\caption{$ G_{1212}$ versus $\log_{10}{[m^2]}$ for $g = 1$. Solid
line is variational estimate, dots are Monte-Carlo measurements.}
\label{f-G1212}
\end{figure}
\begin{figure}
\centerline{\epsfxsize=5.truecm\epsfbox{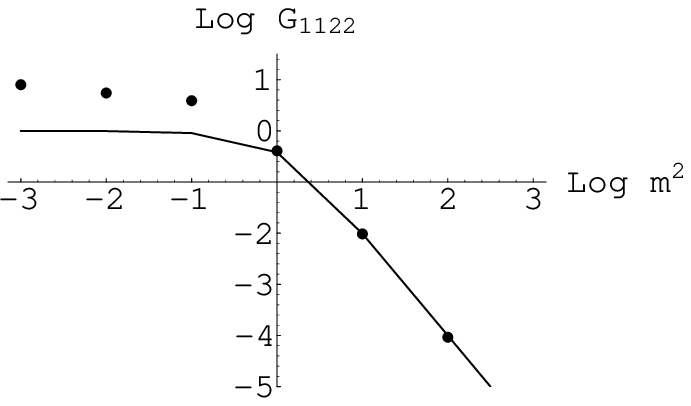}}
\caption{$ \log_{10}{[G_{1122}]}$ versus $\log_{10}{[m^2]}$ for $g =
1$. Solid line is variational estimate, dots are Monte-Carlo
measurements. } \label{f-G1122}
\end{figure}
Figures \ref{f-G1111}, \ref{f-G1212} and \ref{f-G1122} compare
variational estimates (solid lines) and Monte-Carlo measurements
(dots) of $G_{1111}$, $G_{1212}$ and $G_{1122}$ for $10^{-3} \leq
m^2 \leq 10^3$. The $n$ point pure $A_1$ (or $A_2$) correlation is
given by the Catalan numbers
    \beq
    \label{e-catalan}
    G_{111 \cdots 1} = G_{222 \cdots 2} \equiv G_{(n)} = \left\{
    \begin{array}{ll}
    C_{n \over 2} \alpha^{n \over 2} = {n! \over ({n \over 2})!
        ({n \over 2}+1)!} \alpha^{n \over 2},
        & \hbox{if n is even;} \\
    0, & \hbox{if n is odd.} \\
    \end{array}    \right\}.
    \eeq
More generally,
    \beqs \label{useful-correlations}
    G_{11 \cdots 1 22 \cdots 2} &\equiv& G_{(n_1)(n_2)} = G_{(n_1)}
        G_{(n_2)} \cr
    G_{11 \cdots 1 22 \cdots 2 11 \cdots 1 22 \cdots 2} &\equiv&
        G_{(n_1)(n_2)(n_3)(n_4)} = G_{(n_1+n_3)} G_{(n_2)} G_{(n_4)}
        \cr &&
        + G_{(n_1)} G_{(n_3)} G_{(n_2+n_4)} - G_{(n_1)} G_{(n_2)}
        G_{(n_3)}G_{(n_4)}
    \eeqs
We mention these since they are useful in estimating expectation
values of Wilson loop-like operators. The variational estimate for
free energy is
    \beq
     E_{var}= F(S) - F(S_0) = -\log \alpha = -\log{ \bigg[\sqrt{1 +
        {m^4 \over 4}} - {m^2 \over 2} \bigg]}
    \eeq

\subsubsection{Wilson Loop-like Operators}
\label{s-wilson-loop-ops}

It is also interesting to see what the wignerian ansatz says about
the 2-matrix analogue of the expectation of the Wilson loop in the
large-$N$ limit. See \cite{drukker-gross,Akemann-wilson-line} for a
calculation of these observables without any approximations in the
one-matrix models.

\begin{figure}
\centerline{\epsfxsize=5.truecm\epsfbox{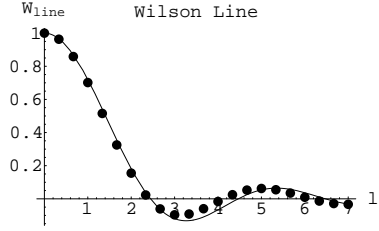}}
\caption{$W_{line}(l)$ for $m^2 = 1, ~ g = 1$. Dots are numerical
and solid line variational estimate.} \label{f-W-line}
\end{figure}

\fl {\bf \large Wilson Line:} The simplest analogue is a ``Wilson
line'' the analogue of the parallel transport along a line of
length $l$ in the $A_1$ direction
    \beq
    W_{line}(l) \equiv \lim_{N \to \infty} <\Ntr e^{i l A_1}>
    \eeq
For the wignerian ansatz, we get (using (\ref{e-catalan}))
    \beqs
    W_{line}(l) &=& \sum_{n=0}^\infty {(i l)^{2k} \over (2k)!} c_{k}
        \alpha^k \cr
    &=& \ov{l \sqrt{\alpha}} J_1(2 l \sqrt{\alpha}) \sim
        \ov{\sqrt{\pi} (l \sqrt{\alpha})^{3 \over 2}} \cos{({3\pi \over 4}
        - 2 l \sqrt{\alpha})} {\rm ~as~} l \to \infty
    \eeqs
Here $J_n(z)$ is the Bessel function of the first kind.
$W_{line}(l)$ is a real-valued function of real $l$ since the odd
order correlations vanish. Thus, for the wignerian ansatz, the
expectation value of the ``Wilson line'' is oscillatory but decays
as a power $l^{-3/2}$. For small $l$, $W_{line}(l) \to 1 - \half
\alpha l^2 + {\alpha^2 l^4 \over 12} - \cdots$. Figure
\ref{f-W-line} compares this ansatz with Monte-Carlo measurements
for $m^2 = 1$. The behavior for small and moderate values of $l$
is well captured by our ansatz.

\begin{figure}
\centerline{\epsfxsize=5.truecm\epsfbox{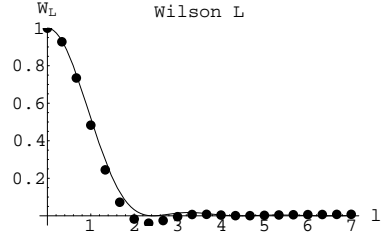}}
\caption{$W_{L}(l)$ for $m^2 =1,~ g = 1$. Solid line is variational
estimate, dots are Monte-Carlo measurements. } \label{f-W-L}
\end{figure}

\fl {\bf \large L shaped Wilson Line:} For an $L$ shaped curve, we
define
    \beq
    W_{L}(l)=\lim_{N \to \infty} <\Ntr e^{ilA_1} e^{ilA_2}>
    \eeq
For the wignerian ansatz (use eq. (\ref{useful-correlations});
$_1F_2(a,{\bf b};z) = \sum_{n=0}^\infty {(a)_n \over (b_1)_n
(b_2)_n} {z^n \over n!}$ is a generalized Hypergeometric function
with $(a)_n$ the Pochhammer symbol),
    \beqs
     W_{L}(l) &=& \sum_{n_1, n_2 =0}^\infty {(il)^{n_1 +n_2}
        \over n_1! n_2!} \lim_{N \to \infty} <\Ntr A_1^{n_1} A_2^{n_2}> \cr
     &=& \sum_{k_1,k_2 =0}^\infty {(-l^2 \alpha)^{k_1 + k_2} \over
        k_1! (k_1 +1)! k_2! (k_2 +1)!} \cr
     &=& \sum_{n=0}^\infty (-l^2 \alpha)^n {4^{n+1} \Gamma(n+ {3 \over 2})
        \over \sqrt{\pi} ~\Gamma(n+1) \Gamma(n+2) \Gamma(n+3)} \cr
     W_L(l) &=& _1F_2({3 \over 2}; \{2,3 \}; -4 l^2 \alpha)
    \eeqs
As before $W_{L}(l)$ is real for real $l$. For small $l$, $W_L(l)
\to 1 - \alpha l^2 + {5 \alpha^2 l^4 \over 12} - \cdots$. This is
compared with the numerical calculation in fig. \ref{f-W-L} for
$m^2 = 1$.


\fl {\bf \large Wilson Square:} The analogue of the parallel
transport around a square of side $l$ in the $A_1 - A_2$ plane is
    \beq
    W_{square}(l) = \lim_{N \to \infty} <\Ntr e^{i l A_1} e^{i l A_2}
        e^{-i l A_1} e^{-i l A_2}>
    \eeq
In the wignerian variational approximation, $W_{square}(l)$ is
real-valued since odd order correlations vanish. Using
eq.(\ref{useful-correlations}) we get
    \beq
    W_{square}(l) = \sum_{n=0}^\infty (-l^2 \alpha)^n T_1(n) + 2
            (l^2 \alpha) \sum_{n=0}^\infty (-l^2 \alpha)^n T_2(n)
    \eeq
    \beqs
        {\rm where~~}
        T_1(n) &=& \sum_{k_i \geq 0, k_1+\cdots +k_4=n} {2 c_{k_1
        + k_3} c_{k_2} c_{k_4}
        - \Pi_{i=1}^4 c_{k_i} \over \Pi_{i=1}^4 (2k_i)!} \cr
    T_2(n) &=& \sum_{k_i \geq 0, k_1+\cdots +k_4=n} {c_{k_1 +k_3 +1} c_{k_2} c_{k_4}
        \over (2k_1 +1)! (2k_2)! (2k_3 +1)! (2k_4)!}
    \eeqs
\begin{figure}
\centerline{\epsfxsize=5.truecm\epsfbox{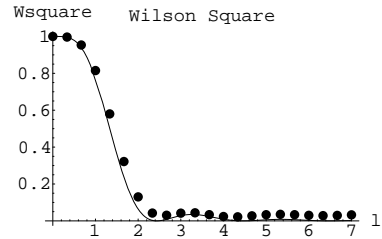}}
\caption{$W_{square}(l)$ for $m^2 = 1,~ g=1$. Solid line is
variational estimate, dots are the Monte-Carlo measurements. }
\label{f-W-square}
\end{figure}
For small $l$, $W_{square}(l) \to 1 - l^4 \alpha^2 + {5 l^6 \alpha^3
\over 6} - \cdots$. The expectation value of the Wilson loop is a
rapidly decaying function for large values of $l$. It would be
interesting to find the asymptotic rate of decay. It is oscillatory
but a positive function, unlike $W_{line}$. These variational
predictions are confirmed by the numerics, in fig. \ref{f-W-square}
for $m^2 = 1, ~ g =1$. Our crude mean-field variational ansatz does
a very good job of estimating the Wilson loop averages over a range
of small and moderate values of $l$ studied, for $m^2$ of order
unity or more and $g = 1$.

\chapter{Summary}
\label{ch-summary}

\thispagestyle{myheadings}

\markboth{}{Chapter \ref{ch-summary}. Summary}

In this thesis, we studied rather simplified models of a non-abelian
gauge theory in the limit of a large number of colors ($N$). The
emphasis has been on the classical nature of this limit, where,
unlike in the usual $\hbar \to 0$ classical limit, where all
variables stop fluctuating, fluctuations vanish only in
gauge-invariant variables.

Part \ref{part-baryon-large-N} focused on the baryon in
two-dimensional QCD, and Part
\ref{part-large-N-mat-mod-class-systems} on euclidean multi-matrix
models.

Chapter \ref{ch-intro} contained historical motivations, and an
introduction to QCD, matrix models, and the large-$N$ limit. We
illustrated a classical limit, different from the $\hbar \to 0$
limit, in the context of the large-dimension limit for
rotation-invariant variables in atomic physics, which provides a
``classical'' explanation for the stability of the atom. The $1/d$
expansion is an accurate way of calculating atomic energy levels in
many electron atoms, as shown by Herschbach and collaborators
\cite{herschbach}.

Chapter \ref{ch-QCD-to-QHD} summarizes Rajeev's reformulation of the
large-$N$ limit of two-dimensional QCD in terms of color-singlet
quark bilinears in the null gauge: two-dimensional Hadrondynamics.
This is an interacting bilocal theory of mesons, since the quark
density matrix is a projection operator. 't Hooft's linear integral
equation for the meson spectrum in two dimensions does not account
for baryons. In Hadrondynamics, free mesons are small fluctuations
around the vacuum, and baryons arise as topological solitons on the
disconnected infinite grassmannian phase space with connected
components labelled by baryon number.

In Chapt. \ref{ch-grnd-state-baryon}, we studied the structure of
baryons in two-dimensional Hadrondynamics. A version of steepest
descent for the curved grassmannian phase space was provided to
determine the ground-state of the baryon. Then, the ground-state of
the baryon was determined variationally by minimizing its energy on
submanifolds of the phase space. A succession of finite but
increasing rank submanifolds of the phase space was found. The
reduced dynamical systems on these submanifolds were shown to
correspond to interacting-parton models. We thus derived the parton
picture as a variational approximation to the soliton description.
The rank-one separable ansatz corresponds to a Hartree-type
approximation for valence-quarks. We determined the exact
two-dimensional baryon form factor in the large-$N$ and chiral
limits. The rank-three ansatz allowed us to estimate the sea and
anti-quark content of the baryon, which turned out to be small. To
model structure functions measured in Deep Inelastic Scattering, we
used the two-dimensional valence-quark distribution as the initial
condition for perturbative $Q^2$ evolution. This allowed us to model
the structure function $F_3$ measured in neutrino-nucleon deep
inelastic scattering. We have two free parameters, the initial
momentum transfer $Q_0^2$, and the fraction of baryon momentum
carried by valence-quarks.

Chapter \ref{ch-free-alg-non-comm-prob} provided background on free
algebras and probability theory for operator-valued random
variables. The tensor algebra, its automorphisms, and derivations
were introduced in the context of non-commutative probability
theory. Voiculescu's notion of non-commutative entropy was defined.

In Chapter \ref{ch-eucl-mat-mod}, we studied euclidean large-$N$
multi-matrix models as toy models for gluon dynamics. We formulated
them as classical mechanical systems for $U(N)$ invariant variables.
We found that the classical configuration space is a space of
non-commutative probability distributions. The classical equations
of motion are the factorized Schwinger-Dyson equations for gluon
correlations. There is an anomalous term in the factorized
Schwinger-Dyson equations arising from the change in measure of
integration. We showed that this term is a closed but not exact
one-form. This universal term is an element of the Lie algebra
cohomology of the automorphism group of the tensor algebra. This
cohomological obstruction prevented us from finding a classical
action principle as a power series in gluon correlations. We showed
that the configuration space can be expressed as a quotient of the
automorphism group of the tensor algebra by the isotropy subgroup of
measure-preserving automorphisms. Probability distributions are
parameterized by the change of variable that maps them to a standard
reference distribution. This provides a classical action on the
group that is invariant under the action of the isotropy subgroup.
The action principle was interpreted as the maximization of
non-commutative entropy, while holding fixed the correlations
conjugate to the coupling tensors. The restriction of observables
from matrix elements to $U(N)$ invariants is the source of this
entropy. The free energy is the negative of the maximum value of
this entropy. By treating the change of variable as a variational
quantity, we obtain variational approximations for the correlations
and free energy of large-$N$ matrix models. A linear automorphism is
the simplest variational ansatz, and leads to an analog of
mean-field theory. This approximation method gives reasonable
estimates for the observables of one and two-matrix models, away
from divergences of the free energy (phase transitions).

\appendix

\chapter{Definition of $\ov{p^2}$ Finite Part Integrals}
\label{a-finite-part}

\thispagestyle{myheadings}

\markboth{}{Appendix \ref{a-finite-part}. Finite Part Integrals}


Here we discuss the definition of the $\ov{p^2}$ singular integrals
defining the potential energy in momentum space. In position space
integrals this singularity is manifested in the linearly rising
$|x-y|$ potential. Let us consider the rank-one case for example:
    \beq
        PE = \half {\cal FP} \int [dp] \int {dr \over
        2\pi} \tilde\psi(p) \tilde\psi^*(p+r) \tilde V(r)
    \eeq
where $[dp] = dp / 2\pi$. In this appendix, we desire to give a
meaning to the above ``finite part'' integral. We call it a ``finite
part'' integral since similar {\it ``part finie''} integrals appear
in the work of Hadamard \cite{hackbusch}. The self-consistent
potential $V(x)$ satisfies Poisson's equation $V''(x) = |\psi(x)|^2$
along with a pair of boundary conditions,
    \beq
        V(0) = \int_{-\infty}^{\infty} |\psi(y)|^2 {|y| \over 2} dy ~~{\rm and}~
        V'(0) = \half \int_{-\infty}^{\infty} |\psi(-y)|^2 {\sgn(y)} dy.
    \eeq
In momentum space Poisson's equation implies that
    \beq
        \tilde V(p) = {\cal FP} ({-1 \over p^2} \int [dq] \tilde \psi(p+q)
        \tilde \psi^{*}(q)) \equiv {\cal FP} (-{1 \over p^2} \W(p)).
    \eeq
The two boundary conditions take the form
    \beq
        -{\cal FP}\int_{-P}^{P} {1 \over p^2} \W(p) [dp] =
        \int |\psi(y)|^2 {|y| \over 2} dy \eeq \beq -i {\cal FP}
        \int_{-P}^{P} {1 \over p} \W(p) [dp] = \half
        \int_0^{\infty} \{-|\psi(y)|^2 + |\psi(-y)|^2 \} dy.
        \label{e-boundary-cond-for-poisson-eqn}
    \eeq
In effect, the boundary conditions are rules for integrating
non-singular functions with respect to the singular measures $dp
\over 2\pi p^{1,2}$. We will use these conditions to define the
``finite part'' integral in terms of Riemann integrals.

For simplicity, let us consider the case where the wave function
$\tilde \psi(p)$ is real. Then $\psi(-x) = \psi^*(x)$, and the
self-consistent potential is even: $\tilde V(-p) = \tilde V(p)$.
Then the second boundary condition
(\ref{e-boundary-cond-for-poisson-eqn}) becomes
    \beq
        {\cal FP} \int_{-P}^{P} {-1 \over p} \tilde W(p) [dp] = 0.
    \eeq
Since the integrand is odd, this condition is automatically
satisfied. Let us restrict attention to wave functions $\tilde
\psi(p)$ such that
    \beq
        \tilde \psi(p) \sim p^a , ~{\rm as}~ p\to 0, ~{\rm for~some}~ a>0.
    \eeq
From the Frobenius analysis (section \ref{s-frobenius}), we know
that this is equivalent to assuming a non-zero current quark mass,
so that superficial infrared divergences are avoided.

Our aim is to define $ -{\cal FP}\int_{-P}^{P} {1 \over s^2} \W(s)
[ds]$ so as to satisfy the first boundary condition. Moreover, the
definition should reduce to the usual one, when this quantity is
finite to begin with.

\noindent {\bf Claim:} If
    \beq
        {\cal FP}\int_{-P}^{P} {1 \over s^2} \W(s) [ds] :=
        \int_{-P}^{P} {\W(s) - \W(0) \over s^2} [ds] -
        {\W(0) \over \pi P}.
    \eeq
Then
    \beq
        {\cal FP}\int_{-P}^{P} {1 \over s^2} \W(s) [ds] = -
        \int_{-\infty}^{\infty} |\psi(x)|^2 {|x| \over 2} dx.
    \eeq
provided $\W'(0) = 0$, which is ensured if $\tilde \psi(p)$ is
real and $C^a$ from the right at $p = 0$ for some $a > 0$.

\noindent {\bf Proof:} The motivation for this definition is
clear: we have essentially subtracted out the divergent terms and
analytically continued the result that we would have got if
$\W(s)$ vanished sufficiently fast at the origin (i.e. like
$s^{1+\eps}, ~ \eps > 0$) to make the integral converge. However,
there is always the danger of mistakenly adding/subtracting some
finite quantity along with the divergent one. We show here that
this definition actually satisfies the boundary conditions.

Recall that $\W(s)$ is the Fourier transform of the charge
density:
    \beq
        \W(s) = \int_{-\infty}^{\infty} |\psi(x)|^2 e^{-isx} dx
    \eeq
so that
    \beq
        \W(s) - \W(0) = \int_{-\infty}^{\infty} dx |\psi(x)|^2
        (e^{-isx} - 1)
    \eeq
Moreover, $\W'(0)$ vanishes, since
    \beq
        \W'(0) = -i \int_{-\infty}^{\infty} x |\psi(x)|^2 dx
    \eeq
and the integrand is odd. Therefore, $\W(s) - \W(0)$ vanishes at
least as fast as $s^{1+\eps}, ~ \eps > 0$ as $s \to 0$. E.g. for
$\tilde \psi(p) = A p^a e^{-p},~ \W(s) = 1 - ~{\rm const~} s^{2a +
1} + O(s^2)$. Therefore,
    \beq
        \int_{-P}^P {\W(s) - \W(0) \over s^2} [ds]
    \eeq
exists as a Riemann integral.  Since the integrand is even it
suffices to consider the case $s > 0$:
    \beq
        \int_{0}^P {\W(s) - \W(0) \over s^2} [ds] = \int_0^P {ds \over
        2\pi s^2} \int_{-\infty}^{\infty} dx |\psi(x)|^2 (e^{-isx} -1)
    \eeq
Using the fact that only the even part of $(e^{-isx} - 1)$
contributes to the integral over $x$, we change the order of
integration,
    \beq
        \int_0^P {\W(s) - \W(0) \over s^2} [ds] =
        \int_{-\infty}^{\infty} dx |\psi(x)|^2 \int_0^P {ds \over 2\pi
        s^2} (\cos{sx} -1).
    \eeq
The inner integral can be performed in terms of the sine integral
function $Si$. Assuming that $\psi(x)$ is normalized to one, we
have
    \beq
        \int_0^P {\W(s) - \W(0) \over s^2} [ds] - {\W(0) \over 2\pi P}
        = -\int_{-\infty}^{\infty} dx |\psi(x)|^2 v(x)
    \eeq
Where,
    \beq
        v(x) = {1 \over 2\pi P} (\cos{(Px)} + Px Si({Px}))
    \eeq
The asymptotic expansion of $Si(t)$ for large $t$ is
    \beq
        Si(t) \sim {\pi \over 2} + \cos{t} ({-1 \over t} + O({1 \over
        t^3})) + \sin{t} ({-1 \over t^2} + O({1 \over t^4}))
    \eeq
Since the sine integral is odd, we have
    \beq
        v(x) = {|x| \over 4} + {1 \over 2\pi P} (Px Si(Px) - {P|x|\pi
        \over 2} + \cos(Px)) = {|x| \over 4} + {{\rm ~Remainder}(Px) \over
        2\pi P}
    \eeq
We have the desired result, except for the remainder term:
    \beqs
        \int_{-P}^{P} {\W(s) - \W(0) \over s^2} [ds] - {\W(0) \over
        \pi P} &=& - \int_{-\infty}^{\infty} |\psi(x)|^2 {|x| \over 2} dx
        \cr && - {1 \over \pi P} \int_{-\infty}^{\infty} |\psi(x)|^2 ~{\rm
        Remainder}(Px) dx.
    \eeqs
In the $P = \infty$ limit, the remainder term vanishes since $-1
\leq {\rm Remainder}(t) = t Si(t) - {\pi|t| \over 2} \leq 1$. For
finite $P$, Remainder$(t) \sim {-\sin{t} \over t}, |t| \to \infty$
is an oscillatory function, and we expect the remainder term to be
small due to cancellations. In fact, it is zero. Consider
    \beqs
        & & \int_{-\infty}^{\infty} dx |\psi(x)|^2 ~{\rm
        Remainder}(Px) \cr &&  = \int_0^P [dq] \int_{-q}^{P-q} [dr]
        \tilde \psi(q+r) \tilde \psi^*(q) \int_{-\infty}^{\infty} dx e^{i r x} ~{\rm
        Remainder}(Px).
    \eeqs
Here Remainder(t) is an even function and
    \beq
        \int_{-\infty}^{\infty} dx e^{i r x} ~{\rm Remainder}(Px) = 2
        \int_{0}^{\infty} dx \cos(rx) ~{\rm Remainder}(Px) = 0,
    \eeq
from the properties of the sine integral, provided $|r| < P$,
which is the region of interest. Thus the ``Remainder'' term
vanishes, and we have shown that our definition of the ``finite
part'' integral satisfies the boundary conditions. This justifies
our definition
    \beq
        {\cal FP} \int_{-P}^{P} {\tilde W(s) \over s^2} [ds] :=
        \int_{-P}^P {\tilde W(s) - \tilde W(0) \over s^2} [ds] -
        {\tilde W(0) \over \pi P}
    \eeq
when $\tilde W'(0) =0$.

\chapter{One Matrix Model}
\label{a-one-mat-mod}

\thispagestyle{myheadings}

\markboth{}{Appendix \ref{a-one-mat-mod}. One Matrix Model}


\section{Equations of Motion and Classical Action}
\label{a-one-mat-mod-eqns-motion-action}

The observables of the single matrix model are the $U(N)$ invariants
$\Phi_n = \Ntr A^n$ of a single $N \times N$ hermitian random matrix
$A$. Their expectation values are obtained by averaging with respect
to the Boltzmann measure ${dA \over Z} e^{-N \tr S(A)}$ for a
polynomial action $S(A) = \sum_{n = 1}^m S_n A^n$. The large-$N$
limit is a classical limit where the fluctuations vanish $ \langle
\Phi_n \Phi_m \rangle = \langle \Phi_n\rangle\langle\Phi_m \rangle +
{\cal O}(\ov{N^2})$. In this limit, the large-$N$ moments $G_n$ are
a complete set of observables.
    \beqs
        G_n &=& \lim_{N \to \infty} \langle \Ntr A^n \rangle = \lim_{N \to \infty} \ov{Z_N(S)} \int
            dA e^{-N \tr S(A)} \Ntr A^n
        \cr Z_N(S) &=& \int dA e^{-N \tr S(A)}; ~~ F(S)= -\lim_{N \to
        \infty} \ov{N^2} Z_N(S)
    \eeqs
$G_n$ parameterize the configuration space of the theory, which is
the space of non-commutative probability distributions in a single
operator-valued random variable: ${\cal P}_1$. The classical
equations of motion are the factorized Schwinger Dyson or loop
equations
    \beq
    \sum_{l} l ~S_{l} G_{k+l} = \sum_{p+q=k} G_p G_q := \eta_k , ~~
        k = 0, 1, 2,\cdots
        \label{e-fSD-1-mat}
    \eeq
These are recursion relations for the higher moments $G_n,~ n>m$
given the moments, $G_1,G_2, \cdots G_m$ conjugate to the coupling
constants $S_1, \cdots S_m$. The factorized loop equations are
obtained by requiring the partition function to be invariant under
the infinitesimal non-linear change of variables $L_n : A \mapsto
A + \eps A^{k+1}, ~ k = 0,1,\cdots$. (The case $k = -1$ also leads
to a valid equation, the RHS of (\ref{e-fSD-1-mat}) vanishes in
this case, from the translation invariance of the measure.) The
LHS comes from the change in $S(A)$ and the RHS $\eta_k$ is an
anomaly coming from the change in the measure. The vector fields
$L_k$ act on the moments
    \beq
        L_k G_l = l G_{k+l}; ~~~~
        L_k = \sum_p p G_{k+p} \dd{}{G_p}
    \eeq
and satisfy commutation relations of the Virasoro algebra
$[L_p,L_q] = (q-p) L_{p+q}$

Since $\langle \Ntr f^\dag(A) f(A) \rangle \geq 0$ for any
polynomial $f(A) = f_n A^n$, the Hankel matrix $G_{p,q} = G_{p+q}$
must be non-negative. This ensures that there is a probability
density on the real line $\rho(x)$ whose moments are $G_n = \int
~\rho(x)~ x^n ~dx$. In fact, $\rho(x)$ is the density of
eigenvalues of the random matrix $A$, as discussed in \S
\ref{s-chi-as-entropy}. The factorized loop equations turn into a
linear integral equation for $\rho(x)$, which we call the
Mehta-Dyson equation
    \beq
        S'(z) = 2 {\cal P} \int {\rho(x) \over z-x}.
    \eeq
The factorized loop equations can be obtained by multiplying
either side by $z^{k+1} \rho(z)$ and integrating with respect to
$z$. The Mehta Dyson equation follows from maximizing the action
    \beq
        \Omega(\rho) = \chi - \int \rho(x) S(x) dx; ~~~
        \chi = {\cal P}  \int \log{|x-y|} \rho(x) \rho(y) dx dy
    \eeq
with respect to $\rho$. As explained in \S \ref{s-chi-as-entropy},
$\chi$ is the entropy due to our lack of knowledge of matrix
elements of $A$ when we restrict measurements to the $U(N)$
invariants of $A$. The simplest possibility is a quadratic action
$S(A) = \half A^2$ which leads to the Wigner semicircular
distribution $\rho(x) = \ov{2\pi} \sqrt{[4-x^2]} \theta(|x|<2)$
discussed further in Appendix \ref{a-wigner-distr}. The
maximization of $\Omega$ involves a balance: $\chi$ tends to
spread the eigenvalues out due to the logarithmic repulsion, while
$-\int \rho(x) S(x) dx$ is maximized if the eigenvalues are
clustered near the minima of $S(x)$. For polynomial actions such
that $S(x) \to \infty$ as $|x| \to \infty$, the eigenvalue
distribution is supported over a finite number of intervals on the
real line, located around the minima of $S(x)$. These intervals
are the branch cuts of the moment generating function $G(z) =
\sum_{n \geq 0} G_n z^{-n-1}$. $\rho(x)$ is the discontinuity of
$-\ov{\pi}G(z)$ across its branch cuts.

Can $\Omega(\rho)$ be expressed in terms of the moments in order to
serve as a classical action principle for the factorized loop
equations (\ref{e-fSD-1-mat})? i.e. $L_k \Omega(G) = \eta_k -
\sum_{l \geq 0}l S_l G_{k+l}$? Clearly, $L_k \bigg(\sum_{l \geq 1}
S_l G_l \bigg) = \sum_{l \geq 1} l S_l G_{k+l}$. However, there is
no power series in the moments $\chi(G)$ such that $L_k \chi(G) =
\eta_k(G)~$. This is because $\log|x-y|$ cannot be expressed as a
power series in both $x$ and $y$. This is a symptom of a
cohomological obstruction, $\eta(G)$ is not exact, but is a closed
one-form valued in the space of power series in the moments. The
components of the one-form $\eta$ are obtained by acting on the
basis vector fields $\eta_k = \eta(L_k)$. Now,
    \beqs
        d\eta (L_p,L_q) &=& L_p \eta (L_q) - L_q \eta(L_p) -
        \eta([L_p,L_q]) \cr
        &=& L_p \eta_q - L_q \eta_p - (q-p) \eta_{p+q}
    \eeqs
and it is a straightforward calculation to check that $d\eta=0$.

We can get around this obstruction by expressing both the moments
$G_k$ and $\chi$ in terms of auxiliary variables. Suppose
$\rho_\Gamma$ is a reference probability distribution and $\phi$
an invertible change of variables that maps $\rho_\Gamma$ to
$\rho_G$
    \beqs
        \rho_\Gamma(x) &=& \rho_G(x) \phi'(x) \cr
        G_k &=& \int \rho_\Gamma(y) \phi^k(y) dy \cr
        \chi(G) &=& \chi(\Gamma) + {\cal P} \int dx dy
        \rho_\Gamma(x)\rho_\Gamma(y) \log \bigg|{\phi(x)
            - \phi(y) \over x-y} \bigg| \cr
            &:=& \chi(\Gamma) + \bigg\langle \log \bigg|{\phi(x)
            - \phi(y) \over x-y} \bigg| \bigg\rangle_\Gamma
    \eeqs
If $\phi$ is a formal power series
    \beqs
        \phi(x) &=& \sum_{n=1}^\infty \phi_n x^n; \phi_1 > 0 \cr
        \phi(x) &=& \phi_1 [x + \tilde \phi(x)]; \tilde \phi(x) =
        \sum_{k=2}^\infty \tilde \phi_k x^k
    \eeqs
Then,
    \beqs
        \chi(G) &=& \chi(\Gamma) + \log \phi_1 + \sum_{n=1}^\infty
            {(-1)^{n+1} \over n} \sum_{k_i + l_i > 0} \tilde
            \phi_{k_1 + 1+ l_1} \cdots \tilde \phi_{k_n + 1 + l_n}
            \Gamma_{k_1 + \cdots k_n} \Gamma_{l_1 + \cdots + l_n}
            \cr
        G_k &=& \sum_{p = k}^\infty \Gamma_p \sum_{n_1+ \cdots+ n_k =p,~ n_i \geq
            1} \phi_{n_1} \cdots \phi_{n_k} \cr
        \Omega(G) &=& \Omega(\phi,\Gamma) = \chi(G) - \sum_{k \geq 1} S_k
            G_k
            \label{e-entropy-1-mat-series}
    \eeqs
What we have done is to parameterize the configuration space ${\cal
P}_1$ by automorphisms of the algebra of power series in one
variable ${\cal G}_1 = {\rm Aut} {\cal T}_1$ which transform between
probability distributions. Actually, we should restrict to the
subgroup $\tilde {{\cal G}}_1$ of automorphisms that preserve the
moment inequalities, i.e. the positivity of the probability
distributions. In the case of a single generator that we are
considering, the only measure-preserving automorphism is the
identity, i.e. the isotropy subgroup ${\cal SG}_1$ is trivial. So we
may identify ${\cal P}_1$ with ${\tilde {\cal G}}_1$. In other
words, the moments $G_k$ and the Taylor coefficients of the
automorphism $\phi_n$ both serve as coordinates on the classical
configuration space. The left action of the vector fields $L_k$ on
$\phi_n$ is (see \S \ref{s-der-automor-tensor-alg})
    \beqs
        {\cal L}_k \phi(x) &=& \phi(x)^{k+1} \cr
        [{\cal L}_k \phi]_n &=& \sum_{l_1 + l_2 + \cdots + l_{k+1}}
            \phi_{l_1} \cdots \phi_{l_{k+1}}
    \eeqs
Now we can check explicitly that $L_k \chi(G) = \eta_k(G)$
    \beqs
        L_k \chi &=& \sum_{n} [{\cal L}_k \phi]_n \dd{\chi}{\phi_n}
            ~=~ \sum_{n} [{\cal L}_k \phi]_n \dd{}{\phi_n} \bigg\langle \log \bigg|{\phi(x)
            - \phi(y) \over x-y} \bigg| \bigg\rangle_\Gamma \cr
        &=& \sum_n [{\cal L}_k \phi]_n \bigg\langle {x^n - y^n \over
            \phi(x)-\phi(y)}\bigg\rangle_\Gamma
        ~=~ \sum_n \sum_{l_1 + \cdots + l_{k+1}=n} \phi_{l_1} \cdots
            \phi_{l_{k+1}} \bigg\langle {x^n - y^n \over
            \phi(x)-\phi(y)}\bigg\rangle_\Gamma \cr
        &=& \sum_n \bigg\langle {\phi^{k+1}(x) - \phi^{k+1}(y)
            \over \phi(x) - \phi(y)} \bigg\rangle_\Gamma
        ~=~ \sum_{p+q=k} \langle \phi^p(x) \phi^q(y) \rangle_\Gamma \cr
        &=& \sum_{p+q=k} G_p G_q = \eta_k
        \label{e-pf-lk-chi-equals-eta-k}
    \eeqs
Thus, we have a classical action $\Omega(\phi,\Gamma) = \chi(G) -
\sum_n S_n G_n$ whose extremization leads to the factorized loop
equations. This leads to a variational principle to solve the
classical equations of motion: pick a reference distribution
$\Gamma$ such as the Wigner distribution (see Appendix
\ref{a-wigner-distr}); maximize entropy $\chi$ with respect to the
automorphism $\phi$, while holding the moments $G_n$ conjugate to
the coupling constants fixed (via the Lagrange multipliers $S_n$).
The free energy $F$ is the negative of the maximum value of entropy.
As a variational ansatz, we may pick $\phi$ to be a polynomial, its
Taylor coefficients are the variational parameters. They
parameterize a finite dimensional space of probability distributions
in the neighborhood of $\Gamma$. The extremization of $\Omega$ will
give the probability distribution in this finite dimensional family
that best approximates the exact moments of the action $S(A)$. We
give below a variational approximation for the quartic one-matrix
model.

\section{Mean Field Theory for Quartic One Matrix Model}
\label{a-mft-quartic-one-mat}

\begin{figure}
\centerline{\epsfxsize=6.truecm\epsfbox{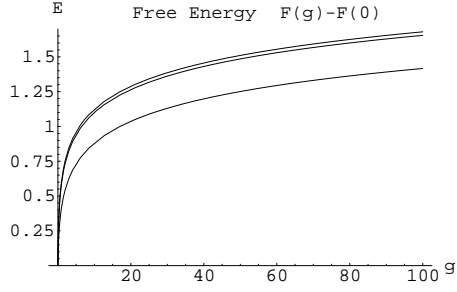}}
\caption{Free energy $E$ versus $g$. Curves top to bottom are:
mean-field ansatz, $\phi(x)=\phi_1 x + \phi_3 x^3$ ansatz calculated
to first order beyond mean-field theory, and the exact energy.}
\label{f-1-mat-1-loop-free-egy}
\end{figure}

\begin{figure}
\centerline{\epsfxsize=6.truecm\epsfbox{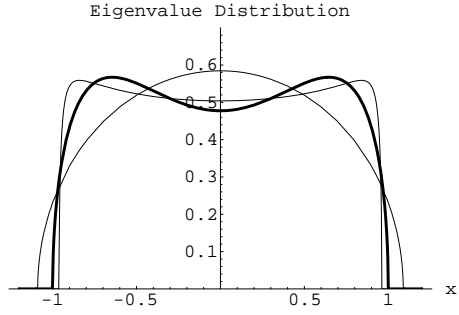}}
\caption{ Eigenvalue Distribution. Dark curve is exact, semicircle
is mean-field and bi-modal light curve is cubic ansatz at $1$-loop.}
\label{f-1-mat-1-loop-rho}
\end{figure}

The exact moments and free energy for the quartic one-matrix model
were obtained by Brezin et. al. \cite{brezin-et-al}. We use this as
a way of calibrating our variational approach
    \beqs
    Z(g) &=& \int dA e^{-N \tr [\half A^2 + g A^4]} \cr
    E(g) &=& F(g) - F(0) = -\lim_{N \to \infty} {1 \over N^2} \log{Z(g) \over Z(0)}
    \eeqs
We take the Wigner distribution with $\Gamma_2=1$ as our reference
(see Appendix \ref{a-wigner-distr}). Pick a linear automorphism
$\phi(x) = \phi_1 x$, which merely scales the width of the Wigner
distribution. The $\phi_1$ that maximizes $\Omega$ represents the
Wigner distribution that best approximates the quartic matrix
model.
    \beq
    \Omega(\phi_1) = \log \phi_1 - \half G_2(\phi_1) - g G_4(\phi_1)
    \eeq
Here $G_{2k} = \phi_1^{2k} \Gamma_{2k}$. Let $\alpha = \phi_1^2$.
$\Omega(\alpha) = \half \log \alpha - {\alpha \over 2} - 2g\alpha^2$
is bounded above only for $g \geq 0$. The maximum occurs at
$\alpha(g) = {-1 + \sqrt{1+32g} \over 16 g}$. Notice that $\alpha$
is determined by a non-linear equation. The linear change of
variable ansatz from a wignerian reference is an analogue of
mean-field theory for large-$N$ matrix models. Our variational
estimates are ($C_k$ are Catalan numbers, see Appendix
\ref{a-wigner-distr}):
    \beqs
    E(g) &=& -\half \log{{-1 + \sqrt{1+32g} \over 16 g}} \cr
    G_{2k}(g) &=& ({-1 + \sqrt{1+32g} \over 16 g})^k C_{k}.
    \eeqs
The exact results from \cite{brezin-et-al} are:
    \beqs
    E_{ex}(g) &=& {1 \over 24} (a^2(g) - 1)(9-a^2(g)) - \half
                    \log{(a^2(g))} \cr
    G_{ex}^{2k}(g) &=& {(2k)! \over k! (k+2)!} a^{2k}(g) [2k + 2
                        -ka^2(g)].
    \eeqs
where $a^2(g) = {1 \over 24 g} [-1 + \sqrt{1+48 g}]$. In both cases,
the free energy is analytic at $g = 0$ with a square root branch
point at a negative critical coupling. The mean-field critical
coupling $g_c^{MF} = -{1 \over 32}$ is $50\%$ more than the exact
value $g_c^{ex} = -{1 \over 48}$. The variational and exact free
energies are compared in figure \ref{f-1-mat-1-loop-free-egy}. In
the weak coupling region
    \beqs
    E_{ex}(g) = 2g -18 g^2 + \ldots &&
    E(g) = 4g - 48g^2 + \ldots \cr
    G_{ex}^2(g) = 1 - 8g + 144 g^2 + \ldots &&
    G^2(g) = 1 - 8 g + 128 g^2 + \ldots \cr
    G_{ex}^4(g) = 2 - 36g + \ldots &&
    G^4(g) = 2 - 32 g + \ldots
    \eeqs
and asymptotically, as $g \to \infty$
    \beqs
    E_{ex}(g)  \sim {1 \over 4} \log{g} + {\log{(144)} - 3 \over 8} - {1 \over
                        3\sqrt{3g}} + \ldots &&
    E(g) = {1 \over 4} \log{g} + {1 \over 4} \log{8} + {1 \over 8
              \sqrt{2 g}} + \ldots \cr
    G_{ex}^2(g) \sim {2 \over 3 \sqrt{3g}} - {1 \over 12 g} +
                        \ldots &&
    G^2(g) = {1 \over \sqrt{8 g}} - {1 \over 16 g} + \ldots \cr
    G_{ex}^4(g) \sim {1 \over 4 g} - {1 \over 6 \sqrt{3}} {1 \over g^{3 \over
                        2}} + \ldots &&
    G^4(g) = {1 \over 4 g} - {1 \over 8 \sqrt{2}} {1 \over g^{3 \over 2}} + \ldots
    \eeqs
The distribution of eigenvalues of the best wignerian
approximation is given by $\rho_G(x) = \phi_1^{-1}
\rho_\Gamma(\phi_1^{-1} x)$ where $\rho_\Gamma(x) = {1 \over 2\pi}
\sqrt{4 - x^2}, ~ |x| \leq 2$ is the reference Wigner
distribution. The exact distribution
    \beq
    \rho_{ex}(x,g) = {1 \over \pi} (\half + 4 g a^2(g) + 2 g x^2)
    \sqrt{4 a^2(g) - x^2}, ~ |x| \leq 2 a(g).
    \eeq
is compared with the best wignerian approximation in figure
\ref{f-1-mat-1-loop-rho}. The latter does not capture the bimodal
property of the former. The wignerian ansatz is like mean-field
theory, but is not restricted to small values of the coupling $g$.
To improve on this, get a non-trivial estimate for the higher
cumulants (connected moments) and capture the bimodal distribution
of eigenvalues, we need to make a non-linear change of variable.

\section{Beyond Mean Field Theory: Non-linear Ansatz}
\label{a-non-lin-change-of-var-1-mat}

To go beyond mean-field theory while sticking to the standard Wigner
distribution as the reference, we must make a cubic change of
variable. $\phi(x) = \phi_1 x + \phi_3 x^3$. A quadratic ansatz will
not increase $\Omega(\phi)$ since $S(A)$ is even. $\phi_{1,3}$ are
determined by the condition that $\Omega$ is maximal
    \beqs
    \Omega[\phi] &=& \chi - \half G_2 - g G_4 \cr
    \chi[\phi] &=& \int dx dy \rho_\Gamma(x) \rho_\Gamma(y)
        \log\bigg|{\phi(x) - \phi(y) \over x-y} \bigg| \cr
    G_4 &=& \int dx \rho_\Gamma(x) \phi^4(x) = \phi_1^4 \Gamma^4 +
        4 \phi_1^3 \phi_3 \Gamma_6 + 6 \phi_1^2 \phi_3^2 \Gamma_8
        + 4 \phi_1 \phi_3^3 \Gamma_10 + \phi_3^4 \Gamma_{12}\cr
    G_2 &=& \int dx \rho_\Gamma(x) \phi^2(x) = \phi_1^2 \Gamma_2 + 2
        \phi_1 \phi_3 \Gamma_4 + \phi_3^2 \Gamma_6
    \eeqs
We were not able to evaluate the integral for $\chi$ exactly. For
multi-matrix models, the problem is worse, since we do not have such
a closed form expression for $\chi$ and have to rely on the power
series (\ref{e-entropy-multi-mat}). However, considering the success
of the linear change of variable, we expect the deviations of
$\phi_{1,3}$ from their mean-field values $(\sqrt{\alpha},0)$ to be
small. We evaluate $\Omega$ using the power series
(\ref{e-entropy-1-mat-series}) and solve for the corrections to
$\phi_{1,3}$ by linearizing around their mean-field values. This may
be regarded as a kind of 1-loop correction beyond mean-field theory,
though it is not restricted to small values of the coupling $g$.
Within this approximation, we get (with $\alpha = {-1 + \sqrt{1+32g}
\over 16 g}$)
    \beqs
    \phi_1 &=& \sqrt{\alpha} - {\sqrt{\alpha}(-3 + 2 \alpha + (1-32 g) \alpha^2
        + 48 g \alpha^3 + 144 g^2 \alpha^4) \over 3 + 4\alpha +
        (1+96g)\alpha^2 + 48g \alpha^3 + 432g^2 \alpha^4} \cr \cr
    \phi_3 &=& {8 g \alpha^{5 \over 2} (-2 + \alpha) \over 3
        + 4\alpha + (1+96g)\alpha^2 + 48g \alpha^3 + 432g^2 \alpha^4}.
    \eeqs
from which we calculate the variational moments and free energy.
Comparing with the exact results of \cite{brezin-et-al}, we find the
following qualitative improvements over the mean-field ansatz.

In addition to the mean-field branch cut from $-\infty$ to
$g_c^{MF}$, the vacuum energy now has a double pole at $g_c^{MF} <
g_c^{1-\rm{loop}} = {-346 - 25 \sqrt{22} \over 15138} < g_c^{ex}$.
We can understand this double pole as a sort of Pad$\acute{\rm e}$
approximation to a branch cut that runs all the way up to
$g_c^{ex}$. The vacuum energy variational estimate is lowered for
all $g$.

Figure \ref{f-1-mat-1-loop-rho} demonstrates that the cubic ansatz
is able to capture the bimodal nature of the exact eigenvalue
distribution. If $\psi(x) = \phi^{-1}(x)$, then $\rho(x) =
\rho_0(\psi(x)) \psi'(x)$, where $\rho_0(x) = {1 \over 2\pi}
\sqrt{4-x^2}, ~ |x| \leq 2$.

\begin{figure}
\centerline{\epsfxsize=5.5truecm\epsfbox{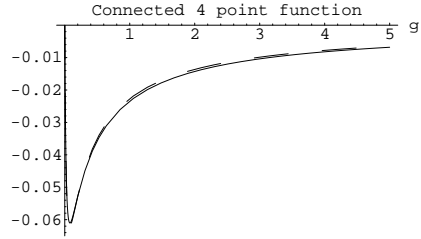}}
\caption{ Connected $4^{\rm th}$ moment (cumulant). Dashed curve is
estimate from cubic ansatz. Solid curve is exact. Mean field
estimate was identically zero} \label{f-1-mat-1-loop-G4c}
\end{figure}

The moments $G_2, G_4$ are now within a percent of their exact
values, for all $g$. More significantly, the connected 4-point
function $G_4^c = G_4 - 2 (G_2)^2$ which vanished for the
wignerian ansatz, is non-trivial, and within $10\%$ of its exact
value, across all values of $g$ (figure \ref{f-1-mat-1-loop-G4c}).

\chapter{Single and Multivariate Wigner Distribution}
\label{a-wigner-distr}

\thispagestyle{myheadings}

\markboth{}{Appendix \ref{a-wigner-distr}. Wigner Distribution}


The analogue of the gaussian for operator-valued random variables is
the Wigner semi-circular distribution. Indeed, the Wigner
distribution of matrices is realized when the matrix elements are
distributed as independent gaussians. The Wigner distribution
derives its importance from the fact that it maximizes entropy (see
\S \ref{s-free-entropy}, \ref{s-chi-as-entropy}) among all
non-commutative probability distributions with given variance. We
collect some facts about the Wigner distribution here. This section
relies on some results on matrix models which can be found in
Appendix \ref{a-one-mat-mod} and Chapter \ref{ch-eucl-mat-mod}.

Let us first consider the case of a single hermitian $N \times N$
matrix $A$ whose elements are gaussian distributed with zero mean
and unit variance. The moments of $A$ in the large-$N$ limit are
    \beq
        \Gamma_n = \lim_{N \to \infty} \ov{Z_N} \int dA
            e^{-N \tr \half A^2} \Ntr A^n; ~~~
        Z_N = \int dA e^{-N \tr \half A^2}
    \eeq
The Wigner moments satisfy the so-called factorized loop equations
(Appendix \ref{a-one-mat-mod}) which are recursion relations
    \beq
        \Gamma_{n+1} = \sum_{p+1+q=n} \Gamma_p \Gamma_q
    \eeq
They may be solved in terms of Catalan numbers $C_n$
    \beq
        \Gamma_{2n} = C_n = {(2n)! \over n! (n+1)!}; ~~ \Gamma_{2n+1} =
        0; ~~ n = 0, 1, 2 \cdots
    \eeq
The first few are $\Gamma_0 = 1, ~\Gamma_2 = 1,~ \Gamma_4 = 2,~
\Gamma_6 = 5$. The Wigner moments are realized by the semicircular
distribution of eigenvalues $\Gamma_k = \int \rho(x) x^k dx$.
Maximizing non-commutative entropy (\S \ref{s-free-entropy},
\ref{s-chi-as-entropy}) holding the variance fixed
    \beqs
        0 &=& \delta \bigg( \int dx dy \rho(x) \rho(y) \log|x-y| - \half \int dx
        \rho(x) x^2 \bigg) \cr
        \Rightarrow~~ z &=& 2 {\cal P} \int_{\mathrm R} {\rho(x) \over z-x} dx
    \eeqs
gives the Wigner semi-circular distribution $\rho(x) =
\theta(|x|<2) \ov{2\pi} \sqrt{4 - x^2}$ \cite{brezin-et-al}.


The analogue of the gaussian process and heat equation of
real-valued random variables are the semi-circular process and
complex Burger equation in non-commutative probability theory
\cite{Voiculescu}. A semi-circular process starting at $X$ is a time
dependent operator-valued random variable $Y(t) = X + \sqrt{t} S$
where $S$ is semicircular $<S^n> = \Gamma_n$. Let $G(z,t) = \sum_{n
\geq 0} {<Y(t)^n> \over z^{n+1}}$ be the moment generating function
of $Y(t)$. Then, $G(z, t)$ satisfies the analogue of the heat
equation, the complex Burger equation:
    \beq
        \dd{G}{t} + G \dd{G}{z} = 0
    \eeq

The standard multivariate Wigner distribution corresponds to a
gaussian $M$ matrix model with action $S_0(A) = \half \sum_{i=1}^M
A_{i}^2$ in the large-$N$ limit
    \beqs
        \Gamma_{I} &=& \lim_{N \to \infty} \int dA e^{-\half N \delta^{ij} \tr A_i
            A_j} {dA \over Z_N} \cr
        Z_N &=& \int dA e^{-\half N \delta^{ij} \tr A_i  A_j}
    \eeqs
The odd moments vanish as before and the second moment is
    \beq
        \Gamma_{ij} = \delta_{ij}
    \eeq
The factorized loop equations (\ref{e-fSD}) are again recursion
relations
    \beq
        \Gamma_{jI} = \Gamma_{ji} \delta_I^{I_1 i I_2}
        \Gamma_{I_1}\Gamma_{I_2}
    \eeq
involving the order-preserving partition of the original string
$I$ into sub-strings. By iterating, we may reduce the evaluation
of an arbitrary moment to a non-crossing sum over products of the
second moment
    \beq
        \Gamma_{i_1 \cdots i_n} = \sum_{\pi \in {\rm NCP}_2(i_1 \cdots i_n)}
        \prod_{\{i_a, i_b\} \in \pi} \Gamma_{i_a i_b}
    \eeq
This is Wick's theorem for Feynman diagrams of planar topology.
NCP$_2(i_1, \cdots ,i_n)$ is the set of non-crossing partitions of
the cyclically symmetric set of indices $(i_1 \cdots i_n)$ into
pairs. For example
    \beqs
    \Gamma_{i j} &=& \delta_{i j} \cr
    \Gamma_{i j k l} &=& \Gamma_{ij}\Gamma_{kl} + \Gamma_{il}
        \Gamma_{jk} \cr
    \Gamma_{ijklmn} &=& \Gamma_{ij}\Gamma_{kl}\Gamma_{mn} + \Gamma_{ni}\Gamma_{jk} \Gamma_{lm}
        + \Gamma_{il}\Gamma_{jk}\Gamma_{mn} + \Gamma_{ij}\Gamma_{kn}\Gamma_{lm}
        + \Gamma_{jm}\Gamma_{ni}\Gamma_{kl}
    \eeqs
The Wigner distribution serves as a candidate for a reference
distribution from which other distributions may be obtained by a
change of variable. For example, a Wigner distribution with
positive covariance $G_{ij} = [K^{-1}]_{ij}$ which arises from the
action $S(A) = \half K^{ij} A_i A_j$ can be obtained from the
standard one. Use the invertible linear transformation $A_i
\mapsto \phi_i^j A_j$ where $G_{ij} = \Gamma_{kl} \phi_i^k
\phi_j^l$. By the polar decomposition, the space of Wigner
distributions is the coset space $GL(M) / O(M)$.

We can use a similar change of variable along with our variational
principle to find the best wignerian variational approximation to a
matrix model. It is a kind of mean-field theory, as discussed in \S
\ref{s-var-approx-mat-mod}.

\chapter{Anomaly is a Closed $1$-form}
\label{a-pf-integ-condition}

\thispagestyle{myheadings}

\markboth{}{Appendix \ref{a-pf-integ-condition}. Anomaly is a
Closed $1$-form}


The conditions for $\eta$ to be a closed $1$-form are
(\ref{e-integ-cond-multi-mat})
    \beq
    L^I_i \eta^J_j - L^J_j \eta^I_i = \delta^J_{J_1 i J_2} \eta^{J_1 I J_2}_j -
    \delta^I_{I_1 j I_2} \eta^{I_1 J I_2}_i.
    \eeq
where $\eta^I_i = \delta^I_{I_1 i I_2} G^{I_1} G^{I_2}$. Let us
check that these are satisfied. The RHS of
(\ref{e-integ-cond-multi-mat}) is:
    \beqs
 {\rm RHS} &=& G^{K_1} G^{K_2} \bigg[\delta^J_{J_1 i J_2} \delta^{J_1 I J_2}_{K_1 j K_2}
             -  ({i \leftrightarrow j}, {I \leftrightarrow J}) \bigg] \cr
           &=& G^{K_1} G^{K_2} \bigg[\delta^J_{J_1 i J_2}
               (\delta^{J_1}_{K_1 j P} \delta^{P I J_2}_{K_2}
               + \delta^I_{PjQ} \delta^{J_1 P}_{K_1} \delta^{Q J_2}_{K_2}
               + \delta^{J_2}_{PjK_2} \delta^{J_1 I P}_{K_1})
               -  ({i \leftrightarrow j},{I \leftrightarrow J}) \bigg] \cr
           &=& G^{K_1} G^{K_2} \bigg[ \delta^J_{K_1 j J_1 i J_2} \delta^{J_1 I J_2}_{K_2}
                + \delta^J_{J_1 i J_2 j K_2} \delta^{J_1 I J_2}_{K_1}
                -  ({i \leftrightarrow j},{I \leftrightarrow J}) \bigg]
    \label{e-integ-cond-multi-mat-rhs}
    \eeqs
We cancelled the middle terms using acyclicity of $G^K$. On the
other hand,
    \beqs
    L^I_i \eta^J_j &=& L^I_i \delta^J_{J_1 j J_2} G^{J_1} G^{J_2} \cr
                   &=& \delta^J_{J_1 j J_2} [\delta^{J_1}_{K_1 i K_2} G^{K_1 I K_2} G^{J_2}
                        + \delta^{J_2}_{K_1 i K_2} G^{J_1} G^{K_1 I K_2}] \cr
                   &=& \delta^J_{J_1i J_2 j J_3} G^{J_1 I J_2} G^{J_3} +
                        \delta^{J}_{J_1 j J_2 i J_3} G^{J_1} G^{J_2 I
                        J_3} \cr
                   &=& G^{J_1 I J_2} G^{J_3} [\delta^J_{J_1 i J_2 j J_3} +
                        \delta^J_{J_3 j J_1 i J_2}] \cr
                   &=& G^{K_1} G^{K_2} [\delta^J_{J_1 i J_2 j K_2} \delta^{J_1 I J_2}_{K_1} +
                        \delta^J_{K_1 j J_1 i J_2} \delta^{J_1 i
                        J_2}_{K_2}]
    \eeqs
Thus the LHS of (\ref{e-integ-cond-multi-mat}) is:
    \beqs
     L^I_i \eta^J_j - L^J_j \eta^I_i &=& G^{K_1} G^{K_2} [\delta^J_{J_1 i J_2 j K_2}
                                    \delta^{J_1 I J_2}_{K_1} + \delta^J_{K_1 j J_1 i J_2}
                                    \delta^{J_1 i J_2}_{K_2}
                                    -  ({i \leftrightarrow j},{I \leftrightarrow J})]
    \label{e-integ-cond-multi-mat-lhs}
    \eeqs
Comparing (\ref{e-integ-cond-multi-mat-lhs}) and
(\ref{e-integ-cond-multi-mat-rhs}), we see that the integrability
condition (\ref{e-integ-cond-multi-mat}) is satisfied. Thus,
$\eta^I_i$ is a closed one-form.

\chapter{Group Cohomology}
\label{a-gp-lie-alg-cohomology}

\thispagestyle{myheadings}

\markboth{}{Appendix \ref{a-gp-lie-alg-cohomology}. Group
Cohomology}


It is natural in mathematics to study the cohomology of a group
twisted by a representation \cite{mickelsson,group-cohomology}.
Given a group $G$ and a representation of $G$ on a vector space
$V$ ($V$ is called a $G$ module), we can define a cohomology
theory. We will sometimes call this the cohomology of $G$ valued
in $V$. The $r$-cochains are functions
    \beq
    c : G^r \to V.
    \eeq
The coboundary $d$ is defined on $r$-cochains as
    \beqs
    dc(g_1,g_2,\cdots, g_{r+1}) &=&
    g_1c(g_2,\cdots, g_{r+1}) \cr && + \sum_{s=1}^r(-1)^s c(g_1,g_2,\cdots,
    g_{s-1},g_{s}g_{s+1},g_{s+2},\cdots, g_{r+1}) \cr
    && +(-1)^{r+1} c(g_1,\cdots, g_{r}).
    \eeqs
and satisfies $d^2 = 0$. A closed cochain ($dc = 0$) is called a
cocycle. An exact cochain ($b = dc$ for some $c$) is called a
coboundary. The $r^{\rm th}$ cohomology of $G$ twisted by the
representation of $G$ on $V$, $H^r(G,V)$ is the space of closed
$r$-cochains modulo exact $r$-cochains.

For example, $H^0(G,V)$ is the space of all $G$-invariant elements
of $V$, i.e. the space of $v \in V$ satisfying $g v - v=0$ for all
$g \in G$. A $1$-cocycle is a function $c: G \to V$ that is
closed,
    \beqs
    && dc(g_1,g_2)~= ~ g_1 c(g_2) - c(g_1g_2) + c(g_1) = 0 \cr
    {\rm i.e.} & & c(g_1 g_2)~=~ g_1 c(g_2) + c(g_1).
    \label{e-1cocycle-cond}
    \eeqs
The space of solutions to this equation modulo 1-coboundaries
(which are of the form $b(g) = gv - v {\rm ~for ~some~} v \in V$)
is the first cohomology $H^1(G,V)$. If $G$ acts trivially on $V$,
a cocycle is just a homomorphism from $G$ to the additive group of
$V$: $c(g_1g_2)=c(g_2)+c(g_1)$.

As an example consider the loop group $G = S^1 G' = \{g: S^1
\rightarrow G' \}$ of a Lie group $G'$. And let $V = S^1
\underline{G'}$ be the loop of the Lie algebra $\underline{G}$.
Then there is the adjoint representation of $G$ on $V$: ${\rm
ad}_g v = g v g^{-1}$. A non-trivial 1-cocycle is $c : G
\rightarrow V$, $c(g) = g d g^{-1}$. It satisfies the co-cycle
condition (\ref{e-1cocycle-cond})
    \beq
        c(g h) = gh d(h^{-1} g^{-1}) = g [h dh^{-1}] g^{-1}  + g h
        h^{-1} dg^{-1} = {\rm ad}_g c(h) + c(g).
    \eeq

\chapter{Formula for Entropy}
\label{a-formula-for-entropy}

\thispagestyle{myheadings}

\markboth{}{Appendix \ref{a-formula-for-entropy}. Formula for
Entropy}


Here we show that $\sigma(\phi,A) = \ov{N^2} \log \det J(\phi,A)$
can be expressed in terms of the loop variables $\Phi_I$ and the
coefficients of the automorphism $\phi_i^I$. The jacobian matrix
of the change of variable
    \beqs
        \phi(A)_i &=& [\phi_1]_i^j \tilde \phi(A)_j; ~~~
        \tilde \phi(A)_i = A_i + \sum_{|I| \geq 2} \tilde \phi_i^I
        A_I \cr
    {\rm is ~~~}    J^{~a~jd}_{ib~c}(\phi,A) & =& \dd{\phi(A)^{~a}_{ib}}{A^{~c}_{jd}}
        = [\phi_1]^k_i \dd{[\tilde\phi(A)]_{kb}^{~a}}{A_{jd}^{~c}} \cr
        &=& [\phi_1]^k_i \bigg\{\delta_k^j \delta_c^a \delta_b^d +
            \sum_{m+n \geq 1} \tilde \phi_k^{i_1 \cdots i_m j j_1 \cdots j_n}
            [A_{i_1 \cdots i_m}]^a_c [A_{j_1 \cdots j_n}]^d_b
            \bigg\} \cr
        &=& [\phi_1]^k_i \bigg\{\delta_k^j \delta_c^a \delta_b^d +
        + K^{~ajd}_{kc~b}(A) \bigg\}
    \eeqs
Or suppressing color indices,
    \beq
        J^j_i(\phi,A) = [\phi_1]^k_i \bigg\{ \delta_k^j 1 \otimes
        1  + \sum_{|I|+|J|\geq 1} \tilde \phi_k^{IjJ} A_I \otimes A_J \bigg\}
        \equiv [\phi_1]^k_i \bigg\{ \delta_k^j 1 \otimes
        1  + K^j_k(A) \bigg\}
    \eeq
Then
    \beqs
        \sigma(\phi,A) &=& \ov{N^2} \log \det [\phi_1 (1 + K(A))]
            \cr
        &=& \log \det \phi_1 + \sum_{n \geq 1} {(-1)^{n+1} \over
        n}\ov{N^2} \tr K^n
    \eeqs
So we calculate
    \beqs
        {1\over N^2}\tr K^n(A) &=& {1\over N^2}K^{a_1\ i_2b_2}_{i_1b_1\
        a_2} K^{a_2\ i_3b_3}_{i_2b_2\ a_2}\cdots K^{a_n\ i_1b_1}_{i_nb_n\
            a_1} \cr
        &=&\tilde\phi_{i_1}^{K_1i_2L_1}\tilde\phi_{i_2}^{K_2i_3L_2}\cdots
            \tilde\phi_{i_n}^{K_ni_1L_n} {1\over N}\left[A_{K_1}\right]^{a_1}_{a_2}\cdots
            \left[A_{K_n}\right]^{a_n}_{a_1} {1\over N}\left[A_{L_1}\right]^{b_2}_{b_1}\cdots
            \left[A_{L_n}\right]^{b_1}_{b_n}\cr {1\over N^2}\tr K^n(A)
        &=&\tilde\phi_{i_1}^{K_1i_2L_1}\tilde\phi_{i_2}^{K_2i_3L_2}\cdots
            \tilde\phi_{i_n}^{K_ni_1L_n}\Phi_{K_1\cdots K_n}\Phi_{L_n\cdots L_1}
    \eeqs
Thus
    \beq
        \sigma(\phi,A) = \log \det \phi_1 + \sum_{n \geq 1} {(-1)^{n+1} \over
        n} \tilde\phi_{i_1}^{K_1i_2L_1}\tilde\phi_{i_2}^{K_2i_3L_2}\cdots
            \tilde\phi_{i_n}^{K_ni_1L_n}\Phi_{K_1\cdots K_n}\Phi_{L_n\cdots L_1}
    \eeq



\end{document}